\documentclass[longauth]{aa}

\usepackage{euclid}
\usepackage{amssymb}
\usepackage[fleqn]{amsmath}

\usepackage{times}
\usepackage{graphicx}
\usepackage{epstopdf}
\usepackage[T1]{fontenc}
\usepackage{aecompl} 
\usepackage{multirow}
\usepackage{pdflscape}
\usepackage{rotating}
\usepackage{float}

\usepackage{anyfontsize}

\usepackage{txfonts}
\usepackage{url}

\usepackage[pdfencoding=auto,psdextra]{hyperref}
\hypersetup{
    colorlinks=true,
    linkcolor=blue,
    filecolor=magenta,      
    urlcolor=blue,
    citecolor=blue
}
\usepackage{cleveref}

\usepackage{xcolor}

\makeatletter
\renewcommand*\aa@pageof{, page \thepage{} of \pageref*{LastPage}}
\makeatother

\usepackage[utf8]{inputenc}

\usepackage[switch, modulo]{lineno}

\def\gs{\mathrel{\lower0.6ex\hbox{$\buildrel {\textstyle >}\over{\scriptstyle \sim}$}}}
\def\ls{\mathrel{\lower0.6ex\hbox{$\buildrel {\textstyle <}\over{\scriptstyle \sim}$}}}
\newcommand{\simgt}{\lower.5ex\hbox{$\; \buildrel > \over \sim \;$}}
\newcommand{\simlt}{\lower.5ex\hbox{$\; \buildrel < \over \sim \;$}}

\begin{document}

\title{\Euclid preparation}
\subtitle{XLII. A unified catalogue-level reanalysis of weak lensing by galaxy clusters in five imaging surveys}    

\newcommand{\orcid}[1]{} 
\author{Euclid Collaboration: M.~Sereno\orcid{0000-0003-0302-0325}\thanks{\email{mauro.sereno@inaf.it}}\inst{\ref{aff1},\ref{aff2}}
\and S.~Farrens\orcid{0000-0002-9594-9387}\inst{\ref{aff3}}
\and L.~Ingoglia\orcid{0000-0002-7587-0997}\inst{\ref{aff4}}
\and G.~F.~Lesci\orcid{0000-0002-4607-2830}\inst{\ref{aff4},\ref{aff1}}
\and L.~Baumont\orcid{0000-0002-1518-0150}\inst{\ref{aff5}}
\and G.~Covone\orcid{0000-0002-2553-096X}\inst{\ref{aff6},\ref{aff7},\ref{aff8}}
\and C.~Giocoli\inst{\ref{aff1},\ref{aff2}}
\and F.~Marulli\orcid{0000-0002-8850-0303}\inst{\ref{aff4},\ref{aff1},\ref{aff2}}
\and S.~Miranda~La~Hera\inst{\ref{aff3}}
\and M.~Vannier\inst{\ref{aff9}}
\and A.~Biviano\inst{\ref{aff10},\ref{aff11}}
\and S.~Maurogordato\inst{\ref{aff9}}
\and L.~Moscardini\orcid{0000-0002-3473-6716}\inst{\ref{aff4},\ref{aff1},\ref{aff2}}
\and N.~Aghanim\inst{\ref{aff12}}
\and S.~Andreon\orcid{0000-0002-2041-8784}\inst{\ref{aff13}}
\and N.~Auricchio\orcid{0000-0003-4444-8651}\inst{\ref{aff1}}
\and M.~Baldi\orcid{0000-0003-4145-1943}\inst{\ref{aff4},\ref{aff1},\ref{aff2}}
\and S.~Bardelli\orcid{0000-0002-8900-0298}\inst{\ref{aff1}}
\and F.~Bellagamba\inst{\ref{aff14},\ref{aff1}}
\and C.~Bodendorf\inst{\ref{aff15}}
\and D.~Bonino\inst{\ref{aff16}}
\and E.~Branchini\orcid{0000-0002-0808-6908}\inst{\ref{aff17},\ref{aff18}}
\and M.~Brescia\orcid{0000-0001-9506-5680}\inst{\ref{aff6},\ref{aff7}}
\and J.~Brinchmann\inst{\ref{aff19}}
\and S.~Camera\orcid{0000-0003-3399-3574}\inst{\ref{aff20},\ref{aff21},\ref{aff16}}
\and V.~Capobianco\orcid{0000-0002-3309-7692}\inst{\ref{aff16}}
\and C.~Carbone\orcid{0000-0003-0125-3563}\inst{\ref{aff22}}
\and V.~F.~Cardone\inst{\ref{aff23},\ref{aff24}}
\and J.~Carretero\orcid{0000-0002-3130-0204}\inst{\ref{aff25},\ref{aff26}}
\and S.~Casas\orcid{0000-0002-4751-5138}\inst{\ref{aff27}}
\and M.~Castellano\orcid{0000-0001-9875-8263}\inst{\ref{aff23}}
\and S.~Cavuoti\orcid{0000-0002-3787-4196}\inst{\ref{aff7},\ref{aff8}}
\and A.~Cimatti\inst{\ref{aff28}}
\and R.~Cledassou\orcid{0000-0002-8313-2230}\thanks{Deceased}\inst{\ref{aff29},\ref{aff30}}
\and G.~Congedo\orcid{0000-0003-2508-0046}\inst{\ref{aff31}}
\and C.~J.~Conselice\inst{\ref{aff32}}
\and L.~Conversi\orcid{0000-0002-6710-8476}\inst{\ref{aff33},\ref{aff34}}
\and Y.~Copin\orcid{0000-0002-5317-7518}\inst{\ref{aff35}}
\and L.~Corcione\orcid{0000-0002-6497-5881}\inst{\ref{aff16}}
\and F.~Courbin\orcid{0000-0003-0758-6510}\inst{\ref{aff36}}
\and H.~M.~Courtois\orcid{0000-0003-0509-1776}\inst{\ref{aff37}}
\and M.~Cropper\orcid{0000-0003-4571-9468}\inst{\ref{aff38}}
\and A.~Da~Silva\orcid{0000-0002-6385-1609}\inst{\ref{aff39},\ref{aff40}}
\and H.~Degaudenzi\orcid{0000-0002-5887-6799}\inst{\ref{aff41}}
\and A.~M.~Di~Giorgio\orcid{0000-0002-4767-2360}\inst{\ref{aff42}}
\and J.~Dinis\inst{\ref{aff40},\ref{aff39}}
\and F.~Dubath\orcid{0000-0002-6533-2810}\inst{\ref{aff41}}
\and C.~A.~J.~Duncan\inst{\ref{aff43},\ref{aff32}}
\and X.~Dupac\inst{\ref{aff34}}
\and S.~Dusini\orcid{0000-0002-1128-0664}\inst{\ref{aff44}}
\and M.~Farina\inst{\ref{aff42}}
\and S.~Ferriol\inst{\ref{aff35}}
\and M.~Frailis\orcid{0000-0002-7400-2135}\inst{\ref{aff10}}
\and E.~Franceschi\orcid{0000-0002-0585-6591}\inst{\ref{aff1}}
\and M.~Fumana\orcid{0000-0001-6787-5950}\inst{\ref{aff22}}
\and S.~Galeotta\orcid{0000-0002-3748-5115}\inst{\ref{aff10}}
\and B.~Garilli\orcid{0000-0001-7455-8750}\inst{\ref{aff22}}
\and B.~Gillis\orcid{0000-0002-4478-1270}\inst{\ref{aff31}}
\and A.~Grazian\orcid{0000-0002-5688-0663}\inst{\ref{aff45}}
\and F.~Grupp\inst{\ref{aff15},\ref{aff46}}
\and S.~V.~H.~Haugan\orcid{0000-0001-9648-7260}\inst{\ref{aff47}}
\and W.~Holmes\inst{\ref{aff48}}
\and I.~Hook\orcid{0000-0002-2960-978X}\inst{\ref{aff49}}
\and F.~Hormuth\inst{\ref{aff50}}
\and A.~Hornstrup\orcid{0000-0002-3363-0936}\inst{\ref{aff51},\ref{aff52}}
\and P.~Hudelot\inst{\ref{aff53}}
\and K.~Jahnke\orcid{0000-0003-3804-2137}\inst{\ref{aff54}}
\and B.~Joachimi\orcid{0000-0001-7494-1303}\inst{\ref{aff55}}
\and E.~Keih\"anen\orcid{0000-0003-1804-7715}\inst{\ref{aff56}}
\and S.~Kermiche\orcid{0000-0002-0302-5735}\inst{\ref{aff57}}
\and A.~Kiessling\orcid{0000-0002-2590-1273}\inst{\ref{aff48}}
\and B.~Kubik\inst{\ref{aff35}}
\and M.~Kunz\orcid{0000-0002-3052-7394}\inst{\ref{aff58}}
\and H.~Kurki-Suonio\orcid{0000-0002-4618-3063}\inst{\ref{aff59},\ref{aff60}}
\and S.~Ligori\orcid{0000-0003-4172-4606}\inst{\ref{aff16}}
\and P.~B.~Lilje\orcid{0000-0003-4324-7794}\inst{\ref{aff47}}
\and V.~Lindholm\orcid{0000-0003-2317-5471}\inst{\ref{aff59},\ref{aff60}}
\and I.~Lloro\inst{\ref{aff61}}
\and D.~Maino\inst{\ref{aff62},\ref{aff22},\ref{aff63}}
\and E.~Maiorano\orcid{0000-0003-2593-4355}\inst{\ref{aff1}}
\and O.~Mansutti\orcid{0000-0001-5758-4658}\inst{\ref{aff10}}
\and O.~Marggraf\orcid{0000-0001-7242-3852}\inst{\ref{aff64}}
\and K.~Markovic\orcid{0000-0001-6764-073X}\inst{\ref{aff48}}
\and N.~Martinet\orcid{0000-0003-2786-7790}\inst{\ref{aff65}}
\and R.~Massey\orcid{0000-0002-6085-3780}\inst{\ref{aff66}}
\and E.~Medinaceli\orcid{0000-0002-4040-7783}\inst{\ref{aff1}}
\and S.~Mei\orcid{0000-0002-2849-559X}\inst{\ref{aff67}}
\and Y.~Mellier\inst{\ref{aff68},\ref{aff53}}
\and M.~Meneghetti\orcid{0000-0003-1225-7084}\inst{\ref{aff1},\ref{aff2}}
\and E.~Merlin\orcid{0000-0001-6870-8900}\inst{\ref{aff23}}
\and G.~Meylan\inst{\ref{aff36}}
\and M.~Moresco\orcid{0000-0002-7616-7136}\inst{\ref{aff4},\ref{aff1}}
\and E.~Munari\orcid{0000-0002-1751-5946}\inst{\ref{aff10}}
\and S.-M.~Niemi\inst{\ref{aff69}}
\and T.~Nutma\inst{\ref{aff70},\ref{aff71}}
\and C.~Padilla\orcid{0000-0001-7951-0166}\inst{\ref{aff25}}
\and S.~Paltani\inst{\ref{aff41}}
\and F.~Pasian\inst{\ref{aff10}}
\and K.~Pedersen\inst{\ref{aff72}}
\and V.~Pettorino\inst{\ref{aff73}}
\and S.~Pires\inst{\ref{aff3}}
\and G.~Polenta\orcid{0000-0003-4067-9196}\inst{\ref{aff74}}
\and M.~Poncet\inst{\ref{aff29}}
\and L.~A.~Popa\inst{\ref{aff75}}
\and F.~Raison\orcid{0000-0002-7819-6918}\inst{\ref{aff15}}
\and R.~Rebolo\inst{\ref{aff76},\ref{aff77}}
\and A.~Renzi\orcid{0000-0001-9856-1970}\inst{\ref{aff78},\ref{aff44}}
\and J.~Rhodes\inst{\ref{aff48}}
\and G.~Riccio\inst{\ref{aff7}}
\and E.~Romelli\orcid{0000-0003-3069-9222}\inst{\ref{aff10}}
\and M.~Roncarelli\orcid{0000-0001-9587-7822}\inst{\ref{aff1}}
\and E.~Rossetti\inst{\ref{aff14}}
\and R.~Saglia\orcid{0000-0003-0378-7032}\inst{\ref{aff46},\ref{aff15}}
\and D.~Sapone\orcid{0000-0001-7089-4503}\inst{\ref{aff79}}
\and B.~Sartoris\inst{\ref{aff46},\ref{aff10}}
\and M.~Schirmer\orcid{0000-0003-2568-9994}\inst{\ref{aff54}}
\and P.~Schneider\orcid{0000-0001-8561-2679}\inst{\ref{aff64}}
\and T.~Schrabback\orcid{0000-0002-6987-7834}\inst{\ref{aff80}}
\and A.~Secroun\orcid{0000-0003-0505-3710}\inst{\ref{aff57}}
\and G.~Seidel\orcid{0000-0003-2907-353X}\inst{\ref{aff54}}
\and S.~Serrano\orcid{0000-0002-0211-2861}\inst{\ref{aff81},\ref{aff82}}
\and C.~Sirignano\orcid{0000-0002-0995-7146}\inst{\ref{aff78},\ref{aff44}}
\and G.~Sirri\orcid{0000-0003-2626-2853}\inst{\ref{aff2}}
\and L.~Stanco\orcid{0000-0002-9706-5104}\inst{\ref{aff44}}
\and J.-L.~Starck\orcid{0000-0003-2177-7794}\inst{\ref{aff5}}
\and P.~Tallada-Cresp\'{i}\orcid{0000-0002-1336-8328}\inst{\ref{aff83},\ref{aff26}}
\and A.~N.~Taylor\inst{\ref{aff31}}
\and I.~Tereno\inst{\ref{aff39},\ref{aff84}}
\and R.~Toledo-Moreo\orcid{0000-0002-2997-4859}\inst{\ref{aff85}}
\and F.~Torradeflot\orcid{0000-0003-1160-1517}\inst{\ref{aff26},\ref{aff83}}
\and I.~Tutusaus\orcid{0000-0002-3199-0399}\inst{\ref{aff86}}
\and E.~A.~Valentijn\inst{\ref{aff70}}
\and L.~Valenziano\orcid{0000-0002-1170-0104}\inst{\ref{aff1},\ref{aff87}}
\and T.~Vassallo\orcid{0000-0001-6512-6358}\inst{\ref{aff10},\ref{aff46}}
\and A.~Veropalumbo\orcid{0000-0003-2387-1194}\inst{\ref{aff13}}
\and Y.~Wang\orcid{0000-0002-4749-2984}\inst{\ref{aff88}}
\and J.~Weller\orcid{0000-0002-8282-2010}\inst{\ref{aff46},\ref{aff15}}
\and A.~Zacchei\orcid{0000-0003-0396-1192}\inst{\ref{aff10},\ref{aff11}}
\and G.~Zamorani\orcid{0000-0002-2318-301X}\inst{\ref{aff1}}
\and J.~Zoubian\inst{\ref{aff57}}
\and E.~Zucca\orcid{0000-0002-5845-8132}\inst{\ref{aff1}}
\and A.~Boucaud\orcid{0000-0001-7387-2633}\inst{\ref{aff67}}
\and E.~Bozzo\orcid{0000-0002-8201-1525}\inst{\ref{aff41}}
\and C.~Cerna\inst{\ref{aff53},\ref{aff89}}
\and C.~Colodro-Conde\inst{\ref{aff76}}
\and D.~Di~Ferdinando\inst{\ref{aff2}}
\and R.~Farinelli\inst{\ref{aff1}}
\and H.~Israel\orcid{0000-0002-3045-4412}\inst{\ref{aff90}}
\and N.~Mauri\orcid{0000-0001-8196-1548}\inst{\ref{aff28},\ref{aff2}}
\and C.~Neissner\inst{\ref{aff25},\ref{aff26}}
\and V.~Scottez\inst{\ref{aff68},\ref{aff91}}
\and M.~Tenti\orcid{0000-0002-4254-5901}\inst{\ref{aff87}}
\and M.~Wiesmann\inst{\ref{aff47}}
\and Y.~Akrami\orcid{0000-0002-2407-7956}\inst{\ref{aff92},\ref{aff93},\ref{aff94},\ref{aff95},\ref{aff96}}
\and V.~Allevato\orcid{0000-0001-7232-5152}\inst{\ref{aff7},\ref{aff97}}
\and C.~Baccigalupi\orcid{0000-0002-8211-1630}\inst{\ref{aff98},\ref{aff11},\ref{aff10},\ref{aff99}}
\and M.~Ballardini\orcid{0000-0003-4481-3559}\inst{\ref{aff100},\ref{aff101},\ref{aff1}}
\and D.~Benielli\inst{\ref{aff57}}
\and S.~Borgani\orcid{0000-0001-6151-6439}\inst{\ref{aff10},\ref{aff102},\ref{aff99},\ref{aff11}}
\and A.~S.~Borlaff\orcid{0000-0003-3249-4431}\inst{\ref{aff103},\ref{aff104}}
\and C.~Burigana\orcid{0000-0002-3005-5796}\inst{\ref{aff105},\ref{aff87}}
\and R.~Cabanac\orcid{0000-0001-6679-2600}\inst{\ref{aff86}}
\and A.~Cappi\inst{\ref{aff1},\ref{aff9}}
\and C.~S.~Carvalho\inst{\ref{aff84}}
\and G.~Castignani\orcid{0000-0001-6831-0687}\inst{\ref{aff4},\ref{aff1}}
\and T.~Castro\orcid{0000-0002-6292-3228}\inst{\ref{aff10},\ref{aff99},\ref{aff11}}
\and G.~Ca\~{n}as-Herrera\orcid{0000-0003-2796-2149}\inst{\ref{aff69},\ref{aff106}}
\and K.~C.~Chambers\orcid{0000-0001-6965-7789}\inst{\ref{aff107}}
\and A.~R.~Cooray\orcid{0000-0002-3892-0190}\inst{\ref{aff108}}
\and J.~Coupon\inst{\ref{aff41}}
\and S.~Davini\inst{\ref{aff18}}
\and G.~De~Lucia\orcid{0000-0002-6220-9104}\inst{\ref{aff10}}
\and G.~Desprez\inst{\ref{aff109}}
\and S.~Di~Domizio\orcid{0000-0003-2863-5895}\inst{\ref{aff110}}
\and H.~Dole\orcid{0000-0002-9767-3839}\inst{\ref{aff12}}
\and J.~A.~Escartin~Vigo\inst{\ref{aff15}}
\and S.~Escoffier\orcid{0000-0002-2847-7498}\inst{\ref{aff57}}
\and I.~Ferrero\orcid{0000-0002-1295-1132}\inst{\ref{aff47}}
\and L.~Gabarra\inst{\ref{aff78},\ref{aff44}}
\and E.~Gaztanaga\orcid{0000-0001-9632-0815}\inst{\ref{aff111},\ref{aff81},\ref{aff112}}
\and K.~George\orcid{0000-0002-1734-8455}\inst{\ref{aff46}}
\and F.~Giacomini\orcid{0000-0002-3129-2814}\inst{\ref{aff2}}
\and G.~Gozaliasl\orcid{0000-0002-0236-919X}\inst{\ref{aff59}}
\and H.~Hildebrandt\orcid{0000-0002-9814-3338}\inst{\ref{aff113}}
\and J.~J.~E.~Kajava\orcid{0000-0002-3010-8333}\inst{\ref{aff114}}
\and V.~Kansal\inst{\ref{aff5}}
\and C.~C.~Kirkpatrick\inst{\ref{aff56}}
\and L.~Legrand\orcid{0000-0003-0610-5252}\inst{\ref{aff58}}
\and P.~Liebing\inst{\ref{aff38}}
\and A.~Loureiro\orcid{0000-0002-4371-0876}\inst{\ref{aff115},\ref{aff96}}
\and J.~Macias-Perez\inst{\ref{aff116}}
\and M.~Magliocchetti\orcid{0000-0001-9158-4838}\inst{\ref{aff42}}
\and G.~Mainetti\inst{\ref{aff117}}
\and R.~Maoli\orcid{0000-0002-6065-3025}\inst{\ref{aff118},\ref{aff23}}
\and M.~Martinelli\orcid{0000-0002-6943-7732}\inst{\ref{aff23},\ref{aff24}}
\and C.~J.~A.~P.~Martins\orcid{0000-0002-4886-9261}\inst{\ref{aff119},\ref{aff19}}
\and S.~Z.~Matthew\inst{\ref{aff31}}
\and M.~Maturi\inst{\ref{aff120},\ref{aff121}}
\and L.~Maurin\orcid{0000-0002-8406-0857}\inst{\ref{aff12}}
\and R.~B.~Metcalf\orcid{0000-0003-3167-2574}\inst{\ref{aff4}}
\and P.~Monaco\inst{\ref{aff102},\ref{aff10},\ref{aff99},\ref{aff11}}
\and G.~Morgante\inst{\ref{aff1}}
\and S.~Nadathur\inst{\ref{aff112}}
\and A.~A.~Nucita\inst{\ref{aff122},\ref{aff123},\ref{aff124}}
\and L.~Patrizii\inst{\ref{aff2}}
\and A.~Peel\orcid{0000-0003-0488-8978}\inst{\ref{aff36}}
\and M.~P{\"o}ntinen\orcid{0000-0001-5442-2530}\inst{\ref{aff59}}
\and V.~Popa\inst{\ref{aff75}}
\and C.~Porciani\orcid{0000-0002-7797-2508}\inst{\ref{aff64}}
\and D.~Potter\orcid{0000-0002-0757-5195}\inst{\ref{aff125}}
\and P.~Reimberg\orcid{0000-0003-3410-0280}\inst{\ref{aff68}}
\and Z.~Sakr\orcid{0000-0002-4823-3757}\inst{\ref{aff86},\ref{aff126},\ref{aff120}}
\and A.~G.~S\'anchez\orcid{0000-0003-1198-831X}\inst{\ref{aff15}}
\and A.~Schneider\orcid{0000-0001-7055-8104}\inst{\ref{aff125}}
\and E.~Sefusatti\orcid{0000-0003-0473-1567}\inst{\ref{aff10},\ref{aff11},\ref{aff99}}
\and P.~Simon\inst{\ref{aff64}}
\and A.~Spurio~Mancini\orcid{0000-0001-5698-0990}\inst{\ref{aff38}}
\and J.~Stadel\orcid{0000-0001-7565-8622}\inst{\ref{aff125}}
\and S.~A.~Stanford\orcid{0000-0003-0122-0841}\inst{\ref{aff127}}
\and J.~Steinwagner\inst{\ref{aff15}}
\and R.~Teyssier\orcid{0000-0001-7689-0933}\inst{\ref{aff128}}
\and J.~Valiviita\orcid{0000-0001-6225-3693}\inst{\ref{aff59},\ref{aff60}}
\and M.~Viel\orcid{0000-0002-2642-5707}\inst{\ref{aff98},\ref{aff11},\ref{aff10},\ref{aff99}}}
										   
\institute{INAF-Osservatorio di Astrofisica e Scienza dello Spazio di Bologna, Via Piero Gobetti 93/3, 40129 Bologna, Italy\label{aff1}
\and
INFN-Sezione di Bologna, Viale Berti Pichat 6/2, 40127 Bologna, Italy\label{aff2}
\and
Universit\'e Paris-Saclay, Universit\'e Paris Cit\'e, CEA, CNRS, AIM, 91191, Gif-sur-Yvette, France\label{aff3}
\and
Dipartimento di Fisica e Astronomia "Augusto Righi" - Alma Mater Studiorum Universit\`a di Bologna, via Piero Gobetti 93/2, 40129 Bologna, Italy\label{aff4}
\and
AIM, CEA, CNRS, Universit\'{e} Paris-Saclay, Universit\'{e} de Paris, 91191 Gif-sur-Yvette, France\label{aff5}
\and
Department of Physics "E. Pancini", University Federico II, Via Cinthia 6, 80126, Napoli, Italy\label{aff6}
\and
INAF-Osservatorio Astronomico di Capodimonte, Via Moiariello 16, 80131 Napoli, Italy\label{aff7}
\and
INFN section of Naples, Via Cinthia 6, 80126, Napoli, Italy\label{aff8}
\and
Universit\'e C\^{o}te d'Azur, Observatoire de la C\^{o}te d'Azur, CNRS, Laboratoire Lagrange, Bd de l'Observatoire, CS 34229, 06304 Nice cedex 4, France\label{aff9}
\and
INAF-Osservatorio Astronomico di Trieste, Via G. B. Tiepolo 11, 34143 Trieste, Italy\label{aff10}
\and
IFPU, Institute for Fundamental Physics of the Universe, via Beirut 2, 34151 Trieste, Italy\label{aff11}
\and
Universit\'e Paris-Saclay, CNRS, Institut d'astrophysique spatiale, 91405, Orsay, France\label{aff12}
\and
INAF-Osservatorio Astronomico di Brera, Via Brera 28, 20122 Milano, Italy\label{aff13}
\and
Dipartimento di Fisica e Astronomia, Universit\`a di Bologna, Via Gobetti 93/2, 40129 Bologna, Italy\label{aff14}
\and
Max Planck Institute for Extraterrestrial Physics, Giessenbachstr. 1, 85748 Garching, Germany\label{aff15}
\and
INAF-Osservatorio Astrofisico di Torino, Via Osservatorio 20, 10025 Pino Torinese (TO), Italy\label{aff16}
\and
Dipartimento di Fisica, Universit\`a di Genova, Via Dodecaneso 33, 16146, Genova, Italy\label{aff17}
\and
INFN-Sezione di Genova, Via Dodecaneso 33, 16146, Genova, Italy\label{aff18}
\and
Instituto de Astrof\'isica e Ci\^encias do Espa\c{c}o, Universidade do Porto, CAUP, Rua das Estrelas, PT4150-762 Porto, Portugal\label{aff19}
\and
Dipartimento di Fisica, Universit\`a degli Studi di Torino, Via P. Giuria 1, 10125 Torino, Italy\label{aff20}
\and
INFN-Sezione di Torino, Via P. Giuria 1, 10125 Torino, Italy\label{aff21}
\and
INAF-IASF Milano, Via Alfonso Corti 12, 20133 Milano, Italy\label{aff22}
\and
INAF-Osservatorio Astronomico di Roma, Via Frascati 33, 00078 Monteporzio Catone, Italy\label{aff23}
\and
INFN-Sezione di Roma, Piazzale Aldo Moro, 2 - c/o Dipartimento di Fisica, Edificio G. Marconi, 00185 Roma, Italy\label{aff24}
\and
Institut de F\'{i}sica d'Altes Energies (IFAE), The Barcelona Institute of Science and Technology, Campus UAB, 08193 Bellaterra (Barcelona), Spain\label{aff25}
\and
Port d'Informaci\'{o} Cient\'{i}fica, Campus UAB, C. Albareda s/n, 08193 Bellaterra (Barcelona), Spain\label{aff26}
\and
Institute for Theoretical Particle Physics and Cosmology (TTK), RWTH Aachen University, 52056 Aachen, Germany\label{aff27}
\and
Dipartimento di Fisica e Astronomia "Augusto Righi" - Alma Mater Studiorum Universit\`a di Bologna, Viale Berti Pichat 6/2, 40127 Bologna, Italy\label{aff28}
\and
Centre National d'Etudes Spatiales -- Centre spatial de Toulouse, 18 avenue Edouard Belin, 31401 Toulouse Cedex 9, France\label{aff29}
\and
Institut national de physique nucl\'eaire et de physique des particules, 3 rue Michel-Ange, 75794 Paris C\'edex 16, France\label{aff30}
\and
Institute for Astronomy, University of Edinburgh, Royal Observatory, Blackford Hill, Edinburgh EH9 3HJ, UK\label{aff31}
\and
Jodrell Bank Centre for Astrophysics, Department of Physics and Astronomy, University of Manchester, Oxford Road, Manchester M13 9PL, UK\label{aff32}
\and
European Space Agency/ESRIN, Largo Galileo Galilei 1, 00044 Frascati, Roma, Italy\label{aff33}
\and
ESAC/ESA, Camino Bajo del Castillo, s/n., Urb. Villafranca del Castillo, 28692 Villanueva de la Ca\~nada, Madrid, Spain\label{aff34}
\and
Universit\'e Claude Bernard Lyon 1, CNRS/IN2P3, IP2I Lyon, UMR 5822, Villeurbanne, F-69100, France\label{aff35}
\and
Institute of Physics, Laboratory of Astrophysics, Ecole Polytechnique F\'ed\'erale de Lausanne (EPFL), Observatoire de Sauverny, 1290 Versoix, Switzerland\label{aff36}
\and
UCB Lyon 1, CNRS/IN2P3, IUF, IP2I Lyon, 4 rue Enrico Fermi, 69622 Villeurbanne, France\label{aff37}
\and
Mullard Space Science Laboratory, University College London, Holmbury St Mary, Dorking, Surrey RH5 6NT, UK\label{aff38}
\and
Departamento de F\'isica, Faculdade de Ci\^encias, Universidade de Lisboa, Edif\'icio C8, Campo Grande, PT1749-016 Lisboa, Portugal\label{aff39}
\and
Instituto de Astrof\'isica e Ci\^encias do Espa\c{c}o, Faculdade de Ci\^encias, Universidade de Lisboa, Campo Grande, 1749-016 Lisboa, Portugal\label{aff40}
\and
Department of Astronomy, University of Geneva, ch. d'Ecogia 16, 1290 Versoix, Switzerland\label{aff41}
\and
INAF-Istituto di Astrofisica e Planetologia Spaziali, via del Fosso del Cavaliere, 100, 00100 Roma, Italy\label{aff42}
\and
Department of Physics, Oxford University, Keble Road, Oxford OX1 3RH, UK\label{aff43}
\and
INFN-Padova, Via Marzolo 8, 35131 Padova, Italy\label{aff44}
\and
INAF-Osservatorio Astronomico di Padova, Via dell'Osservatorio 5, 35122 Padova, Italy\label{aff45}
\and
Universit\"ats-Sternwarte M\"unchen, Fakult\"at f\"ur Physik, Ludwig-Maximilians-Universit\"at M\"unchen, Scheinerstrasse 1, 81679 M\"unchen, Germany\label{aff46}
\and
Institute of Theoretical Astrophysics, University of Oslo, P.O. Box 1029 Blindern, 0315 Oslo, Norway\label{aff47}
\and
Jet Propulsion Laboratory, California Institute of Technology, 4800 Oak Grove Drive, Pasadena, CA, 91109, USA\label{aff48}
\and
Department of Physics, Lancaster University, Lancaster, LA1 4YB, UK\label{aff49}
\and
von Hoerner \& Sulger GmbH, Schlossplatz 8, 68723 Schwetzingen, Germany\label{aff50}
\and
Technical University of Denmark, Elektrovej 327, 2800 Kgs. Lyngby, Denmark\label{aff51}
\and
Cosmic Dawn Center (DAWN), Denmark\label{aff52}
\and
Institut d'Astrophysique de Paris, UMR 7095, CNRS, and Sorbonne Universit\'e, 98 bis boulevard Arago, 75014 Paris, France\label{aff53}
\and
Max-Planck-Institut f\"ur Astronomie, K\"onigstuhl 17, 69117 Heidelberg, Germany\label{aff54}
\and
Department of Physics and Astronomy, University College London, Gower Street, London WC1E 6BT, UK\label{aff55}
\and
Department of Physics and Helsinki Institute of Physics, Gustaf H\"allstr\"omin katu 2, 00014 University of Helsinki, Finland\label{aff56}
\and
Aix-Marseille Universit\'e, CNRS/IN2P3, CPPM, Marseille, France\label{aff57}
\and
Universit\'e de Gen\`eve, D\'epartement de Physique Th\'eorique and Centre for Astroparticle Physics, 24 quai Ernest-Ansermet, CH-1211 Gen\`eve 4, Switzerland\label{aff58}
\and
Department of Physics, P.O. Box 64, 00014 University of Helsinki, Finland\label{aff59}
\and
Helsinki Institute of Physics, Gustaf H{\"a}llstr{\"o}min katu 2, University of Helsinki, Helsinki, Finland\label{aff60}
\and
NOVA optical infrared instrumentation group at ASTRON, Oude Hoogeveensedijk 4, 7991PD, Dwingeloo, The Netherlands\label{aff61}
\and
Dipartimento di Fisica "Aldo Pontremoli", Universit\`a degli Studi di Milano, Via Celoria 16, 20133 Milano, Italy\label{aff62}
\and
INFN-Sezione di Milano, Via Celoria 16, 20133 Milano, Italy\label{aff63}
\and
Universit\"at Bonn, Argelander-Institut f\"ur Astronomie, Auf dem H\"ugel 71, 53121 Bonn, Germany\label{aff64}
\and
Aix-Marseille Universit\'e, CNRS, CNES, LAM, Marseille, France\label{aff65}
\and
Department of Physics, Institute for Computational Cosmology, Durham University, South Road, DH1 3LE, UK\label{aff66}
\and
Universit\'e Paris Cit\'e, CNRS, Astroparticule et Cosmologie, 75013 Paris, France\label{aff67}
\and
Institut d'Astrophysique de Paris, 98bis Boulevard Arago, 75014, Paris, France\label{aff68}
\and
European Space Agency/ESTEC, Keplerlaan 1, 2201 AZ Noordwijk, The Netherlands\label{aff69}
\and
Kapteyn Astronomical Institute, University of Groningen, PO Box 800, 9700 AV Groningen, The Netherlands\label{aff70}
\and
Leiden Observatory, Leiden University, Einsteinweg 55, 2333 CC Leiden, The Netherlands\label{aff71}
\and
Department of Physics and Astronomy, University of Aarhus, Ny Munkegade 120, DK-8000 Aarhus C, Denmark\label{aff72}
\and
Universit\'e Paris-Saclay, Universit\'e Paris Cit\'e, CEA, CNRS, Astrophysique, Instrumentation et Mod\'elisation Paris-Saclay, 91191 Gif-sur-Yvette, France\label{aff73}
\and
Space Science Data Center, Italian Space Agency, via del Politecnico snc, 00133 Roma, Italy\label{aff74}
\and
Institute of Space Science, Str. Atomistilor, nr. 409 M\u{a}gurele, Ilfov, 077125, Romania\label{aff75}
\and
Instituto de Astrof\'isica de Canarias, Calle V\'ia L\'actea s/n, 38204, San Crist\'obal de La Laguna, Tenerife, Spain\label{aff76}
\and
Departamento de Astrof\'isica, Universidad de La Laguna, 38206, La Laguna, Tenerife, Spain\label{aff77}
\and
Dipartimento di Fisica e Astronomia "G. Galilei", Universit\`a di Padova, Via Marzolo 8, 35131 Padova, Italy\label{aff78}
\and
Departamento de F\'isica, FCFM, Universidad de Chile, Blanco Encalada 2008, Santiago, Chile\label{aff79}
\and
Universit\"at Innsbruck, Institut f\"ur Astro- und Teilchenphysik, Technikerstr. 25/8, 6020 Innsbruck, Austria\label{aff80}
\and
Institut d'Estudis Espacials de Catalunya (IEEC),  Edifici RDIT, Campus UPC, 08860 Castelldefels, Barcelona, Spain\label{aff81}
\and
Institut de Ciencies de l'Espai (IEEC-CSIC), Campus UAB, Carrer de Can Magrans, s/n Cerdanyola del Vall\'es, 08193 Barcelona, Spain\label{aff82}
\and
Centro de Investigaciones Energ\'eticas, Medioambientales y Tecnol\'ogicas (CIEMAT), Avenida Complutense 40, 28040 Madrid, Spain\label{aff83}
\and
Instituto de Astrof\'isica e Ci\^encias do Espa\c{c}o, Faculdade de Ci\^encias, Universidade de Lisboa, Tapada da Ajuda, 1349-018 Lisboa, Portugal\label{aff84}
\and
Universidad Polit\'ecnica de Cartagena, Departamento de Electr\'onica y Tecnolog\'ia de Computadoras,  Plaza del Hospital 1, 30202 Cartagena, Spain\label{aff85}
\and
Institut de Recherche en Astrophysique et Plan\'etologie (IRAP), Universit\'e de Toulouse, CNRS, UPS, CNES, 14 Av. Edouard Belin, 31400 Toulouse, France\label{aff86}
\and
INFN-Bologna, Via Irnerio 46, 40126 Bologna, Italy\label{aff87}
\and
Infrared Processing and Analysis Center, California Institute of Technology, Pasadena, CA 91125, USA\label{aff88}
\and
CEA Saclay, DFR/IRFU, Service d'Astrophysique, Bat. 709, 91191 Gif-sur-Yvette, France\label{aff89}
\and
Ernst-Reuter-Str. 4e, 31224 Peine, Germany\label{aff90}
\and
Junia, EPA department, 41 Bd Vauban, 59800 Lille, France\label{aff91}
\and
Instituto de F\'isica Te\'orica UAM-CSIC, Campus de Cantoblanco, 28049 Madrid, Spain\label{aff92}
\and
CERCA/ISO, Department of Physics, Case Western Reserve University, 10900 Euclid Avenue, Cleveland, OH 44106, USA\label{aff93}
\and
Laboratoire de Physique de l'\'Ecole Normale Sup\'erieure, ENS, Universit\'e PSL, CNRS, Sorbonne Universit\'e, 75005 Paris, France\label{aff94}
\and
Observatoire de Paris, Universit\'e PSL, Sorbonne Universit\'e, LERMA, 750 Paris, France\label{aff95}
\and
Astrophysics Group, Blackett Laboratory, Imperial College London, London SW7 2AZ, UK\label{aff96}
\and
Scuola Normale Superiore, Piazza dei Cavalieri 7, 56126 Pisa, Italy\label{aff97}
\and
SISSA, International School for Advanced Studies, Via Bonomea 265, 34136 Trieste TS, Italy\label{aff98}
\and
INFN, Sezione di Trieste, Via Valerio 2, 34127 Trieste TS, Italy\label{aff99}
\and
Dipartimento di Fisica e Scienze della Terra, Universit\`a degli Studi di Ferrara, Via Giuseppe Saragat 1, 44122 Ferrara, Italy\label{aff100}
\and
Istituto Nazionale di Fisica Nucleare, Sezione di Ferrara, Via Giuseppe Saragat 1, 44122 Ferrara, Italy\label{aff101}
\and
Dipartimento di Fisica - Sezione di Astronomia, Universit\`a di Trieste, Via Tiepolo 11, 34131 Trieste, Italy\label{aff102}
\and
NASA Ames Research Center, Moffett Field, CA 94035, USA\label{aff103}
\and
Kavli Institute for Particle Astrophysics \& Cosmology (KIPAC), Stanford University, Stanford, CA 94305, USA\label{aff104}
\and
INAF, Istituto di Radioastronomia, Via Piero Gobetti 101, 40129 Bologna, Italy\label{aff105}
\and
Institute Lorentz, Leiden University, Niels Bohrweg 2, 2333 CA Leiden, The Netherlands\label{aff106}
\and
Institute for Astronomy, University of Hawaii, 2680 Woodlawn Drive, Honolulu, HI 96822, USA\label{aff107}
\and
Department of Physics \& Astronomy, University of California Irvine, Irvine CA 92697, USA\label{aff108}
\and
Department of Astronomy \& Physics and Institute for Computational Astrophysics, Saint Mary's University, 923 Robie Street, Halifax, Nova Scotia, B3H 3C3, Canada\label{aff109}
\and
Dipartimento di Fisica, Universit\`a degli studi di Genova, and INFN-Sezione di Genova, via Dodecaneso 33, 16146, Genova, Italy\label{aff110}
\and
Institute of Space Sciences (ICE, CSIC), Campus UAB, Carrer de Can Magrans, s/n, 08193 Barcelona, Spain\label{aff111}
\and
Institute of Cosmology and Gravitation, University of Portsmouth, Portsmouth PO1 3FX, UK\label{aff112}
\and
Ruhr University Bochum, Faculty of Physics and Astronomy, Astronomical Institute (AIRUB), German Centre for Cosmological Lensing (GCCL), 44780 Bochum, Germany\label{aff113}
\and
Department of Physics and Astronomy, Vesilinnantie 5, 20014 University of Turku, Finland\label{aff114}
\and
Oskar Klein Centre for Cosmoparticle Physics, Department of Physics, Stockholm University, Stockholm, SE-106 91, Sweden\label{aff115}
\and
Univ. Grenoble Alpes, CNRS, Grenoble INP, LPSC-IN2P3, 53, Avenue des Martyrs, 38000, Grenoble, France\label{aff116}
\and
Centre de Calcul de l'IN2P3/CNRS, 21 avenue Pierre de Coubertin 69627 Villeurbanne Cedex, France\label{aff117}
\and
Dipartimento di Fisica, Sapienza Universit\`a di Roma, Piazzale Aldo Moro 2, 00185 Roma, Italy\label{aff118}
\and
Centro de Astrof\'{\i}sica da Universidade do Porto, Rua das Estrelas, 4150-762 Porto, Portugal\label{aff119}
\and
Institut f\"ur Theoretische Physik, University of Heidelberg, Philosophenweg 16, 69120 Heidelberg, Germany\label{aff120}
\and
Zentrum f\"ur Astronomie, Universit\"at Heidelberg, Philosophenweg 12, 69120 Heidelberg, Germany\label{aff121}
\and
Department of Mathematics and Physics E. De Giorgi, University of Salento, Via per Arnesano, CP-I93, 73100, Lecce, Italy\label{aff122}
\and
INAF-Sezione di Lecce, c/o Dipartimento Matematica e Fisica, Via per Arnesano, 73100, Lecce, Italy\label{aff123}
\and
INFN, Sezione di Lecce, Via per Arnesano, CP-193, 73100, Lecce, Italy\label{aff124}
\and
Department of Astrophysics, University of Zurich, Winterthurerstrasse 190, 8057 Zurich, Switzerland\label{aff125}
\and
Universit\'e St Joseph; Faculty of Sciences, Beirut, Lebanon\label{aff126}
\and
Department of Physics and Astronomy, University of California, Davis, CA 95616, USA\label{aff127}
\and
Department of Astrophysical Sciences, Peyton Hall, Princeton University, Princeton, NJ 08544, USA\label{aff128}}    



\abstract{
Precise and accurate mass calibration is required to exploit galaxy clusters as astrophysical and cosmological probes in the \Euclid era. 
Systematic errors in lensing signals by galaxy clusters can be empirically estimated by comparing different surveys with independent and uncorrelated systematics. To assess the robustness of the lensing results to systematic errors, we carried out end-to-end tests across different data sets. We performed a unified analysis at the catalogue level by leveraging the \Euclid combined cluster and weak-lensing pipeline (\texttt{COMB-CL}).
\texttt{COMB-CL} will measure weak lensing cluster masses for the Euclid Survey. Heterogeneous data sets from five independent, recent, lensing surveys (CHFTLenS, DES~SV1, HSC-SSP~S16a, KiDS~DR4, and RCSLenS), which exploited different shear and photometric redshift estimation algorithms, were analysed with a consistent pipeline under the same model assumptions. We performed a comparison of the amplitude of the reduced excess surface density and of the mass estimates using lenses from the \Planck PSZ2 and SDSS redMaPPer cluster samples. Mass estimates agree with literature results collected in the LC2 catalogues. Mass accuracy was further investigated considering the AMICO detected clusters in the HSC-SSP XXL North field. The consistency of the data sets was tested using our unified analysis framework. We found agreement between independent surveys, at the level of systematic noise in Stage-III surveys or precursors. This indicates successful control over systematics. If such control continues in Stage-IV, \Euclid will be able to measure the weak lensing masses of around $\num{13000}$ (considering shot noise only) or $\num{3000}$ (noise from shape and large-scale-structure) massive clusters with a signal-to-noise ratio greater than 3.
}


\authorrunning{M. Sereno et al.}
\titlerunning{Cluster weak lensing in Stage-III surveys}

\maketitle

\section{Introduction}

Mass calibration is crucial to exploit galaxy clusters as cosmological probes \citep{ser02, voi05, ett+al09, man+al10, jul+al10, lub+al14, planck_2013_XX, spt_boc+al19,des_abb+al20,les+al22} or astrophysical laboratories \citep{chexmate+al21}. The theory of weak gravitational lensing (WL) by galaxy clusters is well understood \citep{ba+sc01,ume20} and it has emerged as one of the most reliable tools to accurately and precisely measure cluster masses \citep[see, e.g.,][]{hoe+al12, wtg_I_14, wtg_III_14, hoe+al15, ume+al14, ume+al16b,ok+sm16,die+al19}.

The prominence of WL mass calibration has increased in the era of large and deep photometric surveys. These surveys are usually conceived mainly for cosmic shear or galaxy-galaxy lensing analyses, or statistical analyses of ensembles of clusters, but the ever increasing depth of modern surveys have also made it possible to measure the masses of single clusters \citep[see, e.g., ][]{ser+al17_psz2lens,ume+al20}. Until relatively recently, measuring the mass of a single cluster was only possible with dedicated, targeted observations. Now, we can measure individual cluster masses directly from current WL survey data \citep{mel+al15,hsc_med+al18b,ser+al17_psz2lens,ser+al18_psz2lens,mur+al22}.

Large cosmological surveys allow us to study large and homogeneous samples of clusters. While our understanding of calibration issues has made significant progress \citep{gra+al21}, some areas of concern still persist. In particular, WL mass calibration is still seen as a possible source of systematic error for stacked cluster analyses \citep{des_cos+al21}.

In preparation for the Euclid Survey \citep{eucl_lau_11,euclid_pre_sca+al22}, we discuss how well recent and ongoing optical surveys can measure the masses of individual clusters and groups. 
Measurement accuracy can be estimated by checks with simulations or reference samples. Shear calibration requires expensive simulations \citep{hsc_man+al18}. Photometric redshift (photo-$z$) calibration requires deep and unbiased calibration samples \citep{hil+al12,euclid_pre_sag+al22}. Assessment of the total error budgets requires end-to-end testing. These products can be difficult to generate or acquire for deep and large galaxy surveys.

Data-driven approaches can offer an alternative path to assess robustness. Comparison of independent results can unveil and quantify unknown systematics \citep{cha+al19,lea+al22,lon+al23}. 
Analyses from independent collaborations can differ in many aspects: data sets, shear and photo-$z$ estimation algorithms, theory model assumptions, or inference pipelines. Cross-survey analyses can assess the consistency of lensing signals across different data sets and provide the basis for end-to-end tests of systematic errors. 
They can also offer insight into what we can expect from upcoming and future surveys.

Recent cross-comparisons of surveys have shown that cosmic shear and galaxy-galaxy lensing analyses can be robust across different modelling choices and data-sets.
\citet{cha+al19} assessed the robustness of cosmic shear results with a unified analysis at the catalogue level of four cosmic shear surveys. By using a unified pipeline, they showed how the cosmological constraints are sensitive to the various details of the pipeline. The same approach was then used by \citet{lon+al23}, who performed a unified catalogue-level reanalysis of three cosmic shear data sets exploiting and testing the pipeline developed by the LSST (Legacy Survey of Space and Time) Dark Energy Science Collaboration of the Vera C. Rubin Observatory. They found the results from the three surveys to be statistically consistent and the constraints on cosmological parameters to be robust to different small-scale modellings.
\citet{des+kids+23} presented a cosmic shear analysis of two Stage-III surveys in a collaborative effort between the two teams and found consistent cosmological parameters.

\citet{lea+al22} performed a blind comparison of the amplitude of galaxy-galaxy lensing using lens samples from the Baryon Oscillation Spectroscopic survey (BOSS) and six independent lensing surveys and pipelines. They found good agreement between empirically estimated and reported systematic errors.

Here, we extend the cross-survey approach to cluster WL. We perform a uniform analysis with the combined clusters and weak-lensing pipeline (\texttt{COMB-CL}). \texttt{COMB-CL} forms part of the global \Euclid data processing pipeline and will measure cluster WL shear profiles and masses for the survey, see App.~\ref{sec_combcl}. This paper is part of a series presenting and discussing WL mass measurements of clusters exploiting \texttt{COMB-CL}.

The analysis serves a double, data-driven validation purpose. On one hand, we cross-validate measurements and calibrations in lensing surveys by comparing results obtained from a unified pipeline. Agreement suggests that data products are compatible and that systematic errors in each survey were corrected to the required level. On the other hand, the pipeline is validated by comparing the WL mass estimates here obtained with previous works and literature results.

In this work we consider that two independent estimates agree and systematic errors are under control if differences between the results are smaller than the nominal statistical uncertainties. Using cross-validation, we can assess the accuracy and precision of WL shear profiles and mass estimates in Stage-III and precursor surveys. In the following text, accuracy is defined as how close measured estimates are to their true values. Small systematic errors imply high accuracy. In a cross-comparison, we can assess the accuracy of a measurement by quantifying the statistical agreement of independent results from different surveys.

Precision is a measure of how close the estimates are to each other. Small statistical uncertainties imply high precision. Precision can be assessed by measuring the signal-to-noise ratio ($\text{S/N}$) or the statistical uncertainties.



\subsection{Notation}

In this paper we adopt a flat $\Lambda$CDM model with (total) present day matter density parameter $\Omega_\text{m}=0.30$, baryonic density parameter $\Omega_\text{b}=0.05$, Hubble constant $H_0=70\,\kmsMpc$, power spectrum amplitude $\sigma_8=0.8$, and initial index $n_\text{s}=1.0$. As usual, $H(z)$ is the redshift dependent Hubble parameter, $E_z\equiv H(z)/H_0$, and $h = H_0 / (100\,\kmsMpc)$.

$O_{\Delta\text{c}}$ denotes a cluster property, $O$, measured within the radius $r_{\Delta\text{c}}$ which encloses a mean over-density of $\Delta\text{c}$ times the critical density at the cluster redshift, $\rho_\mathrm{cr}(z)\equiv3H^2(z)/(8\pi G)$. 

`$\log$' is the logarithm in base 10, and `$\ln$' is the natural logarithm. Scale results for natural logarithm are quoted as percents, i.e., 100 times the dispersion in natural logarithm. 

When not stated otherwise, the central location and scale are computed as CBI (Centre BIweight) and SBI (Scale BIweight) \citep{bee+al90}. Probabilities are computed considering the marginalised posterior distributions.


\section{Public lensing surveys}
\label{sec_surveys}

\begin{figure*}
\resizebox{\hsize}{!}{\includegraphics{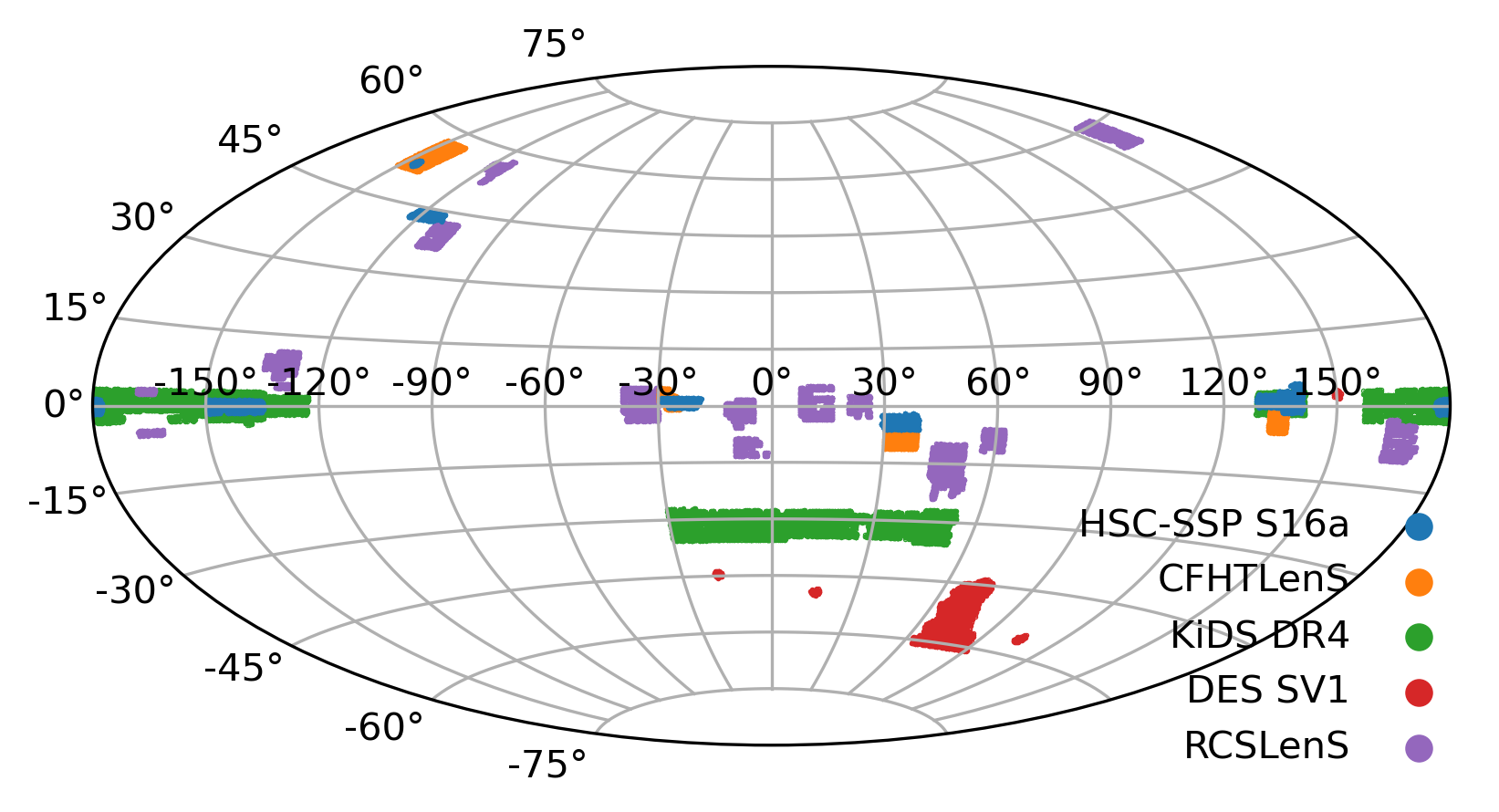}} \\
\caption{Sky coverage in the equatorial coordinate system of the lensing surveys used in this work. We consider only the tiles with sufficiently reliable shape and photo-$z$ measurements.
}
\label{fig_surv_sky}
\end{figure*}

\begin{table*}
\caption{Public lensing surveys considered in the present work. We report the survey name, and the photometric bands, photo-$z$ ($z_\text{p}$), and shape ($e_i$) algorithms we exploit for our analysis, and some primary references.
}
\label{tab_surv_spec}
\centering
\begin{tabular}[c]{l  l l l l}
	\hline
	\noalign{\smallskip}  
	Survey	&	bands	&  $z_\text{p}$ &  $e_i$	& references	  \\
			\noalign{\smallskip}  
	\hline
	\noalign{\smallskip}  
	HSC-SSP S16a 	& $\,\,\,\,\,\,g,\,r,\,i,\,z,\,y$  	& \texttt{Ephor\_AB}	& 	re-Gauss		& \citet{hsc_man+al18,hsc_hik+al19}	\\
	CFHTLenS		& $u,\,g,\,r,\,i,\,z\,\,\,$	 	& \texttt{BPZ}		&	\texttt{lensfit}	& \citet{hey+al12,hil+al12}	\\
	KiDS DR4			& $u,\,g,\,r,\,i\,\,\,$	+NIR	& \texttt{BPZ}		&	\texttt{lensfit}	& \citet{kids_kui+al19} \\
	DES SV1 			& $\,\,\,\,\,\,g,\,r,\,i,\,z,\,Y$		& \texttt{BPZ} 		&	\texttt{IM3SHAPE} 	& \citet{jar+al16} \\
	RCSLenS			& $\,\,\,\,\,\,g,\,r,\,i,\,z\,\,\,$		& \texttt{BPZ}		&	\texttt{lensfit}	& \citet{hil+al16}		\\
	\hline
	\end{tabular}
\end{table*}

\begin{table}
\caption{WL properties of the public lensing surveys computed using a common scheme for comparison. $N_\text{sources}$ is the number of sources with a non-null lensing weight and estimated photometric redshift; $z_\text{s}$ is the median redshift of the considered sources;
$A_\text{eff}$ is an estimate of the effective survey area;
$n_\text{raw}$ and $n_\text{eff}$ are the raw and weighted total density, respectively, of the considered sources.
}
\label{tab_surv_lens}
\centering
\resizebox{\hsize}{!} {
\begin{tabular}[c]{l  r r r r r}
	\hline
	\noalign{\smallskip}  
	Survey	&	$N_\text{sources}$	&  $z_\text{s}$ &  $A_\text{eff}$		&	$n_\text{raw}$			& $n_\text{eff}$ 		  \\
		 	&	 			&	& $[\text{deg}^2]$		&	$[\text{arcmin}^{-2}]$	& $[\text{arcmin}^{-2}]$     \\
	\noalign{\smallskip}  
	\hline
	\noalign{\smallskip}  
	HSC-SSP S16a 	& $\num{12054563}$  	&	 0.83		& 138	& 	23.2  & 	23.4			\\
	CFHTLenS		& $\num{7519395}$	&	 0.72 	& 123	&	16.0  &	14.3			\\
	KiDS DR4			& $\num{21262011}$	&	 0.67		&  715	&	7.8	  &	6.7			\\
	DES SV1 			& $\num{4485202}$	&	0.53		& 142 	&	8.3	  &	8.2			\\
	RCSLenS			& $\num{15277685}$	&	 0.51		& 447	&	7.4	  &	6.1			\\
	\hline
	\end{tabular}
	}
\end{table}

Galaxy imaging surveys have advanced to so-called Stage-III \citep{alb+al06}. Some have been successfully completed while others are still ongoing with very strong intermediate results. 
We list the surveys used for our analysis in Table~\ref{tab_surv_spec}, and we briefly introduce them in the following. We consider the surveys that we are aware of at the time of writing with public calibrated shear, photometric, and photo-$z$ catalogues. We only consider shear catalogues with multiplicative/additive bias corrections. We do not consider public metacalibrated shear catalogues \citep[e.g.,][]{des_gat+al21}, which are relatively rare at the time of writing and whose treatment is not yet fully tested in the version of the \Euclid data processing pipeline used in this work.

The survey data releases under consideration in this work have covered a total effective sky area (after masking) of $\sim 1500\deg^2$ with deep multi-band photometry, see Fig.~\ref{fig_surv_sky}.
For full details of each analysis, we refer to the quoted survey papers. Some surveys use multiple pipelines for photometry, shape, and/or photo-$z$ estimates. In Table~\ref{tab_surv_spec}, we report the estimates we used for our work. We motivate our choices below. 

 We only use photometry for colour estimates, where differences due to slightly differing transmission filters are subdominant for the considered surveys with respect to other effects. Therefore, we do not consider differences in filters or magnitude definitions.

Most of the surveys were designed as dark energy experiments, and were optimised for cosmic shear. These surveys generally quote estimates of, e.g., area, or galaxy density. Here we are interested in cluster lensing and thus, to ease comparison, we recompute some quantities in a common framework of interest for our specific analysis, see Table \ref{tab_surv_lens}. Values for these estimates computed in different ways can be found in the survey papers cited here. The estimates of number density or effective area are for the most part consistent, but there are some minor discrepancies.

As eligible sources for our analysis, $N_\text{sources}$, we consider galaxies with measured shape, i.e., having non-null lensing weight, and photo-$z$. We compute their raw density, $n_\text{raw}$, as the mean density in $\num{1000}$ small random regions of size $2\arcminute\times2\arcminute$.
The effective survey area $A_\text{eff}$ can be estimated as  $N_\text{sources}/ n_\text{raw}$.

Most cluster lenses lie at redshift $\sim 0.3$ and sources for cluster lensing are usually recovered from a field of view with proper size of 3--4 Mpc, i.e., half a degree at $z\sim 0.3$. We compute the effective density of sources for cluster lensing, $n_\text{eff}$, as the median of the weighted densities \citep[see equation 1 in][]{hey+al12} in $\num{1000}$ random regions of size $30\arcminute\times30\arcminute$. We discard the 100 smallest values to mimic selection effects, with clusters less likely to be detected, e.g., in less populated areas, near borders, or near masked regions.

The raw and effective source densities defined above are defined in a cluster lensing context. They may differ from the nominal values reported in the reference survey papers and in the following subsections for the different definitions or the different source selections. For our analysis, we exclusively use our homogenised estimates.


\subsection{HSC-SSP S16a}
\label{sec_HSC-SSP}

The Hyper Suprime-Cam Subaru Strategic Program \citep[HSC-SSP,][]{hsc_miy+al18,hsc_aih+al18} is an ongoing program to carry out a multi-band imaging survey in five optical bands ($g$, $r$, $i$, $z$, $y$) with HSC, an optical wide-field imager with a field-of-view of $1.77\deg^2$  mounted on the prime focus of the $8.2\,\text{m}$ Subaru telescope \citep{hsc_miy+al18,hsc_kom+al18,hsc_fur+al18,hsc_kaw+al18}. 

The wide survey aims to observe around $\num{1400}\,\deg^2$ with a depth of $i \sim26~\text{mag}$ at the $5\,\sigma$ limit within a \ang{;;2} diameter aperture \citep{hsc_aih+al18}. The survey design is optimised for WL  studies \citep{hsc_man+al18,hsc_hik+al19,hsc_miy+al19,hsc_ham+al20}.

The catalogue of galaxy shape measurements from the first-year data release (S16a) is presented in \citet{hsc_man+al18}. The catalogue covers an area of $136.9\deg^2$ split into six fields to final depth, with a mean $i$-band seeing of \ang{;;0.58}. The survey overlaps with the XXL survey \citep{xxl_I_pie+al16} in the XXL-North field.

Galaxy shapes were estimated on the co-added $i$-band images using a moments-based shape measurement method along with the re-Gaussianisation PSF correction method \citep{hi+se03}, which fits a Gaussian profile with elliptical isophotes to the image. 

Conservative galaxy selection criteria were implemented to produce the shear catalogue for first-year science, with a magnitude cut of $i < 24.5~\text{mag}$. This results in nominal unweighted and weighted source number densities of 24.6 and 21.8 arcmin$^{-2}$, respectively \citep{hsc_man+al18}.

Photo-$z$s were found to be well determined in the redshift range $0.2 \ls z \ls1.5$, with an accuracy of $\sigma_{z_\text{phot}} \sim 0.05\,(1 + z_\text{p})$, and an outlier rate of $\sim15\,\%$ for galaxies down to $i$ = 25 \citep{hsc_tan+al18}. For the brighter sample of $i < 24$, performance improves to $\sigma_{z_\text{phot}} \sim 0.04\, (1 + z_\text{p})$ and $\sim8\,\%$ outliers. 

We retrieve catalogues from the public archive.\footnote{\url{https://hsc-release.mtk.nao.ac.jp/datasearch}} The collaboration provides a variety of estimates for photometric magnitude \citep{hsc_hua+al18} and redshift \citep{hsc_tan+al18}. Our choices are informed from previous WL cluster analysis of data release S16a \citep{hsc_chi+al20,ume+al20}. We consider photo-$z$s based on the \texttt{EPHOR\_AB} code, which delivers estimates based on PSF-matched aperture photometry \citep{hsc_tan+al18}. A photo-$z$ risk parameter $R(z_\text{p})$ is provided to represent the expected loss for a given choice of $z_\text{p}$ as the point estimate  \citep{hsc_tan+al18}. For the photometry, we consider the forced \texttt{cmodel}  \citep{hsc_hua+al18}.


\subsection{CFHTLenS}

The Canada France Hawaii Telescope Legacy Survey \citep[CFHTLS,][]{hey+al12} is a completed photometric survey performed with MegaCam. The final footprint covers four independent fields for a total of $154\deg^2$ in five optical bands $u^*$, $g$, $r$, $i$, $z$ \citep{hey+al12}.

The survey was designed for WL analysis, with the deep $i$-band data taken in sub-arcsecond seeing conditions \citep{erb+al13}. The nominal total unmasked area suitable for cosmic shear analysis covers $125.7\deg^2$. The nominal raw number density of lensing sources, including all objects with a measured shape, is 17.8 galaxies per arcmin$^2$ \citep{hil+al16}. The nominal weighted density is 15.1 galaxies per arcmin$^2$ \citep{hil+al16}.

The CFHTLenS team provided WL data processed with \texttt{THELI} \citep{erb+al13}, and shear measurements obtained with \texttt{lensfit} \citep{mil+al13}, a likelihood based model-fitting method.\footnote{The public archive is available through the Canadian Astronomy Data Centre at \url{http://www.cfht.hawaii.edu/Science/CFHLS}.} Photo-$z$s were determined with the $\texttt{BPZ}$ algorithm \citep{ben00}, and the $\texttt{ODDS}$ quantifies the prominence of the most likely redshift \citep{hil+al12}. The photo-$z$s were measured with accuracy $\sigma_{z_\text{phot}} \sim 0.04\,(1+z)$ and a catastrophic outlier rate of about $4\,\%$ \citep{hil+al12,ben+al13}.

\subsection{KiDS DR4}

The Kilo-Degree Survey \citep[KiDS,][]{dej+al13} is a European Southern Observatory (ESO) public survey performed with the OmegaCAM wide-field camera mounted at the VLT Survey Telescope (VST). KiDS was designed to observe a total area of $\num{1350}\,\deg^2$ in the $u$, $g$, $r$, $i$ bands.  

The survey area has been observed to full depth, and the analysis is ongoing. The fourth data release (hereafter referred to as KiDS~DR4 or KiDS~1000) covered approximately $\num{1000}\,\deg^2$ in all four survey filters, with complementary aperture-matched $Z$, $Y$, $J$, $H$, $K_\text{s}$ photometry from the partner VIKING survey on the VISTA telescope \citep{kids_kui+al19}. The mean limiting magnitudes in the four bands are, respectively, 24.23, 25.12, 25.02, and 23.68 ($5\,\sigma$ in a 2\arcsec$\,$ aperture). 

The survey area is divided into the Southern (KiDS-S) and the Northern (KiDS-N) fields. KiDS-N contains two additional smaller areas: KiDS-N-W2, which coincides with the G9 patch of the GAMA survey, and KiDS-N-D2, a single pointing on the COSMOS field. 

The Astro-WISE information system was used for data processing and catalogue extraction. Shear measurements were done with \texttt{lensfit}, similar to CFHTLenS. Shape measurements were performed on $r$-band images, as these images exhibit better seeing properties and higher source density \citep{fen+al17,kan+al19}. The $r$-band images were separately reduced with the \texttt{THELI} pipeline for WL science.  

We use the gold sample, which includes only galaxies with reliable shape and redshift measurements up to $z_\text{p}=1.2$ \citep{kids_kui+al19,kids_hil+al21,kids_gib+al21}. Cosmological parameter constraints from cosmic shear or galaxy clustering have been presented in, e.g., \citet{kids_asg+al21}, \citet{kids_hey+al21}, \citet{kids_tro+al21}, \citet{kids_joa+al21}.

\subsection{DES SV1}

The Dark Energy Survey (DES) is expected to cover approximately $\num{5000}\,\deg^2$ in the South Galactic Cap region in five optical bands,  $g$, $r$, $i$, $z$, and $Y$, in a five-year span with the Dark Energy Camera (DECam) \citep{des_gat+al21}. The DES Science Verification (SV) survey mimicked the number of visits and total image depth ($10\,\sigma$ limiting magnitude of $24.1$ in the $i$ band) planned for the full DES survey \citep{jar+al16}.

The largest portion of the SV area, known as SPT-East (SPT-E for short), covers the Eastern part of the region observed by the South Pole Telescope ($63\deg^2$). For the present analysis, we consider the SVA1 Gold Catalogue\footnote{\url{http://des.ncsa.illinois.edu/releases/sva1}}. The nominal area for the shear catalogues covers $139\deg^2$.

We consider the shear measurements from the \texttt{IM3SHAPE} shear pipeline on $r$-band images based on a maximum likelihood fit using a bulge-or-disc galaxy model. For the photo-$z$s, we consider the \texttt{BPZ} estimates. 

Photo-$z$ reliabilities were not provided in the public catalogues. We estimate the confidence in the redshift point estimate $z_\text{p}$ from the probability density function, in a manner similar to \texttt{ODDS}, as
\be
C(z_\text{p}) = \int ^{z_\text{p} + \Delta z_\text{p}}_{z_\text{p}- \Delta z_\text{p}} P(z)\;{\rm d} z \;,
\ee
where we set $\Delta z_\text{p} = 0.12\,(1+z_\text{p})$, i.e., three times the typical photo-$z$ uncertainty.

\subsection{RCSLenS}

The RCSLenS is a large public survey performed with MegaCam for WL analyses\footnote{The data products are publicly available at \url{http://www.cadc-ccda.hia-iha.nrc-cnrc.gc.ca/en/community/rcslens/query.html}.} \citep{hil+al16}. The parent survey, i.e., the Red-sequence Cluster Survey 2 \citep[RCS2,][]{gil+al11} is a sub-arcsecond seeing, multi-band imaging survey in the $g$, $r$, $i$, $z$ bands initially designed to optically detect galaxy clusters. 

The survey covers a nominal total unmasked area of $571.7\deg^2$ down to a magnitude limit of $r\sim24.3$ (for a point source at $7\,\sigma$). Photo-$z$s are available for a nominal unmasked area covering $383.5\deg^2$, where the nominal raw (weighted) number density of lensing sources is 7.2 (4.9) galaxies per arcmin$^2$.  The survey area is divided into 14 patches, the largest being $10\times10\deg^2$ and the smallest $6\times6\deg^2$. 

The shape measurement and data-analysis were performed with tools developed for the CFHTLenS pipeline. A detailed presentation of imaging data, data reduction, masking, multi-colour photometry, photo-$z$s, shape measurements, tests for systematic errors, and the blinding scheme for objective measurements can be found in \citet{hil+al16}.




\section{Cluster samples}

\begin{table}
\caption{Main properties of the PSZ2 cluster lenses per survey. We list the number of clusters in the considered survey footprint, and their typical redshift, mass proxy value, here the catalogue SZ mass in units of $10^{14}M_\odot$, $(\text{S/N})_\text{WL}$, and the source galaxy density per cluster (in units of arcmin$^{-2}$). Location and scale are computed as biweight estimators of the sample distribution.
}
\label{tab_sample_psz2}
\centering
\resizebox{\hsize}{!}{
\begin{tabular}[c]{l  r  r@{$\,\pm\,$}l  r@{$\,\pm\,$}l  r@{$\,\pm\,$}l  r@{$\,\pm\,$}l}
	\hline
	\noalign{\smallskip}  
	Survey			&	$N_\text{cl}$ 	&	 \multicolumn{2}{c}{$z$} & \multicolumn{2}{c}{$M_\text{SZ}$}	&\multicolumn{2}{c}{$(\text{S/N})_\text{WL}$}  &\multicolumn{2}{c}{$n_\text{sources}$} \\
	\noalign{\smallskip}  
    \hline
    \noalign{\smallskip}
        &   & \multicolumn{2}{c}{}  &  \multicolumn{2}{c}{[$10^{14}M_\odot$]} & \multicolumn{2}{c}{} & \multicolumn{2}{c}{[arcmin$^{-2}$]} \\
	\hline
	\noalign{\smallskip}  
	HSC-SSP S16a	&	5	&  0.20 & 0.10  	&	4.45&1.34	& 	2.45& 1.01	& 12.1	& 4.9 \\
	CFHTLenS		&	8	&   0.16&   0.06	&   	3.88& 1.28&   	2.69	&   1.40	&    6.8	&   2.4 \\
	KiDS DR4			&	37	&   0.21&   0.12	&   	4.36&2.14	&   	2.45	&   1.47	&   2.8	&   1.0 \\
	DES SV1  		&	11	&   0.37&   0.19	&   	6.27&3.87	&   	0.99	&   1.28	&   1.9	&   1.3 \\
	RCSLenS			& 	28	&   0.23&   0.17 &  	4.69&1.79	&   	1.58	& 2.17	&   1.9	&   1.4 \\
	\hline
	\end{tabular}
	}
\end{table}

\begin{table}
\caption{Same as Table~\ref{tab_sample_psz2} but for the redMaPPer sample. Here, the mass proxy is the catalogue richness $\lambda$.
}
\label{tab_sample_redmapper}
\centering
\resizebox{\hsize}{!} {
\begin{tabular}[c]{l  r  r@{$\,\pm\,$}l  r@{$\,\pm\,$}l  r@{$\,\pm\,$}l  r@{$\,\pm\,$}l}
	\hline
	\noalign{\smallskip}  
	Survey			&	$N_\text{cl}$ 	&	 \multicolumn{2}{c}{$z$} & \multicolumn{2}{c}{$\lambda$}	&\multicolumn{2}{c}{$(\text{S/N})_\text{WL}$}  &\multicolumn{2}{c}{$n_\text{sources}$} \\
	\noalign{\smallskip}  
    \hline
    \noalign{\smallskip}
        &   & \multicolumn{2}{c}{}  &  \multicolumn{2}{c}{} & \multicolumn{2}{c}{} & \multicolumn{2}{c}{[arcmin$^{-2}$]} \\
	\hline
	\noalign{\smallskip}    
	HSC-SSP S16a	&	598		&	0.36	&	0.11		& 34.1	&	14.0	& 0.85	& 0.82	& 8.9	& 4.2 	\\
	CFHTLenS		&	329		&	0.38	&	0.11		& 35.0	&	14.3	& 0.90	& 1.14	& 5.3	& 2.1 	\\
	KiDS DR4			&	1240		&	0.40	&	0.11		& 34.5	&	14.8	& 0.57	& 0.97	& 2.2& 0.9 	\\
	DES SV1  		&	8 		&	0.38	&	0.05		& 40.7	&	15.1	& 0.92	& 0.69	& 1.1	& 0.7 	\\
	RCSLenS			& 	1020		&	0.38	&	0.11		& 34.2	&	14.5	& 0.43	& 0.98	& 1.2& 0.9 	\\
	\hline
	\end{tabular}
	}
\end{table}

We make use of five lens catalogues. Firstly, to compare results from different surveys, assess accuracy, and check for systematic errors, we choose two independent cluster catalogues: i) the second \Planck Catalogue of Sunyaev-Zeldovich Sources \citep[PSZ2,][]{planck_2015_XXVII}, based on a Sunyaev-Zeldovich (SZ) selection, and ii) the red-sequence Matched-filter Probabilistic Percolation (redMaPPer) catalogue based on an optical selection on the Sloan Digital Sky Survey (SDSS) DR8 data \citep{ryk+al16}. These catalogues are extracted from data sets different from the surveys we use for shape measurements. This makes the distribution of lenses uncorrelated with residual systematic effects in galaxy shape measurements \citep{miy+al13,ser+al15_bias}.

For each lens sample, we use all clusters that lie in the survey fields without any further selection, and the lensing clusters we consider are an unbiased subsample of the parent catalogue. The main properties of the lensing cluster samples from PSZ2 and redMaPPer per survey are reported in Tables~\ref{tab_sample_psz2} and \ref{tab_sample_redmapper}, respectively.

Secondly, to study the statistical precision of WL mass measurements, we consider the candidate clusters detected with the Adaptive Matched Identifier of Clustered Objects \citep[AMICO,][]{bel+al18,mat+al19} in the XXL-North field of HSC-SSP.

Thirdly, to check for residual systematic effects in the shear calibration, we consider two samples: the clusters detected in HSC-SSP S16a \citep{ogu+al18} with the Cluster finding algorithm based on Multi-band Identification of Red sequence gAlaxies \citep[CAMIRA,][]{ogu14}; the AMICO clusters in KiDS-DR3 \citep{les+al22}.

Finally, for comparison with literature, we consider the Literature Catalogue of weak Lensing Clusters of galaxies \citep[LC$^2$ or LC2,][]{ser15_comalit_III}.

\subsection{\Planck PSZ2}
\label{sec_psz2}

PSZ2 is based on the 29 month full-mission dataset and contains $\num{1653}$ candidate clusters \citep{planck_2015_XXVII}. It is the largest all-sky, SZ selected sample of galaxy clusters produced to date.\footnote{The union catalogue HFI\_PCCS\_SZ-union\_R2.08.fits is available from the \Planck legacy archive at \url{http://pla.esac.esa.int/pla/}.}

The catalogue includes candidates with S/N above 4.5 located outside the highest-emitting Galactic regions, the Small and Large Magellanic Clouds, and point source masks. 1203 clusters were confirmed with counterparts identified either in external optical or X-ray samples, or by dedicated follow-ups. 

Proxy masses (denoted as $M_\text{SZ}$ or $M_\text{500c}^{Y_z}$) of clusters with known redshift were calibrated with a best fitting scaling relation between $M_\text{500c}$ and  the spherically integrated Compton parameter $Y_\text{500c}$ \citep{planck_2013_XX}. The catalogue spans a nominal mass range from $M_\text{SZ}\sim0.8\times10^{14}M_\odot$ to $16\times10^{14}M_\odot$ over the redshift range $0.01 \ls z \ls 1.0$. The mean redshift is $z\sim 0.25$.

Most of the PSZ2 clusters covered by the surveys here considered are at $z < 0.6$, for which data from Stage-III surveys or precursors can provide reliable masses \citep{mel+al15,ser+al17_psz2lens,hsc_med+al18b}. There are two exceptions. PSZ2~G099.86$+$58.45 at $z=0.63$ can be still detected in CFHTLenS with high $(\text{S/N})_\text{WL}$ \citep{ser+al18_psz2lens}, whereas PSZ2~G297.97$-$67.74 at $z=0.87$, here covered by DES SV1, is not significantly detected.

\subsection{The SDSS redMaPPer catalogue}
\label{sec_redmapper}

The redMaPPer algorithm is a red-sequence cluster finder designed for large photometric surveys \citep{ryk+al14}. Here, we consider the cluster candidates found parsing nearly $\num{10000}\,\deg^2$ of contiguous high quality observations of the SDSS DR8 data \citep{ryk+al16}. The resulting catalogue\footnote{We use the latest version of the catalogue (v6.3), which is publicly available at \url{http://risa.stanford.edu/redMaPPer/}.} contains $\num{26111}$ candidate clusters over the redshift range $0.08 \ls z \ls 0.6$. 

The catalogue provides a richness estimate, $\lambda$, defined as the sum of the probabilities of the galaxies found near a cluster to be actual cluster members. The sum extends over all galaxies above a cut-off luminosity ($0.2L_*$) and below a radial cut which scales with richness \citep{ryk+al14}.

\subsection{AMICO clusters in the XXL-HSC field}
\label{sec_amico_hsc_xxl}

AMICO \citep{bel+al18,mat+al19} is an optimal matched filter that takes advantage of the known statistical properties of field galaxies and cluster galaxy members. AMICO can deal with an arbitrary number of quantities describing galaxies. For the AMICO-built catalogues considered here, galaxy angular coordinates, magnitudes, and photo-$z$ were considered, whereas the information concerning the colours was avoided to be independent of the red-sequence \citep{mat+al19}.

AMICO was selected as one of two cluster selection algorithms for the \Euclid mission \citep{euclid_ada+19}, and has been implemented in the extensively tested \Euclid cluster detection pipeline (\texttt{DET-CL}). Here, we consider the runs in the XXL-North field covered by HSC-SSP S18a observations (Euclid Collaboration: Sartoris et al., in prep.).

The XXL-North field has been covered by multi-wavelength observations. HSC-SSP observations over this area from Year 1 have already reached full depth, which makes this an interesting test-case for the Euclid Survey. \texttt{DET-CL} found 3534 candidate clusters with $(\text{S/N})_\text{det} \ge 3$ in the redshift range $0.03 \le z \le 1.05$ with intrinsic richness, $\lambda^*$, in the range $22.5 \la \lambda^* \la  591.2$ (Euclid Collaboration: Sartoris et al., in prep.). The intrinsic richness is defined as the sum of the probabilities of all galaxies associated with the detection brighter than $m_* + 1.5$ and within the model virial radius.

\subsection{CAMIRA clusters in HSC-SSP S16a}
\label{sec_camira_hsc_ssp}
\citet{ogu+al18} presented a cluster sample from HSC-SSP S16a, optically-selected with CAMIRA \citep{ogu14,ogu+al18}. CAMIRA is a red sequence method which fits each galaxy in the image with a stellar population synthesis model to compute the likelihood to be a red sequence galaxy at a given redshift \citep{ogu14}. 

\citet{ogu+al18} constructed a catalogue of 1921 clusters from HSC-SSP S16a. The images were sufficiently deep to detect clusters at redshift $0.1 < z < 1.1$ with richness $\hat{N}_\text{mem}>15$ that roughly corresponds to $M_{200\rm{m}} \gs 10^{14} h^{-1} M_\odot$.

\subsection{AMICO clusters in KiDS-DR3}
\label{sec_amico_kids_dr3}
The AMICO algorithm was run in KiDS-DR3 to detect 4934 candidate galaxy clusters with intrinsic richness $\lambda_* \geq 15$  and $(\text{S/N})_\text{det} \geq 3.5$ in the redshift range $0.1 \ls z \ls 0.8$ \citep{mat+al19,les+al22}.


\section{Background source selection}
\label{sec_back_sel}

\begin{table}
\caption{Parameters for the photo-$z$ background selection. Column names are explained in Sect.~\ref{sec_back_sel_photoz}.
}
\label{tab_back_photo_z_cuts}
\centering
\resizebox{\hsize}{!} {
\begin{tabular}[c]{l  l l l l l}
	\hline
	\noalign{\smallskip}  
	Survey			&	$ \Delta z_\text{lens}$ &	$z_\text{p,range,min}$ &	$z_\text{p,range,max}$ & $z_\text{p, confidence}$	&	$z_\text{p, confidence,min}$ \\
	\noalign{\smallskip}  
	\hline
	\noalign{\smallskip}  
	HSC-SSP S16a	&	0.1	&	0.2	&	1.5	&	$1 - R(z_\text{p})$	& 0.8 	\\
	CFHTLenS		&	0.1	&	0.2	&	1.2	&	$\texttt{ODDS}$	& 0.8 	\\
	KiDS DR4			&	0.1	&	0.2	&	1.2	&	$\texttt{ODDS}$	& 0.8 	\\
	DES SV1  		&	0.1	&	0.2	&	1.2	&	$C(z_\text{p})$		& 0.8 	\\
	RCSLenS			&	0.1	&	0.4	&	1.2	&	$\texttt{ODDS}$	& 0.8 	\\
	\hline
	\end{tabular}
	}
\end{table}

\begin{figure}
\resizebox{\hsize}{!}{\includegraphics{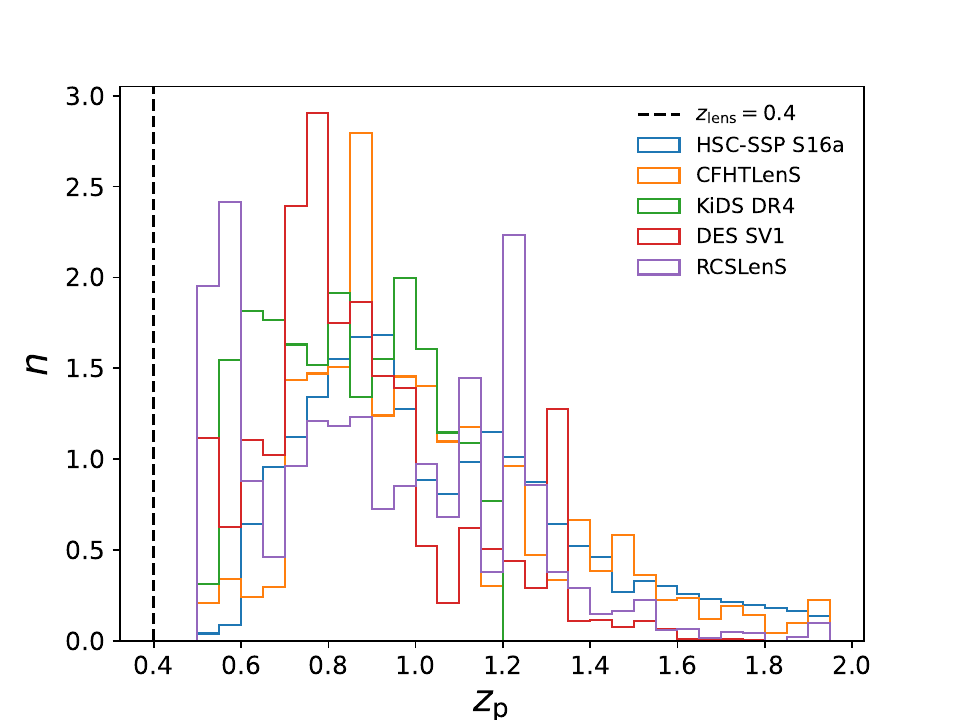}} 
\caption{Normalised distribution of the photo-$z$s up to $z_\text{p}=2$ for selected sources behind a lens plane at $z_\text{lens}=0.4$. The sources are selected either with robust cuts in photo-$z$ or in colour-colour space.}
\label{fig_hist_1D_background_sources_z_040}
\end{figure}

We identify galaxies as background sources for the WL analysis behind the lens at $z_\text{lens}$ based on their photo-$z$ or colours. As a first step, we select galaxies such that
\be
\label{eq_zphot_1}
z_\text{p} > z_\text{lens} +\Delta  z_\text{lens}\;,
\ee
where $z_\text{p}$ is the redshift point-estimate, and $\Delta  z_\text{lens}$ is a threshold above the cluster redshift to lower the contamination. 

On top of this criterion, we require that the sources pass more restrictive cuts in either photo-$z$ or colour properties, which we discuss below. 

\subsection{Photometric redshifts}
\label{sec_back_sel_photoz}

A population of background galaxies can be selected with criteria based on the photo-$z$s \citep{ser+al17_psz2lens}:
\begin{eqnarray}
z_\text{p,2.3\,\%} 		 	&>&	 	z_\text{lens} + \Delta z_\text{lens} \,; \label{eq_zphot_2}  \\
z_\text{p,range,min}		&<&		z_\text{p}\,; \label{eq_zphot_3}  \\
z_\text{p,range,max}		&>&		z_\text{p}\,;   \label{eq_zphot_4} \\
z_\text{p,confidence}		&>&		z_\text{p, confidence,min} \,,\label{eq_zphot_5} 
\end{eqnarray}
where the parameter $z_\text{p,confidence}$ is a measure of the confidence we have in the point estimate of the redshift;
$z_\text{p,2.3\,\%}$ is the lower bound of the region including the 95.4\,\% of the probability density distribution;
$\Delta z_\text{lens}$ is a conservative threshold to better select background galaxies, see Eq.~(\ref{eq_zphot_1});
$z_\text{p,range,min}$ and $z_\text{p,range,max}$ are the boundaries of the redshift range wherein photo-$z$ estimates are thought to be reliable. 

The redshift range can be chosen based on the survey depth or such that the photometric bands straddle the $\num{4000}\,\AA$ break. Employed parameters and cuts for each survey are listed in Table~\ref{tab_back_photo_z_cuts}.

\subsection{Colour–colour}

Selection of background galaxies in colour-colour (CC) space can be highly effective and provide very complete and pure samples \citep{euclid_pre_les+al23}. Here, we adopt the cuts in the $g - i$ vs. $r -z$ CC space proposed in \citet{hsc_med+al18b}, which spans the optical range observed by the considered surveys. Different populations of galaxies are efficiently separated in this space and the level of contamination is consistent with zero within a $0.5\,\%$ uncertainty. The cuts are detailed in \citet[appendix A]{hsc_med+al18b}.

The distribution of sources per survey that would be considered as background for a lens plane at $z_\text{lens}=0.4$ is shown in Fig.~\ref{fig_hist_1D_background_sources_z_040}.




\section{WL signal}
\label{sec_wl_sign}


The lensing signal is recovered from the measured shapes of the background galaxies.

\subsection{Signal definition}

As the main observable, we consider the tangential reduced excess surface density $\Delta \Sigma_{g{\rm t}}$, which can be expressed in terms of the surface density, $\Sigma$, and of the tangential excess surface density $\Delta \Sigma_\text{t}$. For an axially symmetric lens and a single source plane, 
\begin{equation}
\Delta \Sigma_{g{\rm t}}(R) = \frac{\Delta \Sigma_\text{t}(R)}{1  - \Sigma_\text{cr}^{-1}\Sigma (R)}\ ,
\end{equation}
where $R$ is the transverse proper distance from the assumed lens centre, and $\Sigma_\text{cr}$ is the critical density for lensing,
\begin{equation}  
\label{eq_Delta_Sigma_2}
\Sigma_\text{cr} \equiv \frac{c^2}{4\pi G} \frac{D_\text{s}}{D_\text{l} D_\text{ls}}, 
\end{equation}
where $c$ is the speed of light in vacuum, $G$ is the gravitational constant, and $D_\text{l}$, $D_\text{s}$, and $D_\text{ls}$ are the angular diameter distances to the lens, to the source, and from the lens to the source, respectively.

The tangential reduced excess surface density can be rewritten in terms of the reduced tangential shear, \mbox{$g_\text{t}=\gamma_\text{t}/(1 - \kappa)$}, where $\gamma_\text{t}$ is the tangential shear, and $\kappa \equiv \Sigma/\Sigma_\text{cr}$ is the convergence, as
\begin{equation}
\label{eq_signal_1}
\Delta \Sigma_{g{\rm t}}(R) = \Sigma_\text{cr} g_\text{t} (R)  \,.
\end{equation}

For a population of sources distributed in redshift, the source averaged excess surface mass density can be approximated as \citep{ume20}
\begin{equation}
\label{eq_esd_2}
\langle \Delta \Sigma_{g\text{t}}  \rangle \simeq \frac{ \Delta \Sigma_\text{t}}{1- \langle \Sigma_\text{cr}^{-1}  \rangle \Sigma} \;.
\end{equation}

The equations here presented hold for non-axially symmetric lenses too if we consider azimuthally averaged quantities.

\subsection{Signal measurement}

If $\langle e_\alpha \rangle$, i.e., the ensemble average of the shape measurements $e_\alpha$, where $\alpha=1,2$ denotes the shear component in the Cartesian plane, is an unbiased estimator for the reduced shear  \citep{sc+ka95,se+sc97,be+ja02,mil+al13},
\begin{equation} 
\label{eq_shape_2}
\langle e_1, e_2 \rangle =  \langle g_1, g_2 \rangle \;,
\end{equation} 
the reduced excess surface density $\Delta\Sigma_{g{\rm t}}$ in circular annuli can be estimated from the sum over the sources inside each annulus as
\begin{equation}
\label{eq_signal_2}
\Delta \Sigma_{g{\rm t}} (R)  =  \frac{\sum_i  w_{\Delta \Sigma, i}  e_{\text{t},i} \Sigma_{\text{cr},i}} {\sum_i w_{\Delta \Sigma, i} },
\end{equation}
where
\begin{equation}
\label{eq_signal_3}
w_{\Delta \Sigma, i}  =  \Sigma_{\text{cr},i}^{-2} w_i \;,
\end{equation}
and $e_{\text{t},i}$ is the tangential component of the ellipticity of the $i$-th source galaxy, $w_i$ is the weight assigned to the source ellipticity, and $\Sigma_{\text{cr},i}$ is the critical density for the $i$-th source.
The sum runs over the galaxies included in the annulus centred at $R$.

In the following, we use the notation $\langle ... \rangle_{\Delta \Sigma}$ for a weighted average, where $w_{\Delta \Sigma, i}$ are the weights.
With this notation, Eq.~\eqref{eq_signal_2} can be rewritten as
\begin{equation}
\label{eq_signal_4}
\Delta \Sigma_{g{\rm t}} (R)  =  \langle   e_{\text{t},i} \Sigma_{\text{cr},i} \rangle_{\Delta \Sigma} \;.
\end{equation}

Some shape measurements, e.g., the ellipticity \mbox{$|\epsilon|= (1-q) / (1+q)$}, where $q$ is the image axial ratio, from \texttt{lensfit} or \texttt{IM3SHAPE}, fulfil Eq.~\eqref{eq_shape_2}, i.e., $\epsilon_i = e_i$. For other shape estimates, e.g., the distortion $|\delta|= (1-q^2) / (1+q^2)$ measured by the re-Gauss algorithm \citep{man+al08}, \texttt{GALSIM} \citep{row+al15}, or KSB-like algorithms \citep{kai+al95}, one must account for the responsivity ${\cal R}$:
\begin{equation}
e_i = \frac{\delta_i}{2 {\cal R}} \,.    
\end{equation}
The responsivity can be calculated based on the inverse variance weights and the per-object estimates of the RMS distortion $\delta_{\text{RMS},i}$ as \citep{hsc_man+al18}
\begin{equation}
\label{eq_signal_5}
{\cal R} \simeq 1 - \langle \delta^2_{\text{RMS},i}  \rangle_{\Delta \Sigma} \;.
\end{equation}
For our analysis, a responsivity calculation is needed to process the HSC-SSP data.

The source averaged reduced excess surface mass density can be approximated as in Eq.~(\ref{eq_esd_2}) with the $\langle ... \rangle_{\Delta \Sigma}$ average,
\begin{equation}
\label{eq_signal_6}
\langle \Delta \Sigma_{g\text{t}}  \rangle_{\Delta \Sigma} \simeq \frac{ \Delta \Sigma_\text{t}}{1- \langle \Sigma_\text{cr}^{-1}  \rangle_{\Delta \Sigma} \Sigma} \;.
\end{equation}



We compute distances to the sources and critical surface densities in Eq.~\eqref{eq_Delta_Sigma_2} based on the photo-$z$ point-estimator. Methods exploiting the per-source photo-$z$ probability density function, see, e.g., \citet{she+al04}, or the ensemble source redshift distribution, see, e.g., \citet{kids_hil+al20}, have been advocated too. Some methods need inherently unbiased and accurate representations of the redshift probability distribution and of systematic uncertainties, which might be difficult to achieve \citep{hsc_tan+al18,kids_hil+al20}. However, the level of systematic errors introduced by either the point-estimator or the probability density function for quality photo-$z$s and robust selections is usually subdominant for Stage-III surveys, see e.g., \citet{bel+al19}. 

The present version of the \Euclid data processing pipeline for cluster weak lensing relies on photo-$z$ point-estimators. Pros and cons of different approaches are discussed in \citet{lea+al22} and references therein.

We measure the average reduced excess surface density $\Delta\Sigma_{g\text{t}}$ in eight radial circular annuli equally separated in logarithmic space spanning the range between $R_\text{min} = 0.3\,\hMpc$ ($\sim0.43\,\text{Mpc}$) and $R_\text{max} =3.0\,\hMpc$ $(\sim 4.3\,\text{Mpc})$ from the cluster centre. 


\subsection{Calibration}

The raw shape components of the source galaxies, $e_{\text{raw}, 1}$ and $e_{\text{raw}, 2}$, can exhibit a bias that can be parameterised by a multiplicative ($m$) and an additive ($c$) component,
\begin{equation}
\label{eq_calibration_1}
e_{i} = \frac{e_{\text{raw}, i} - c_i}{1 + m}  \, \hspace{1cm} (i=1,2) \, ,
\end{equation}
which must be calibrated.
If needed, we correct each galaxy for the additive bias, whereas the multiplicative bias $m$ is averaged in each annulus \citep{hey+al12,mil+al13,vio+al15},
\begin{equation}
\label{eq_Delta_Sigma_5}
\langle m \rangle = \frac{\sum_i w_{\Delta \Sigma, i} m_i}{\sum_i w_{\Delta \Sigma, i}} \;.
\end{equation}


\subsection{Signal-to-noise}

The S/N of the WL cluster can be defined in terms of the weighted excess surface density $\Delta \Sigma_{g{\rm t}}$ in the relevant radial range $ R_\text{min} <R< R_\text{max}$ \citep{ser+al17_psz2lens},
\begin{equation}
\label{eq_SNR_1}
(\text{S/N})_\text{WL}= \frac{\Delta \Sigma_{g{\rm t}} ( R_\text{min} <R< R_\text{max})}{\delta_\text{t}} \;,
\end{equation}
where the noise $\delta_\text{t}$ includes statistical uncertainties, and cosmic noise (if relevant) added in quadrature.

For our analysis, we consider $(\text{S/N})_\text{WL}$ between $R_\text{min} = 0.3\,\hMpc$ and $R_\text{max} =3.0\,\hMpc$ from the cluster centre.

\section{Mass inference}

Cluster masses can be determined fitting the shear profiles in a fixed cosmological model. The general framework is detailed in, e.g., \citet{ser+al17_psz2lens}, \citet{ume20}, and references therein. Here we only discuss the specific setting adopted for the present analysis.


\subsection{Halo model}
\label{sec:halo_model}

We model the lens either with a simple Navarro, Frenk, White (NFW) profile \citep{nfw96}, characterised by mass, $M_{200\text{c}}$, and concentration,  $c_{200\text{c}}$, or with a composite model consisting of a Baltz, Marshall, Oguri (BMO) profile \citep{bal+al09}, e.g., a truncated NFW profile, parameterised by mass, concentration, and truncation radius, $r_\text{t}$, plus a two-halo term for the correlated matter characterised by the environment bias, $b_\text{e}$ \citep{ser+al18_psz2lens}. We refer to the second model as BMO+2-halo. The NFW model can be seen as a specific form of the BMO model, with $b_\text{e}=0$ and $r_\text{t} \rightarrow \infty$.

For the fitting parameters, we consider the logarithm (base 10) of mass and concentration, $\textbf{p} = (\log\!M_{200\text{c}}, \log\!c_{200\text{c}}$). Here, $\log\!M_{200\text{c}}$ is short for $\logten \left[ M_{200\text{c}} / \left (10^{14} M_\odot \right) \right]$.

The contribution from the uncorrelated large-scale structure (LSS) is treated as a source of noise \citep{ser+al17_psz2lens}.
The cross-correlation between two angular bins $\Delta \theta_i$ and $\Delta \theta_j$ can be written as \citep{sch+al98b,hoe03}
\begin{equation} 
\label{eq_lss_1}
\langle \Delta \Sigma_\text{LSS}(\Delta \theta_i)  \Delta \Sigma_\text{LSS}(\Delta \theta_j) \rangle = 2 \pi  \Sigma_\text{cr}^2 \int_0^{\infty} P_k(l)g(l, \Delta \theta_i) g(l, \Delta \theta_j) \ l \ dl \ ,
\end{equation}
where $P_k(l)$ is the effective projected power spectrum of lensing. We compute the linear matter power spectrum with a semi-analytical fitting function \citep{ei+hu98}, which is adequate for the precision needed in our analysis. The effects of non-linear evolution are accounted for with the revised halofit model \citep{tak+al12}. The function $g$ is the filter. In an angular bin $\theta_1 < \Delta \theta< \theta_2$,
\begin{equation} 
g=\frac{1}{\pi(\theta_1^2 -\theta_2^2)l} \left[ \frac{2}{l} \left( J_0(l \theta_2)  -J_0(l \theta_1) \right) +\theta_2 J_1(l \theta_2) -\theta_1 J_1(l \theta_1) \right].
\end{equation}


\subsection{Inference}
\label{sec:mass_infer}

The lens parameters are measured with a Bayesian analysis, where the posterior probability density function of the parameters, $\textbf{p}$, given the data, $\{{\langle \Delta\Sigma_{g{\rm t}}}\rangle \}$, can be written as
\begin{equation} 
p(\textbf{p}| \{{\langle \Delta\Sigma_{g{\rm t}}} \rangle \})  \propto {\cal L}(\{\langle{\Delta\Sigma_{g{\rm t}}} \rangle\} | \textbf{p}) p_\text{prior}(\textbf{p}) \;,
\end{equation} 
where \textbf{p} is a vector including the model parameters, $\cal L$ is the likelihood, and $p_\text{prior}$ is the prior.

The likelihood is ${\cal L}\propto \exp (-\chi^2/2)$, where  $\chi^2$ is written as
\begin{equation}
\chi^2 = \sum_{i,j}  \left[ 
\langle \Delta\Sigma_{g{\rm t}}\rangle_i  
- \Delta \Sigma_{g{\rm t}} (R_i | \textbf{p}) 
\right]^\text{t} \tens{C}_{ij}^{-1}    
\left[
\langle \Delta\Sigma_{g{\rm t}}\rangle_ij 
- \Delta \Sigma_{g{\rm t}} (R_j | \textbf{p}) 
\right] \;;
\end{equation}
the sum extends over the radial annuli; $\Delta \Sigma_{g{\rm t}} (R_i | \textbf{p})$ is the halo model computed at the lensing weighted radius $R_i$ of the $i$-th bin \citep{ser+al17_psz2lens}; $\langle \Delta\Sigma_{g{\rm t}}\rangle_i$ is the measured reduced excess surface density in the $i$-th bin.

Shape noise, $\delta\Delta \Sigma_\text{Stat}$, and lensing from LSS, $\Delta \Sigma_\text{LSS}$, are treated as uncertainties. The total uncertainty covariance matrix is
\begin{equation}
\label{eq_inf_2}
\tens{C}= \tens{C}^\text{stat}+ \tens{C}^\text{LSS} \;,
\end{equation}
where $\tens{C}^\text{stat}$ accounts for the uncorrelated statistical uncertainties in the measured shear, and $\tens{C}^\text{LSS}_{ij} = \langle \Delta \Sigma_\text{LSS}(\Delta \theta_i)  \Delta \Sigma_\text{LSS}(\Delta \theta_j) \rangle$, where $\Delta \theta_i$ is the $i$-th annular bin, is due to LSS \citep{ser+al17_psz2lens,ume+al20}.

The main source of noise is the intrinsic ellipticity dispersion $\sigma_{e_\alpha}$. For \Euclid, a reference value of $\sigma_{e_\alpha} = 0.26$ was estimated from a sample of galaxies observed with the Hubble Space Telescope and with similar photometric properties to those expected from \Euclid \citep{sch+al18,euc_pre_mar+al19,euclid_pre_aja+al23}.


Here, we do not consider correlated shape noise due to intrinsic alignment of sources, which can be neglected for Stage-III analyses of cluster lensing \citep{mcc+al19,ume+al20} but it might play a role for \Euclid and Stage-IV analyses of clusters and their outskirts \citep{ser+al18_psz2lens}.


\subsection{Priors}

We consider non-informative, uniform priors in log-space, as suitable for positive quantities, with $-1~<~\log\!M_{200\text{c}}~<~2$ and $0 < \log\!c_{200\text{c}} < 1$.

The truncation radius and environment bias are fixed. For the BMO+2-halo model, the truncation radius is set to $r_\text{t} = 3 \, r_{200\text{c}}$, and the environment bias $b_\text{e}$ is fixed with a Dirac delta prior as a function of the peak height $\nu$, $b_\text{e}=b_\text{h}[\nu(M_\text{200c},z)]$ \citep{tin+al10}. 


\section{Consistency tests}

To detect the degree of any potential discrepancy between independent results, we utilise two metrics.  Let $\mathbf{Q}_a$ and $\mathbf{Q}_b$ be two sets of parameters, with total uncertainty expressed through the covariance matrix $\tens{C}_{\mathbf{Q}_{a,b}}$. The $\chi^2$ can be defined as
\be
\label{eq_cons_1}
\chi_{a,b}^2 = (
\vec{Q}_a - \vec{Q}_b)^\text{t}
\tens{C}_{\mathbf{Q}_{a,b}}^{-1}
(\vec{Q}_a - \vec{Q}_b) \;.
\ee
For uncorrelated data, the covariance matrix is diagonal with the diagonal terms given by the sum of the squared uncertainties,
\be
\chi_{a,b}^2 = \frac{(\vec{Q}_a - \vec{Q}_b)^2}{\vec{\delta Q}_a^2 + \vec{\delta Q}_b^2} \;,
\ee
where $\vec{\delta Q}_a$ and $\vec{\delta Q}_b$ are the uncertainties on $\vec{Q}_a$ and $\vec{Q}_b$, respectively.

Given a $\chi^2$ distribution with $N_\text{dof}$, degrees of freedom, we can calculate the probability to exceed a given value, $p(\chi^2 > \chi_{a,b}^2; N_\text{dof})$, and use this probability as a metric for comparison.


When we compare a scalar quantity $Q_i$, we consider
\be
\Delta_{\chi} = \chi_{a,b} = \frac{ Q_a - Q_b}{ \delta_\Delta};,
\ee
where $\delta_\Delta = \sqrt{\delta Q_a^2 + \delta Q_b^2}$.

As a second metric, we consider a function based only on the point-estimates. We compute the CBI of the differences,
\be
\label{eq_cons_2}
\Delta_\text{CBI}  = \text{CBI} (\vec{Q}_a - \vec{Q}_b) \;,
\ee
and the associated SBI, 
\be
\label{eq_cons_3}
\Delta_\text{SBI}  = \text{SBI} (\vec{Q}_a - \vec{Q}_b) \;.
\ee
The estimator $\Delta_\text{SBI}$ quantifies the dispersion of the results. The uncertainty on $\Delta_\text{CBI}$, $\delta\Delta_\text{CBI}$, can be computed by bootstrapping the sample and computing the SBI of the summary CBI statistics.
Any value of $\Delta_\text{CBI}$ in excess of the statistical uncertainty can point to systematic effects or statistical uncertainties biased low. We report the $\Delta$ metric as $\Delta_\text{CBI}(\pm\delta\Delta_\text{CBI})\pm \Delta_\text{SBI}$.
 
The main observable quantities we consider for comparison are the radial profiles of the reduced excess surface density, $\vec{Q}_{a} = \{ \langle \Delta \Sigma_{g\text{t}, \text{survey}}(R_i) \rangle \}$, with $N_\text{dof}$ given by the number of bins, and the lens masses of a sample, $\vec{Q}_{a} = \{ \log\!M_{200\text{c},\,\text{survey}, i} \}$, with $N_\text{dof}$ given by the number of clusters.

When we compare the total signal $\Delta \Sigma_{g{\rm t}} ( R_\text{min} <R< R_\text{max})$, we use $\Delta_\chi$ as the main metric.


\section{WL signal accuracy}
\label{sec_wl_sign_acc}

\begin{figure}
\begin{tabular}{c}
\resizebox{\hsize}{!}{\includegraphics{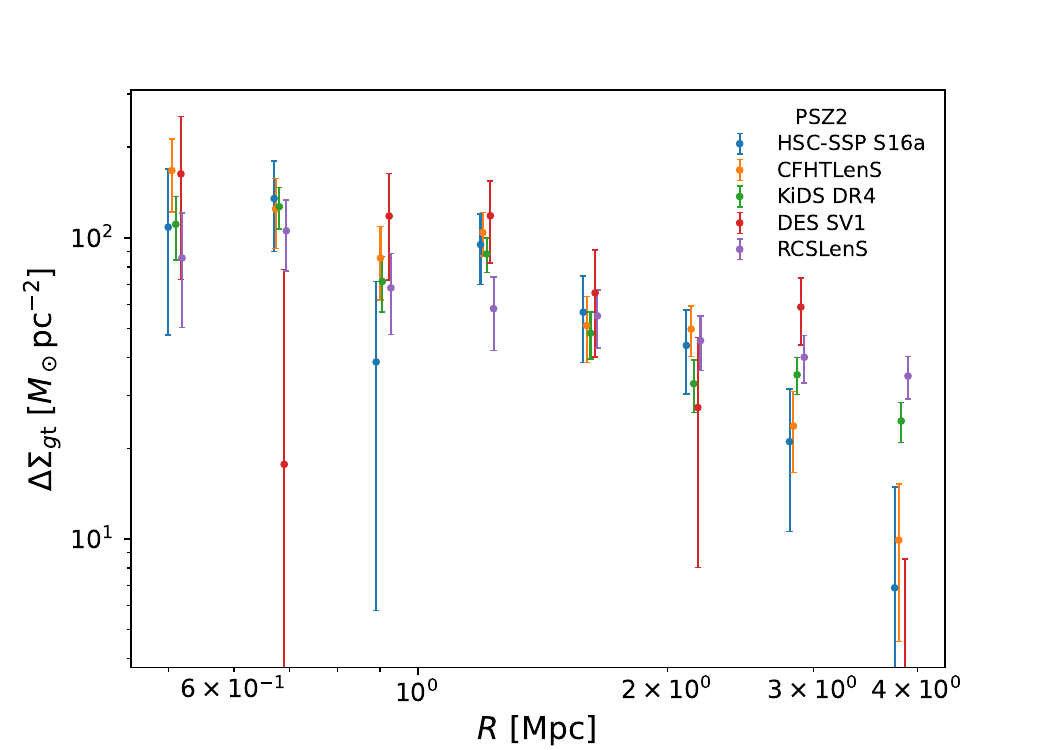}} \\
\noalign{\smallskip}  
\resizebox{\hsize}{!}{\includegraphics{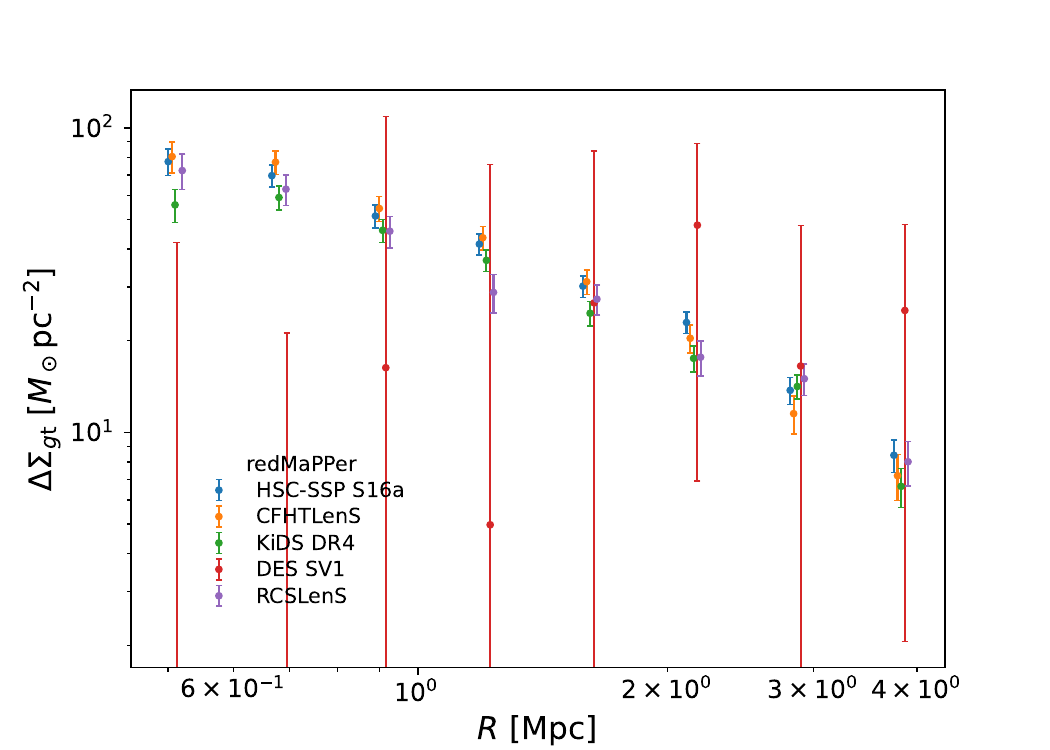}} 
\end{tabular}
\caption{Average reduced excess surface density profiles of clusters detected in public surveys as a function of $R$, the transverse proper distance from the lens centre,
coded by colours as in the legend. \emph{Top}: the PSZ2 clusters. \emph{Bottom}: the redMaPPer sample.}
\label{fig_planck_average_profiles}
\end{figure}

\begin{table}
\caption{
Comparison of the average reduced excess surface density profiles of the \Planck clusters detected in different surveys. We quantify the level of agreement from signals extracted from different surveys. For each pair of surveys, we report the $\chi^2_{a,b}$ between the average profiles ($N_\text{dof} =8$), and (in round brackets) the probability to exceed this value $p_{a,b} = p\left(\chi^2 > \chi_{a,b}^2\right)$. The metric $\Delta$ is reported in the form $\Delta_\text{CBI}\pm \Delta_\text{SBI}$.
}
\label{tab_surv_shear_psz2}
\centering
\resizebox{\hsize}{!} {
\begin{tabular}[c]{l  r r r r}
	\hline
	\noalign{\smallskip}  
					&	CFHTLenS &	KiDS DR4 	&	 DES SV1		&RCSLenS	 \\
	\noalign{\smallskip}  
	\hline
	\noalign{\smallskip}  
	HSC-SSP S16a	&	$\chi^2_{a,b}=3.5$ ($p_{a,b}=0.90$)	&	7.6 (0.48)		&	10.4 (0.24)	& 13.0 (0.11) 	\\
                    &   $\Delta =0.11\pm0.23$               &   $0.35\pm0.21$   &   $0.18\pm0.33$   & $0.59\pm0.23$   \\
	CFHTLenS		&	--		&	11.6 (0.17)	&	9.9 (0.27)		& 19.5 (0.01) 	\\
                    &   --      &   $0.21\pm0.14$   &   $0.06\pm0.27$   & $0.43\pm0.17$   \\
	KiDS DR4		&	--		&  --			&	13.0 (0.11)	& 7.2 (0.52) 	\\
                    &   --      &   --           &   $-0.13\pm0.21$   & $0.17\pm0.12$   \\
	DES SV1  		&	--  	&    --			&	--			& 16.9 (0.03) 	\\
                    &   --      &  --           &   --   & $0.35\pm0.25$   \\
	\hline
	\end{tabular}
	}
\end{table}

\begin{table}
\caption{
Same as Table~\ref{tab_surv_shear_psz2} but for the redMaPPer sample.
}
\label{tab_surv_shear_redmapper}
\centering
\resizebox{\hsize}{!} {
\begin{tabular}[c]{l  r r r r}
	\hline
	\noalign{\smallskip}  
					&	CFHTLenS &	KiDS DR4 	&	 DES SV1		&RCSLenS	 \\
	\noalign{\smallskip}  
	\hline
	\noalign{\smallskip}  
	HSC-SSP S16a	&	4.7 (0.79)          &	28.5 ($\la 0.01$)	&	4.5 (0.81)		& 11.3 (0.19) 	\\
                    &   $-0.06\pm0.0.06$    &   $-0.14\pm0.05$   &   $0.27\pm0.88$   & $-0.09\pm0.06$   \\
	CFHTLenS		&	--		&	18.8 (0.02)	&	5.0 (0.76)		& 14.4 (0.07) 	\\
                    &   --      &   $-0.08\pm0.06$   &   $0.36\pm0.94$   & $-0.03\pm0.07$   \\
    KiDS DR4		&	--		&	--			&	4.2 (0.83)		& 5.9 (0.65) 	\\
                    &   --      &   --          &   $0.48\pm1.03$   & $0.06\pm0.07$   \\
	DES SV1  		&	--  	&	--			&	--			& 4.3 (0.83) 	\\
                    &   --      &   --          &   --          & $-0.28\pm0.70$   \\
	\hline
	\end{tabular}
	}
\end{table}

To assess the accuracy to which the WL signal can be measured, we compare the average radial profiles of the reduced excess surface density of the lenses,  $\Delta \Sigma_{g\text{t, survey}}(R_i)$,  from different surveys, see Fig.~\ref{fig_planck_average_profiles}. We measure the signal in eight radial bins ($N_\text{dof} =  N_\text{bin} = 8$).

In our approach, statistically significant differences between results from different surveys are seen as indications of systematic errors. An analysis with underestimated statistical uncertainties is then conservative as it can inflate differences. When comparing the WL profiles, we only account for the total shape noise including the intrinsic shape dispersion per component and the per-component shape measurement error. We do not account for noise from either correlated or uncorrelated LSS.

Different surveys share lenses and source galaxies in overlapping regions, and intrinsic ellipticity and shape noise are correlated to some degree. This is difficult to quantify at the catalogue level. Ellipticity values measured with a different resolution, in a different galaxy radial range, with different effective radial weight functions, or in different bands might vary, as the spatial distribution of the light emission may not be identical \citep{sch+al18b}. 
We very conservatively quantify the correlation assuming that lensing signal of shared clusters is fully correlated.

We measure the reduced excess surface density which, unlike the excess surface density, brings some residual dependence on source and lens redshift, mostly in the inner regions, see Eq.~(\ref{eq_signal_6}) \citep{ume20}. Notwithstanding our simplifying and conservative assumptions, the agreement between the profiles measured in different surveys is significant both for \Planck, see Table~\ref{tab_surv_shear_psz2}, and redMaPPer clusters, see Table~\ref{tab_surv_shear_redmapper}.

Discrepancies between surveys are smaller than statistical uncertainties. This implies that shear can be accurately calibrated and known systematic effects are properly accounted for. 
For Stage-III surveys and precursors, unknown or residual systematic effects either play a negligible role or very fortuitously counter-balance each other among different surveys and data sets.

The validity of the test relies on the assumption that the cluster subsamples, which are selected from different survey footprints, are homogeneous. This assumption may be invalid if the properties of the parent samples vary spatially \citep{lea+al22}. Even if the samples are spatially homogeneous, we are sampling a small number of clusters per survey, which might not fairly sample the full parent population of dark matter haloes. On the other hand, this choice of the lens samples per survey limits the correlation of sources.
Some concerns on the lens samples can be solved by analysing lenses which are covered by more than one survey. In fact, for the four \Planck clusters covered by a pair of surveys, we found statistical agreement, see App.~\ref{sec_planck_multi}.

In App.~\ref{sec_she_cal}, we perform a cross-check between shear calibrations by comparing the WL profile of clusters extracted by matched source catalogues from different surveys to show how residual effects for Stage-III surveys and precursors are negligible with respect to the statistical uncertainties.


\section{Mass accuracy}
\label{sec_mass_accu}

In this section, we report results on the accuracy we can reach in mass measurements of cluster masses from survey data. 
The section serves a dual purpose. On one hand, we cross-validate Stage-III surveys by comparing results obtained from a uniform, consistent analysis. Agreement can indicate that the different data sets are compatible and that systematic errors in each survey are corrected to the required levels.

On the other hand, fitting procedures here employed for WL profiles and WL mass estimates can be validated by comparison with literature results.

As a reference fitting model, we considered the BMO+2-halo model. For the uncertainty budget, we considered contributions from both shape noise and LSS.

We first look for potential systematic effects by comparing results exploiting different analysis assumptions, see Sects.~\ref{sec_mass_point} and ~\ref{sec_mass_model}. Then we compare results from different surveys, see Sect.~\ref{sec_mass_doubles}, or with estimates from the literature, see Sect.~\ref{sec_mass_comp}.


\subsection{Mass point-estimators}
\label{sec_mass_point}

\begin{table}
\caption{Comparison of mass point-estimators of the PSZ2 clusters. For each estimator pair, we report $\Delta_\text{CBI}\pm \Delta_\text{SBI}$, the biweight location and scale of the difference of the point mass estimates.
}
\label{tab_mass_point_psz2}
\centering
\begin{tabular}[c]{l  r@{$\,\pm\,$}l r@{$\,\pm\,$}l r@{$\,\pm\,$}l}
	\hline
	\noalign{\smallskip}  
	Estimator			&	\multicolumn{2}{c}{CBI} 	&	\multicolumn{2}{c}{Median}	&	\multicolumn{2}{c}{Mean} \\
	\noalign{\smallskip}  
	\hline
	\noalign{\smallskip}  
	ML				&	$0.32$	&	$0.33$		&	$0.30$	&	$0.33$ 	&	$0.37$	&	$0.30$ \\
	CBI				&	\multicolumn{2}{c}{--}			&	$-0.01$	&	$0.03$ 	&	$0.02$	&	$0.05$ \\
	Median			&	\multicolumn{2}{c}{--}			&	\multicolumn{2}{c}{--	}	&	$0.03$	&	$0.08$ \\
	\hline
	\end{tabular}
\end{table}

In the regime of low signal-to-noise, the inferred probability distribution for the mass can be very skewed. The peak may not be prominent and the distribution can show a strong tail to very low values. In the case of negative $(\text{S/N})_\text{WL}$, the distribution could be better summarised by an upper limit rather than an estimate of the central location. 

In Table~\ref{tab_mass_point_psz2}, we compare different mass point-estimates of the PSZ2 clusters. We consider the maximum likelihood (ML) value as well as the biweight location CBI, the median, and the mean of the posterior probability distribution. Point-estimates can significantly differ. As expected for a distribution sampled with a long chain, estimates of the median and the CBI are in very good agreement. The agreement with the mean is also good, though with a larger dispersion. On the other hand, ML estimates are usually larger than CBI or mean estimates, even when considering the large dispersion.

Any mass point-estimate (and a related uncertainty) might fail in summarising the mass posterior, and the full probability distribution should be considered. For example, using the mass location and scale in analyses which implicitly assume a Gaussian distribution can significantly bias the results if the mass probability distribution is not Gaussian. The use of the full probability distribution is recommended in the low $(\text{S/N})_\text{WL}$ regime.

In the following, when comparing our results with literature values, we try to use the most appropriate estimator, i.e., the estimator which most resembles the properties of the comparison sample.


\subsection{Halo model}
\label{sec_mass_model}

A proper modelling of the lens is crucial for an unbiased mass determination \citep{og+ha11}. We estimate the effects of halo modelling in Stage-III surveys and precursors by comparing mass point-estimates of the PSZ2 clusters derived by modelling the lens either with BMO+2-halo or NFW profiles. Results are consistent with $\Delta_\text{CBI}\pm \Delta_\text{SBI} = -0.03\pm0.02$, $-0.05\pm0.24$, $-0.05\pm0.24$, and $-0.05\pm0.22$ for the ML, CBI, median, and mean, respectively.

Halo modelling can play a larger role when fitting the very inner regions, where the inner slope has to be properly accounted for \citep{ser+al16_einasto}, or the outer regions where the matter distribution transits to the infalling region \citep{di+kr14}, or with the better data quality expected for Stage-IV surveys. At the level of the present analysis, where we consider Stage-III surveys and precursors and exclude the inner and outer radial regions from the fitting, the role is minor. We only consider the BMO+2-halo model in the following.


\subsection{Clusters covered by multiple surveys}
\label{sec_mass_doubles}

\begin{table}
\caption{
Comparison of WL masses of redMaPPer clusters covered by multiple surveys. For each pair of surveys, we quote: the number of redMaPPer clusters in common, $N_\text{cl}$; $\chi^2_{a,b}$ and its excess probability for $N_\text{dof} = N_\text{cl}$, $p_{a,b} = p\left(\chi^2 > \chi_{a,b}^2\right)$; $\Delta$ as $\Delta_\text{CBI} (\pm \delta\Delta_\text{CBI}) \pm \Delta_\text{SBI}$, the biweight location (with associated uncertainty) and scale of the difference of the mass estimates.
}
\label{tab_mass_doubles_redmapper}
\centering
\resizebox{\hsize}{!} {
\begin{tabular}[c]{l c c c}
	\hline
	\noalign{\smallskip}
					&	CFHTLenS	                            & KiDS DR4			           & RCSLenS \\
					&	$N_\text{cl}=100$			             &	343				           &		\\
	HSC-SSP S16a	&	$\chi^2_{a,b}=73.5$ $(p_{a,b} = 0.98)$	&	185.9 ($\ls 1$)	 	        &	--	\\
					&	$\Delta =-0.01(\pm0.02)\pm0.23$	         &	$0.00(\pm 0.01)\pm0.20$		&		\\
					&				                              &	13				            &	2	\\
	CFHTLenS		&	--			                              &	5.8 (0.95)		 	     & 	0.1 (0.97)	\\
					&				                              &	$0.00(\pm 0.02)\pm0.09$		&	$-0.06(\pm 0.05)\pm0.08$	\\
					&				                              &					              &	20   	\\
	KiDS DR4		&	--			                              &	-- 				           & 9.3(0.98)\\
			 		&				                             &					          &$-0.05(\pm0.04)\pm0.18$	\\
	\hline
	\end{tabular}
	}
\end{table}

A number of redMaPPer clusters are covered by multiple surveys. For these clusters, we can directly compare the mass estimates.

Any correlation between mass estimates, which is not properly accounted for, can underestimate the statistical significance of mass differences. Three sources of correlation are the galaxies shared by different surveys in overlapping regions, the noise from uncorrelated matter, and the common fitting scheme.

Firstly, different surveys share the same source galaxies in overlapping regions, and intrinsic ellipticity and shape noise are correlated to some degree. As discussed in Sect.~\ref{sec_wl_sign_acc}, this can be quantified with working assumptions. When cross-comparing two surveys, we can assume that the deeper survey detects and measures shear for all the galaxies detected and measured by the shallower survey in the overlapping region. For example, for an hypothetical lens at $z_\text{lens}=0.38$, as is typical for redMaPPer clusters, see Table~\ref{tab_sample_redmapper}, we can assume that all background sources detected in KiDS DR4 are also selected in HSC-SSP S16a. This would account for $\sim 30\,\%$ of the full background source sample in HSC-SSP S16a. If we neglect differences in intrinsic ellipticity due to different bands or spatial extents, this would entail a correlation of $\sim 0.5$ in the measured shear.

Secondly, mass measurements of the same cluster from different surveys experience the same noise from uncorrelated matter. Even though the LSS noise is mostly negligible with respect to the shape noise for Stage-III surveys or precursors, it can still entail some correlation in the shape measurements. For a KiDS-like survey, the LSS noise is $\sim 50\,\%$ of the shape noise for a lens at $z_\text{lens}=0.38$ in the radial range we considered for fitting, which would entail a correlation of $\sim 0.2$.

When correlated shape and LSS noise are considered together, the correlation would be $\sim 0.7$ when comparing mass estimates based on either KiDS DR4 or HSC-SSP S16a.

Finally, if we consider that the use of the same pipeline for mass measurements can further correlate the estimates, we can conservatively consider a total correlation of $\sim 0.8$. We will use this estimate of the correlation for our comparison.

Results for the redMaPPer clusters are summarised in Table~\ref{tab_mass_doubles_redmapper}, where we report results for the mass biweight point-estimator, and we find agreement between different surveys. This can be seen as a consequence of the agreement between the shear profiles, discussed in Sect.~\ref{sec_wl_sign_acc}.


\subsection{Comparison with the WL mass from literature}
\label{sec_mass_comp}

\begin{figure}
\begin{tabular}{c}
\resizebox{\hsize}{!}{\includegraphics{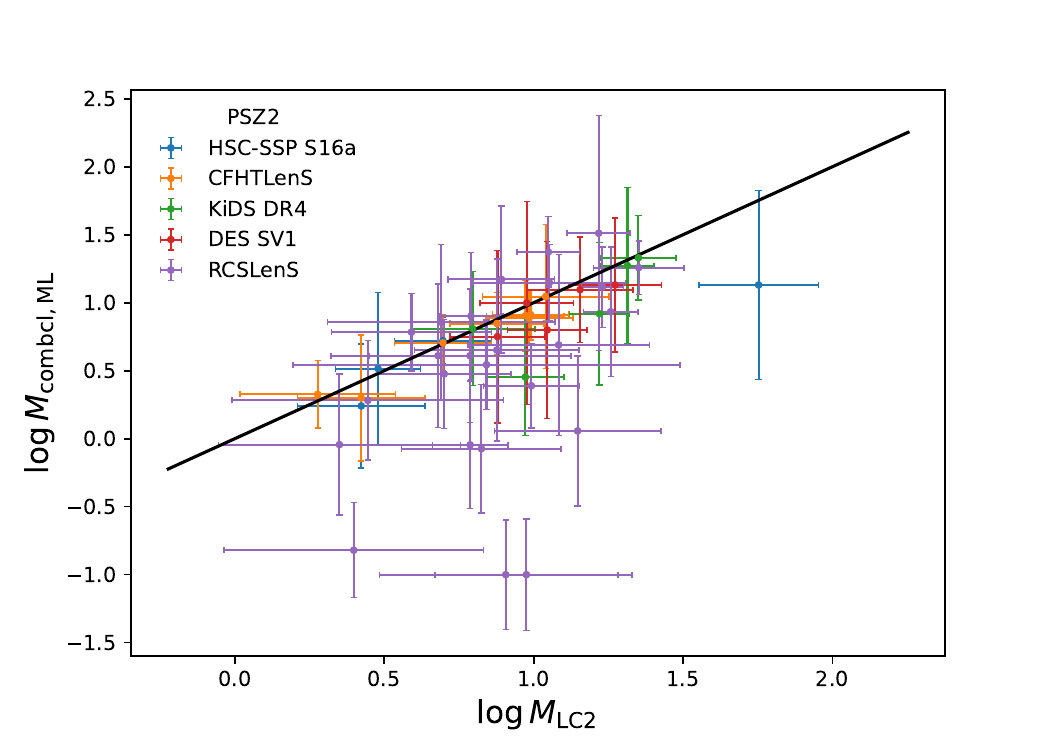}} \\
\resizebox{\hsize}{!}{\includegraphics{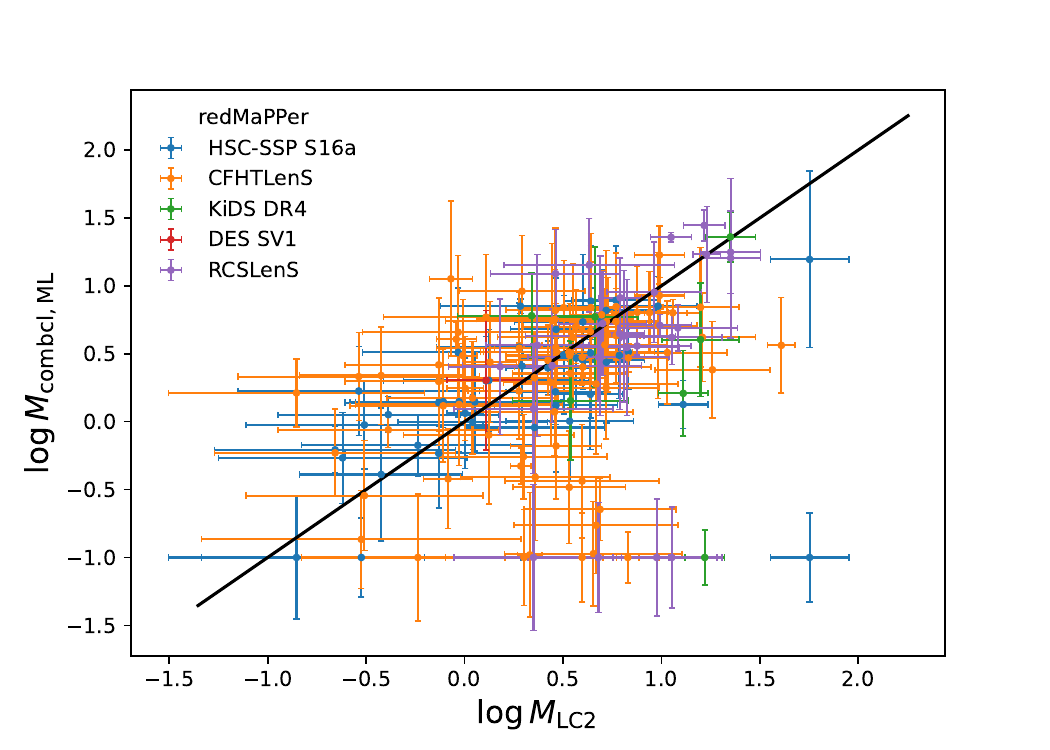}} \\
\end{tabular}
\caption{Maximum likelihood estimates of $M_{200\text{c}}$  of clusters as obtained with the \texttt{COMB-CL} analysis of public surveys (colour coded according to the legend) versus literature masses from the LC2 catalogue.  \emph{Top}: the PSZ2 clusters. \emph{Bottom}: the redMaPPer sample.}
\label{fig_psz2_lc2_combcl}
\end{figure}

\begin{table}
\caption{Comparison of \Planck cluster masses from the \texttt{COMB-CL} analysis of public surveys with literature values. For each survey, we report: the number of clusters $N_\text{cl}$; $\chi^2_{a,b}$ and its excess probability, $p_{a,b} = p\left(\chi^2 > \chi_{a,b}^2\right)$;$\Delta$, the biweight location and scale of the difference of the mass estimates. In the top part of the table, masses are compared with estimates from the LC2 catalogue. In the bottom part, with masses from the PSZ2LenS sample \citep{ser+al17_psz2lens}.
}
\label{tab_mass_lc2_planck}
\centering
\begin{tabular}[c]{l  r r r r@{$\,\pm\,$}l}
	\hline
    \noalign{\smallskip} 
	\multicolumn{6}{c}{vs LC2} \\
	\noalign{\smallskip}  
	Survey			&	$N_\text{cl}$ 	&	 $\chi^2_{a,b}$		& $p_{a,b}$ &\multicolumn{2}{c}{$\Delta_\text{CBI} \pm \Delta_\text{SBI}$} \\
	\noalign{\smallskip}  
	\hline
	\noalign{\smallskip}
	HSC-SSP S16a	&	4	&	1.00	&	 0.91 	& 0.06 & 0.26\\
	CFHTLenS		&	8	&	0.52	&	 $\ls1$ 	& 0.04 & 0.06\\
	KiDS DR4		&	5	&	1.85	&	 0.87 	& 0.03 & 0.14\\
	DES SV1  		&	5	&	0.30	&	 $\ls1 $	&0.11 & 0.09\\
	RCSLenS			&	23	&	25.6	&	 0.37 	&0.19 & 0.43\\
	\hline
	\noalign{\smallskip} 
    \noalign{\smallskip} 
    \multicolumn{6}{c}{vs PSZ2LenS} \\
	\noalign{\smallskip}  
	Survey			&	$N_\text{cl}$ 	&	 $\chi^2_{a,b}$		& $p_{a,b}$ &\multicolumn{2}{c}{$\Delta_\text{CBI} \pm \Delta_\text{SBI}$} \\
	\noalign{\smallskip}  
	\hline
	\noalign{\smallskip}  
	HSC-SSP S16a	&	2	&	0.17	&	 0.92 	& 0.08 & 0.11\\
	CFHTLenS		&	8	&	0.52	&	 $\ls1$ 	& 0.04 & 0.06\\
	KiDS DR4		&	1	&	1.50	&	 0.22 	&\multicolumn{2}{c}{$0.52$}\\
	RCSLenS			&	22	&	20.3	&	 0.56 	&0.21 & 0.33\\
	\hline
	\end{tabular}
\end{table}

\begin{table}
\caption{Comparison of redMaPPer cluster masses from the \texttt{COMB-CL} analysis of public surveys with literature values. For each survey, we report: the number of clusters $N_\text{cl}$; $\chi^2_{a,b}$ and its excess probability, $p_{a,b} = p\left(\chi^2 > \chi_{a,b}^2\right)$; $\Delta$, the biweight location and scale of the difference of the mass estimates. In the top part of the table, masses are compared with estimates from the LC2 catalogue. In the bottom part, with masses from the HSC-XXL sample \citep{ume+al20}.}
\label{tab_mass_lc2_redmapper}
\centering
\begin{tabular}[c]{l  r r r r@{$\,\pm\,$}l}
	\hline
	\noalign{\smallskip} 
    \multicolumn{6}{c}{vs LC2} \\ 
	\noalign{\smallskip}  
	Survey			&	$N_\text{cl}$ 	&	 $\chi^2_{a,b}$		& $p_{a,b}$ &\multicolumn{2}{c}{$\Delta_\text{CBI} \pm \Delta_\text{SBI}$} \\
	\noalign{\smallskip}  
	\hline
	\noalign{\smallskip}  
	HSC-SSP S16a	&	40	&	41.9	&	 0.39 	& $-0.06$ & 0.34\\
	CFHTLenS		&	61	&	72.5	&	 0.15 	& 0.00 & 0.47\\
	KiDS DR4		&	6	&	11.9	&	 0.06 	& 0.21 & 0.48\\
	DES SV1  		&	1	&	0.15	&	 0.70	&\multicolumn{2}{c}{$-0.20$}\\
	RCSLenS			&	23	&	24.7	&	 0.38 	&0.05 & 0.27\\
    \hline
    \noalign{\smallskip} 
	\noalign{\smallskip} 
    \multicolumn{6}{c}{vs HSC-XXL} \\
	\noalign{\smallskip}  
	Survey			&	$N_\text{cl}$ 	&	 $\chi^2_{a,b}$		& $p_{a,b}$ &\multicolumn{2}{c}{$\Delta_\text{CBI} \pm \Delta_\text{SBI}$} \\
	\noalign{\smallskip}  
	\hline
	\noalign{\smallskip}  
	HSC-SSP S16a	&	33	&	40.2	&	 0.18 	& 0.04 & 0.45\\
	CFHTLenS		&	30	&	45.8	&	 0.03 	& 0.04 & 0.45\\
    \hline
	\end{tabular}
\end{table}

To assess the robustness of WL mass estimates, we compare masses of PSZ2 or redMaPPer clusters obtained using \texttt{COMB-CL} on survey data with literature values. We first consider the LC2 meta-catalogues \citep{ser15_comalit_III} in Sect.~\ref{sec_LC2}, and then two smaller, but homogeneous and statistically complete samples, PSZ2LenS \citep{ser+al17_psz2lens} in Sects.~\ref{sec_PSZ2LenS}, and HSC-XXL \citep{ume+al20} in \ref{sec_HSC-XXL}.
LC2 is mostly based on follow-up, targeted observations independent of the survey data we consider here. By comparison, we can test the robustness of the mass estimates to data and analysis systematics. On the other hand, PSZ2LenS or HSC-XXL have to a large extent exploited  the same survey data considered here. By comparing with their results, we can gain insight into the robustness to analysis choices.

The degree of correlation of our mass uncertainties with literature results from a meta-catalogue is difficult to quantity but it is expected to be smaller than for a mass comparison between results exploiting overlapping survey data and a uniform pipeline. Literature results are usually based on targeted observations that cover a smaller radial extent than a survey. This often leads to a small fraction of common source galaxies. Furthermore, methods used to select background galaxies or to measure galaxy ellipticity follow very heterogeneous pipelines, which can further lower the degree of correlation due a shared subsample of source galaxies.

On the other hand, noise from correlated or uncorrelated matter can make the measurements correlated. Following Sect.~\ref{sec_mass_doubles}, we can estimate a correlation of $0.2$ due to LSS noise, and we use this estimate when comparing our masses with masses from LC2.

\subsubsection{LC2}
\label{sec_LC2}

 As a first comparison sample, we consider LC2, a large compilation of WL masses retrieved from the literature and periodically updated \citep{ser15_comalit_III}.\footnote{The catalogues are available at \url{http://pico.oabo.inaf.it/\textasciitilde sereno/CoMaLit/LC2/}.} The latest compilation (v3.9) lists 1501 clusters and groups (806 unique) with measured redshift and WL mass from 119 bibliographic sources. The catalogues report coordinates, redshift, WL masses to over-densities of 2500, 500, 200, and to the virial radius, and spherical WL masses within 0.5, 1.0, and 1.5\,Mpc in a reference cosmological model. 

We identify counterparts in the LC2 catalogue by matching with clusters from the lens samples whose redshifts differ for less than $\Delta z = 0.05\,(1+z_\text{lens})$ and whose projected distance in the sky does not exceed 10\arcminute. We found 47 (44 unique) matches for PSZ2 and 158 (119 unique) matches for redMaPPer. 
We consider a threshold in arcmin since the \Planck positional accuracy is driven by the angular PSF. Matching results do not significantly change considering thresholds in proper lengths, e.g., a projected distance in the sky that does not exceed $1\,\text{Mpc}$.

The ML estimator is often considered for mass point-estimates. To ease comparison with literature values, we adopt this estimator. We find no evidence for disagreement, see Fig.~\ref{fig_psz2_lc2_combcl}, and Tables~\ref{tab_mass_lc2_planck} and \ref{tab_mass_lc2_redmapper}. For the quantitative comparison, we excluded $\num{11}$ clusters with low $(\text{S/N})_\text{WL}$ whose mass estimate collapses on the lower bound of the prior range and for which the median or mean estimator would have been more appropriate. Results do not significantly depend on their exclusion.


Some of the results collected in the LC2 catalogues were based on survey data. For example, \citet{ser+al17_psz2lens} and \citet{hsc_med+al18b} studied PSZ2 clusters covered by CFHTLenS/RCSLenS or HSC-SSP 16A, respectively. This makes the results correlated to some degree. However, the LC2 sample is heterogeneous and mostly-based on follow-up, independent, targeted observations.


\subsubsection{Comparison with PSZ2LenS}
\label{sec_PSZ2LenS}

\citet{ser+al17_psz2lens} studied the PSZ2LenS sample, i.e., the PSZ2 clusters covered by CFHTLenS and RCSLenS. Most of the LC2 matches with PSZ2 are from their analysis. Considering only the matches with PSZ2LenS, we lose some statistical power with respect to the full LC2 vs. PSZ2 comparison, but we can exploit a better defined, statistically complete, and homogeneous comparison sample. We find consistent results, see Table~\ref{tab_mass_lc2_planck}, notwithstanding some notable differences with respect to \citet{ser+al17_psz2lens}.

Here, we are interested in survey results, and we consider lens properties as reported in the cluster catalogues without further elaboration. \citet{ser+al17_psz2lens} re-examined the \Planck candidates, re-centred them to the brightest cluster galaxy (BCG), and re-examined the cluster redshifts reported in the catalogue. However, miscentring effects were found to be small \citep{ser+al17_psz2lens}.

To perform a multi-survey analysis, we select background galaxies with a colour cut in $r-z$ and $g-i$. This is very convenient for deep surveys such as HSC-SSP. For a much shallower survey such as RCSLenS, a colour selection in $g-r$ and $r-i$, as done in \citet{ser+al17_psz2lens}, may be more convenient to collect a larger number of low redshift sources and boost the WL signal. 

Finally, \citet{ser+al17_psz2lens} considered a uniform prior for the mass, which may favour larger masses with respect to our prior which is flat in log-space. Notwithstanding the differences in analyses, agreement is good, see Table~\ref{tab_mass_lc2_planck}.


\subsubsection{HSC-XXL}
\label{sec_HSC-XXL}

\citet{ume+al20} performed a WL analysis of 136 spectroscopically confirmed X-ray detected galaxy groups and clusters selected from the XMM-XXL survey \citep{xxl_I_pie+al16,xxl_XX_ada+al18} in the $25\deg^2$ XXL-North region and covered by the HSC-SSP first-year data. The overlap with the redMaPPer sample is significant, see Table~\ref{tab_mass_lc2_redmapper}, and we can perform a comparison with a homogeneous analysis from the literature. 
WL masses reported in \citet{ume+al20} were rescaled to our reference cosmological model following \citet{ser15_comalit_III}.

\citet{ume+al20} exploited the same HSC-SSP data release used in the present analysis, and we consider the same photo-$z$ estimates, but some noteworthy differences still remain. \citet{ume+al20} considered the radial range within a comoving cluster-centric radius of $3.0\,\hMpc$, whereas we consider proper lengths. They did not perform background selection in the colour-colour space. They adopted a NFW model with log-uniform priors. \citet{ume+al20} centred clusters in the X-ray peak, which can differ from the redMaPPer centre. Notwithstanding the differences, our analyses share the WL data set (for HSC-SSP-16A) and some major assumptions. \citet{ume+al20} considered the biweight location as the mass point-estimate and to be consistent, only for the sake of this comparison, we consider the same estimator. Results are in good agreement, see Table~\ref{tab_mass_lc2_redmapper}.

\subsection{Discussion}

The agreement between mass measurements, estimated with a uniform fitting scheme, of the same clusters covered by multiple surveys further supports the agreement between the different data sets from Stage-III surveys and precursors, and the agreement of the inferred shear profiles. The agreement between shear profiles is a prerequisite for consistent mass measurements. Therefore, if we check for mass agreement, we also check that this prerequisite is met.

The agreement of our mass measurements with literature results derived both from large, heterogeneous compilations, or smaller but homogeneous analyses, demonstrates that the fitting procedures for WL mass estimates are robust at the level required by Stage-III surveys.

We do not correct for miscentring or residual cluster member contamination, which reduce the signal and can make the derived masses systematically lower than what was found in dedicated analyses from the literature. On the other hand, correlation in shear estimates can lower the level of disagreement.
However, our comparison showed that different treatments of the halo model, line-of-sight projections, correlated matter or LSS, triaxiality, contamination and membership dilution, miscentring, or priors still yield WL mass estimates consistent within the statistical uncertainty for Stage-III surveys and precursors.

Our analysis suggests that residual or unknown systematic effects are sub-dominant with respect to known effects for Stage-III surveys and precursors. Comparison of WL masses of redMaPPer clusters covered by multiple surveys, see Table~\ref{tab_mass_doubles_redmapper}, shows a mass accuracy of $\sim 1 \pm 2\,\%$. This result is mostly driven by the cross-comparison of the HSC-SSP S16a and the KiDS DR4 surveys, with 343 redMaPPer clusters in the overlapping fields, but cross-checks are also consistent for other surveys.

The cross-comparison under a unified scheme can show biases at the catalogue level, such as biases due to the calibration of either shear or photo-$z$ measurements, but might be insensitive to modelling assumptions concerning, e.g., miscentring, or background selection.
Dependence on the mass point estimator, see Sect.~\ref{sec_mass_point} and Table~\ref{tab_mass_point_psz2}, or halo model, see Sect.~\ref{sec_mass_model}, is subcritical.
The total level of systematics can be assessed by comparison with literature values. Considering the 130 (with duplicates) redMaPPer clusters with known WL masses covered by the surveys under considerations, see Table~\ref{tab_mass_lc2_redmapper}, we infer an accuracy of $\sim 1 \pm 8\,\%$. This estimate could be inflated due to the heterogeneity of the comparison sample.



\section{Mass precision}

The Euclid Survey will cover about $\num{15000}\,\deg^2$ of the extra-galactic sky \citep{euclid_pre_sca+al22} and \Euclid will deliver an unprecedented number of clusters with high $(\text{S/N})_\text{WL}$. In this section, we want to discuss the statistical precision of WL mass measurements in Stage-III surveys and the expectations for \Euclid. We first discuss the status quo for mass measurements in ongoing or completed surveys. Then, we make a forecast for \Euclid with a semi-analytical approach. Finally, we extrapolate the results from Stage-III surveys and precursors to make a data-driven forecast for \Euclid.


\subsection{Intermediate and massive clusters}
\label{sec_mass_prec_large}

\begin{figure}
\begin{tabular}{c}
\resizebox{\hsize}{!}{\includegraphics{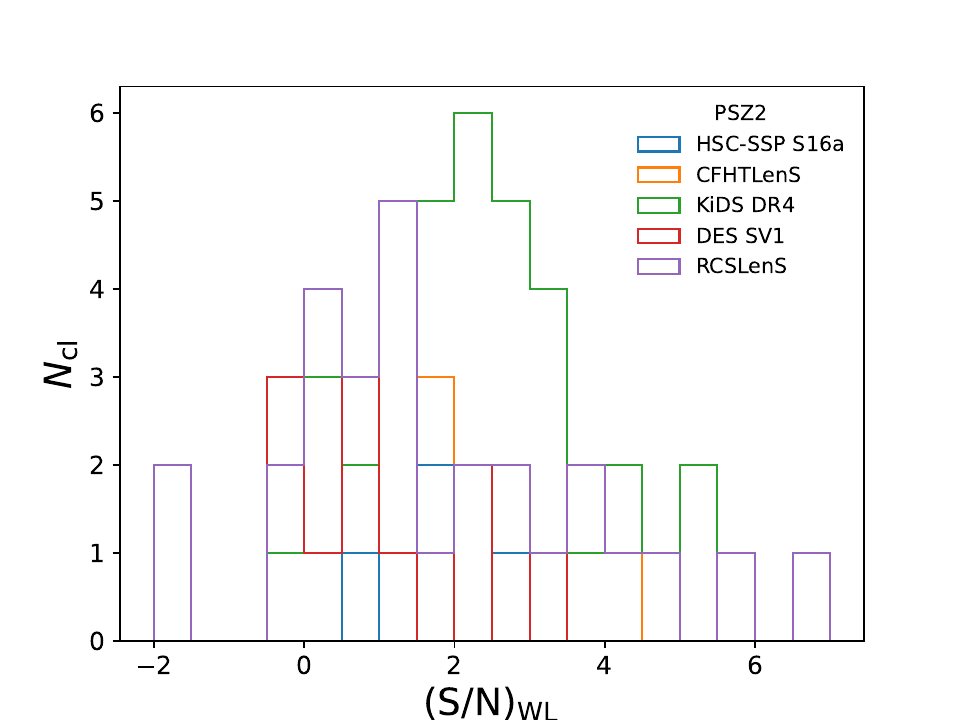}} \\
\resizebox{\hsize}{!}{\includegraphics{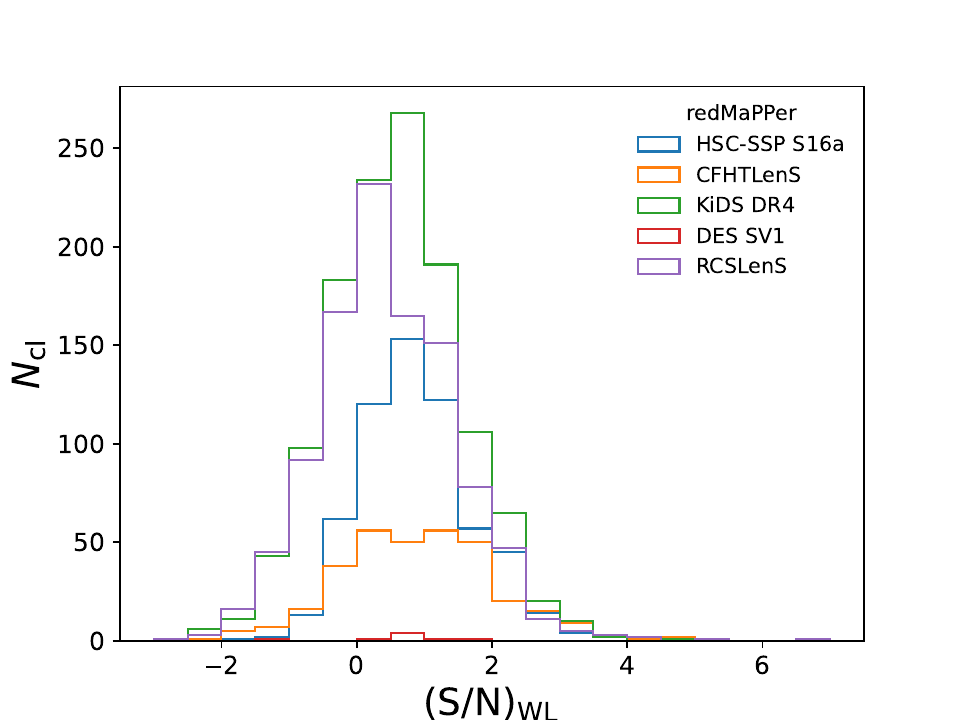}} \\
\end{tabular}
\caption{Binned distribution of $(\text{S/N})_\text{WL}$ of clusters that lie in public surveys, coded by colour as in the legend. \emph{Top}: PSZ2 clusters. \emph{Bottom}: redMaPPer-SDSS sample.}
\label{fig_SNRWL_1D}
\end{figure}

\begin{table}
\caption{Number of PSZ2 clusters per survey with $(\text{S/N})_\text{WL}$ larger than a given threshold. $N_\text{cl}$ is the total number of clusters per survey; columns labelled with $(\text{S/N})_\text{WL} \ge s$ report the number of clusters (the relative fraction is reported in parentheses) with $(\text{S/N})_\text{WL}$ larger than $s$. $(\text{S/N})_\text{Euclid, 3}$ is the survey $(\text{S/N})_\text{WL}$ value which corresponds to $(\text{S/N})_\text{WL} = 3$ in the Euclid Survey, i.e., which gives 3 when rescaled to the \Euclid shape noise.  
}
\label{tab_snr_psz2}
\centering
\resizebox{\hsize}{!} {
\begin{tabular}[c]{l  r r r r r}
	\hline
	\noalign{\smallskip}  
	Survey			&	$N_\text{cl}$ 	&	 $(\text{S/N})_\text{WL} \ge 1$		&  $(\text{S/N})_\text{WL} \ge 2$ &  $(\text{S/N})_\text{WL} \ge 3$ &  $ \ge (\text{S/N})_\text{Euclid, 3}$\\
	\noalign{\smallskip}  
	\hline
	\noalign{\smallskip}  
	HSC-SSP S16a	&	5	& 	4 (0.80)	&	2 (0.40)	&	1 (0.20) & 2 (0.40)\\
	CFHTLenS		&	8	&	8 (1.00)	&	5 (0.63)	&	3 (0.38) & 5 (0.63)\\
	KiDS DR4			&	37	&	31 (0.84)	&	21 (0.57)	&	10 (0.27)& 26 (0.70) \\
	DES SV1  		&	11	&	4 (0.36)	&	3 (0.27)	&	1 (0.09) & 3 (0.27)\\
	RCSLenS			&	28	&	17 (0.61)	&	11 (0.39)	&	7 (0.25) & 12 (0.43)\\
    \hline
	\end{tabular}
	}
\end{table}

\begin{table}
\caption{Same as in Table~\ref{tab_snr_psz2} but for the redMaPPer clusters.
}
\label{tab_snr_redmapper}
\centering
\resizebox{\hsize}{!} {
\begin{tabular}[c]{l  r r r r r}
	\hline
	\noalign{\smallskip}  
	Survey			&	$N_\text{cl}$ 	&	 $(\text{S/N})_\text{WL} \ge 1$		&  $(\text{S/N})_\text{WL} \ge 2$ &  $(\text{S/N})_\text{WL} \ge 3$ &  $ \ge (\text{S/N})_\text{Euclid, 3}$\\
	\noalign{\smallskip}  
	\hline
	\noalign{\smallskip}  
	HSC-SSP S16a	&	598	& 	247 (0.41)	&	68 (0.11)	&	9 (0.02)  & 50(0.08)\\
	CFHTLenS		&	329	&	156 (0.47)	&	50 (0.15)	&	15 (0.05)  & 93(0.28)\\
	KiDS DR4			&	1240	&	396 (0.32)	&	99 (0.08)	&	14 (0.01)  & 383(0.30)\\
	DES SV1  		&	8	&	2 (0.25)	&	0 (0.00)	&	0 (0.00)  & 2(0.25)\\
	RCSLenS			&	1020	&	299 (0.29)	&	70 (0.07)	&	12 (0.01) & 343(0.34)\\	
    \hline
	\end{tabular}
	}
\end{table}

WL measurements of individual clusters are very challenging. For the more massive and well observed clusters, only a mass precision of the order of about $10\,\%$ can be reached \citep{wtg_III_14,ume+al16b}. In Fig.~\ref{fig_SNRWL_1D}, we plot the distribution of the $(\text{S/N})_\text{WL}$ of the PSZ2 and redMaPPer clusters per survey. Some survey specifics are listed in Tables~\ref{tab_sample_psz2} and~\ref{tab_sample_redmapper}. Cumulative statistics for $(\text{S/N})_\text{WL}$ are reported in Tables~\ref{tab_snr_psz2} and \ref{tab_snr_redmapper}.  

The very massive end of the halo mass function is covered by the redMaPPer sample (up to $z\sim 0.6$) and the PSZ2 sample (up to $z\sim 1$). However, $(\text{S/N})_\text{WL}$ exceeds 3 only for a few lenses. We find 50 redMaPPer lenses with $(\text{S/N})_\text{WL} \ge 3$ from a total multi-survey area of $\sim \num{1500}\,\deg^2$ covered with a not homogeneous depth.


\subsection{Small groups}

\label{sec_mass_prec_small}

\begin{figure}
\resizebox{\hsize}{!}{\includegraphics{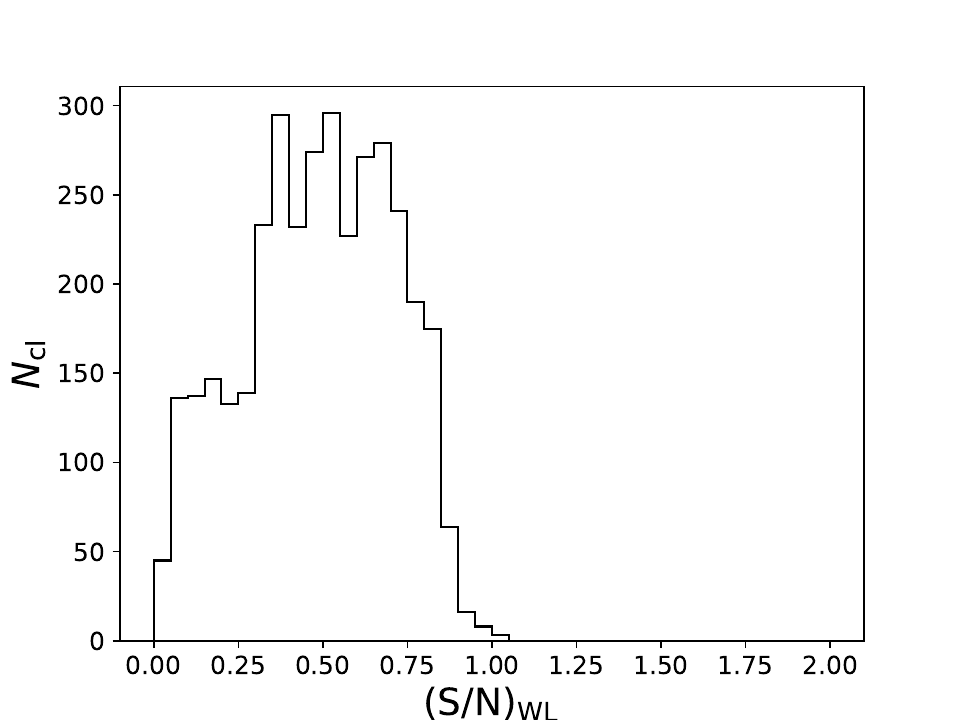}}
\caption{Binned distribution of $(\text{S/N})_\text{WL}$ of the candidate clusters detected by AMICO in the XXL North field covered by HSC-SSP S16a.}
\label{fig_amico_hsc_s18a_xmm_SNRWL_1D_amico_hsc}
\end{figure}

\begin{figure}
\begin{tabular}{c}
\resizebox{\hsize}{!}{\includegraphics{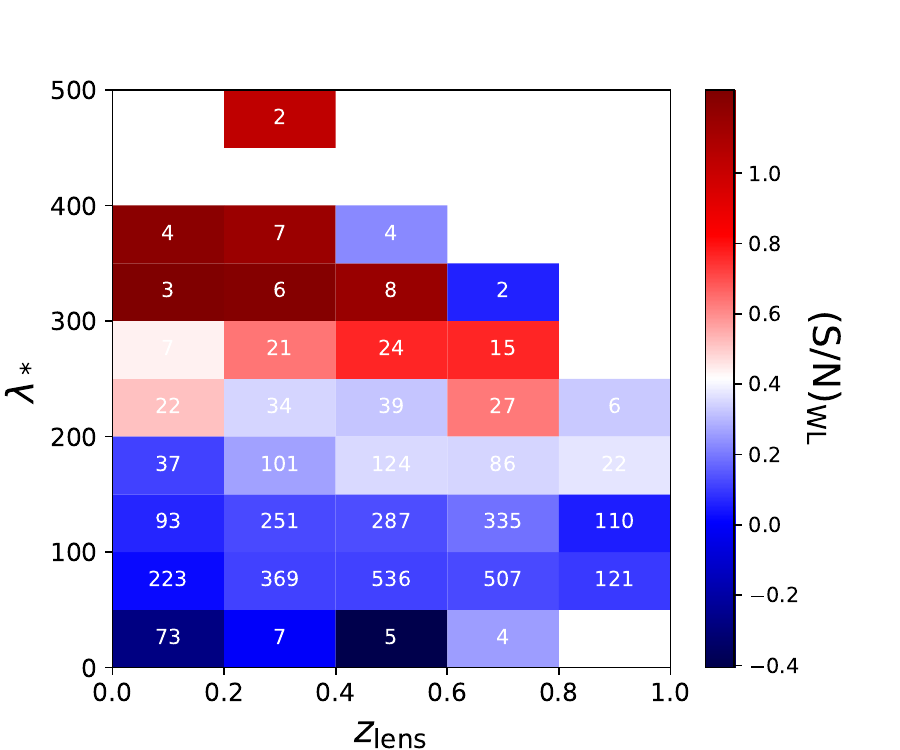}} \\
\resizebox{\hsize}{!}{\includegraphics{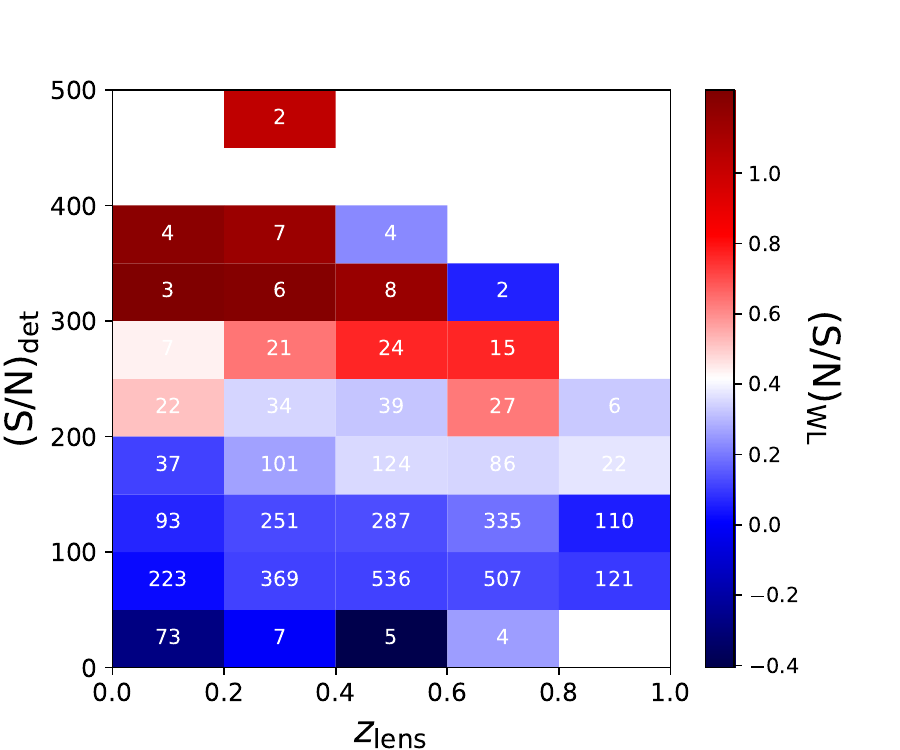}}
\end{tabular}
\caption{$(\text{S/N})_\text{WL}$ of the clusters and groups detected by AMICO in the XXL-North field of HSC-SSP. \emph{Top}: $(\text{S/N})_\text{WL}$ as a function of redshift and richness. $(\text{S/N})_\text{WL}$ is computed as the median value for the candidate groups in the bin. $(\text{S/N})_\text{WL}$ is colour-coded as shown in the right bar. The number of candidates is superimposed in each bin. \emph{Bottom}: same as in the top panel for $(\text{S/N})_\text{WL}$ as a function of redshift and $(\text{S/N})_\text{det}$, i.e. the S/N of the optical detection of the algorithm.}
\label{fig_amico_hsc_s18a_xmm_SNR_2D_amico_hsc}
\end{figure}

In a photometric survey, clusters can be detected from the same data set used for shape measurements. In Figs.~\ref{fig_amico_hsc_s18a_xmm_SNRWL_1D_amico_hsc} and \ref{fig_amico_hsc_s18a_xmm_SNR_2D_amico_hsc}, we consider the distribution of the AMICO clusters detected in XXL-North exploiting the HSC-SSP data, see Sect.~\ref{sec_amico_hsc_xxl}. For 1474 clusters ($\sim42\,\%$ of the total sample), $(\text{S/N})_\text{WL}$ is negative. Based on WL data only, we could not detect them. Mass can be significantly constrained only for a few very rich clusters.

The AMICO-defined S/N for optical detection, $(\text{S/N})_\text{det}$, is nearly one order of magnitude larger than the corresponding $(\text{S/N})_\text{WL}$. $(\text{S/N})_\text{WL}$ exceeds 2 (3) only for 28 (5) out of 3541 detections with $(\text{S/N})_\text{det} \ge 3$.


\subsection{Semi-analytical forecasting}
\label{sec_fore_semi}

\begin{figure}
\resizebox{\hsize}{!}{\includegraphics{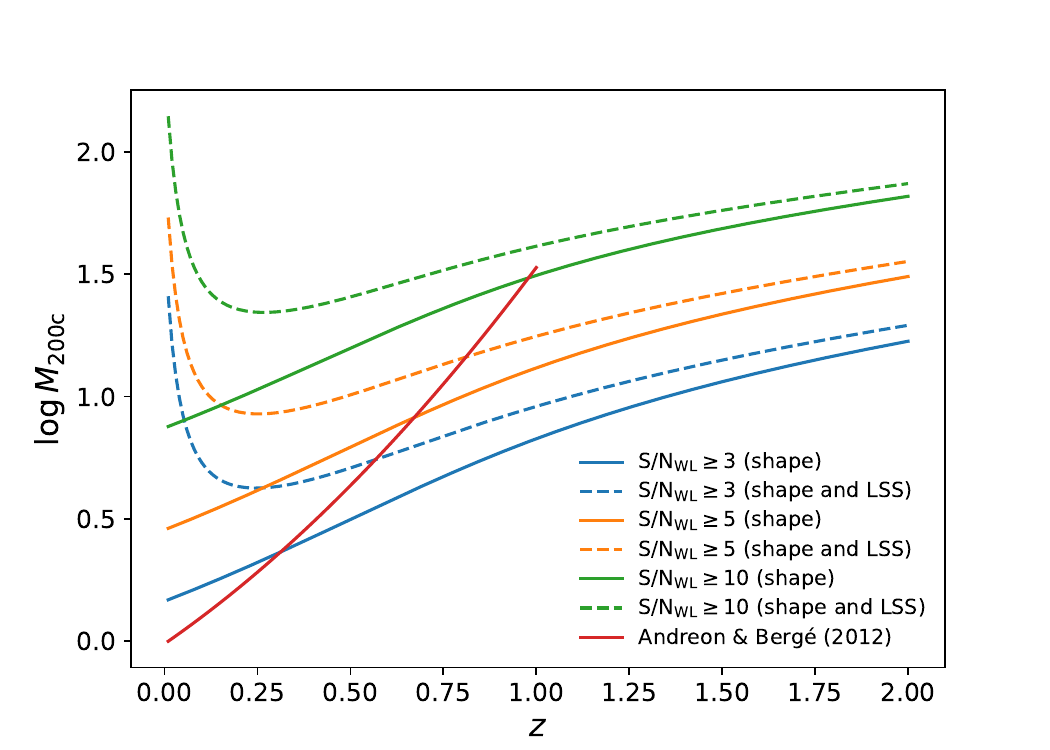}}
\caption{Expected mass thresholds for WL mass precision with \Euclid as a function of redshift. For each redshift, we plot the minimum mass for which the mean $(\text{S/N})_\text{WL}$ is expected to exceed a threshold in the reference cosmology. For comparison we also show the result from \citet{an+be12} for a different $(\text{S/N})_\text{WL}$ definition.}
\label{fig_logM200c_z_SNR_th}
\end{figure}

\begin{figure}
\resizebox{\hsize}{!}{\includegraphics{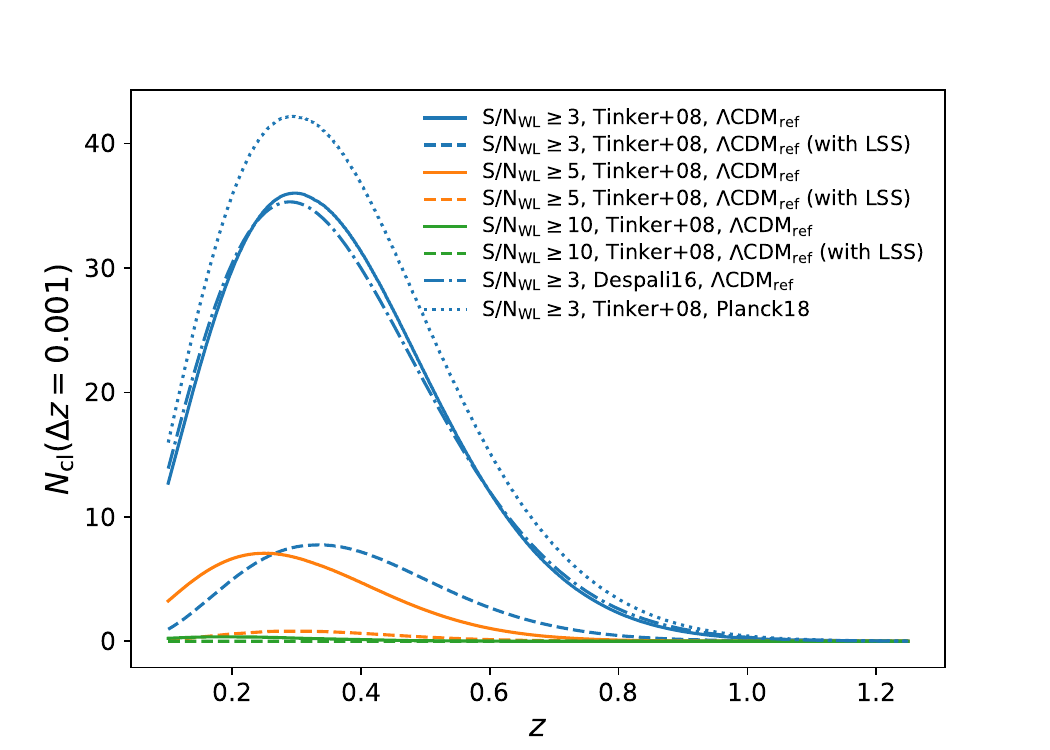}}
\caption{Expected number of clusters (in small redshift bins of $\Delta z = 0.001$) with $(\text{S/N})_\text{WL}$ in excess of a given threshold. We show results for different $(\text{S/N})_\text{WL}$ thresholds (3, 5, 10), halo mass function modelling, either from \citet{tin+al08} or \citet{des+al16}, and cosmological models (the reference one or Planck18).}
\label{fig_Ncl_z_SNRWL}
\end{figure}

The expected number of \Euclid clusters with a well measured WL mass can be estimated using a semi-analytical approach. We first estimate the expected $(\text{S/N})_\text{WL}$ of a lens at a given mass and redshift. Subsequently, we estimate the limiting mass at a given redshift such that the mean $(\text{S/N})_\text{WL}$ is above a given threshold. We finally estimate the number of clusters above this threshold mass.

As in Sect.~\ref{sec_wl_sign}, we consider the signal collected in the radial range between $0.3$ and $3.0\,\hMpc$ from the cluster centre. We model the cluster as a NFW halo whose concentration follows the mass-concentration relation from \citet{di+jo19}. 

For noise, we consider either shape noise and LSS noise or shape noise only. Galaxy shape noise is mainly due to intrinsic ellipticity, which we model as being Gaussian distributed with $\sigma_\text{int} = 0.26$, as measured for a sample of galaxies observed by the Hubble Space Telescope with similar photometric properties to those expected from \Euclid \citep{sch+al18,euclid_pre_aja+al23}.

Given the resolution and depth of \Euclid, we anticipate that the survey can measure galaxy shapes with a density of about $30$ galaxies per arcmin$^2$ \citep{eucl_lau_11}. To model the source distribution, we consider the \Euclid-like distribution of sources approximated from the COSMOS2015 photometric redshift catalogs \citep{lai+al16} by \citet{mar+al21}, and modelled as proposed in \citet{fu+al08}, with parameters from \citet{euclid_pre_aja+al23}.

For the source selection, we consider one third of the \Euclid sky to be masked. Subsamples of sources with high completeness can be extracted in \Euclid with a robust selection at the sub-percent contamination level \citep{euclid_pre_les+al23}. This can be approximated by selecting $80\,\%$ of the galaxies with $z  > z_\text{lens} + 0.1$, see Eq.~(\ref{eq_zphot_1}).

We can then compute the mass threshold for a halo to have mean $(\text{S/N})_\text{WL}$ larger than a given threshold value defined as a function of redshift. The threshold mass is shown in Fig.~\ref{fig_logM200c_z_SNR_th}. Shape noise or noise contribution from LSS to the shear smoothly increase with the lens redshift, as well as the reduced shear, but the different growth rates at low redshift cause the upturn. For shape noise only, the mass threshold for $(\text{S/N})_\text{WL} >3 $ can be approximated as
\begin{equation}
\label{eq_m200c_th}
\log M_{200\text{c,th}} \simeq 0.130 + 0.832\,  z - 0.141\, z^2 \,.
\end{equation}

Finally, we compute how many haloes exceed the mass threshold in a $\Lambda$CDM universe. We model the halo mass function as proposed in \citet{tin+al08}. The mass threshold in Eq.~(\ref{eq_m200c_th}) is much larger than the mass limit for optical detection and we expect the cluster sample to be complete in this mass range. We consider the redshift range $0.1 < z < 2.0$ and we assume that all clusters with $(\text{S/N})_\text{WL} \geq 3$ in this redshift range can be optically detected. 

The total number of clusters with mean $(\text{S/N})_\text{WL}$ larger than a threshold can be found by integration of the halo mass function.

Considering shape noise only, we find $\sim \num{14800}$, $\sim \num{2400}$, $\sim \num{90}$ clusters with $(\text{S/N})_\text{WL} \geq 3$, 5, 10, respectively, in the redshift range $0.1 < z < 2.0$. The redshift distribution of clusters is shown in Fig.~\ref{fig_Ncl_z_SNRWL}. The mean lens redshift is $\sim 0.37$. We expect $\sim \num{1600}$ ($\num{100}$), $\num{250}$ ($\num{8}$), or $\num{20}$ ($\num{0}$) clusters at $z>0.6$, $0.8$, or $1.0$, respectively, with $(\text{S/N})_\text{WL} \geq 3$ (5).

The main source of uncertainty in this forecasting is due to the cosmological model. We estimated an uncertainty of $\pm \num{3000}$ by comparing the expected number of clusters with mean $(\text{S/N})_\text{WL} \geq 3$ with the results obtained using a halo mass function with cosmological parameters derived from the final Planck data release of Cosmic Microwave Background anisotropies \citep{planck_2018_VI}. Uncertainties due to the modelling of the halo mass function ($\pm \num{600}$) were estimated comparing to results based on the halo mass function proposed by \citet{des+al16}. These uncertainties can be added to give a final estimate of $\num{15000} \pm \num{4000}$ expected clusters with $(\text{S/N})_\text{WL} \geq 3$. 

\citet{an+be12} performed a forecast for the expected number of WL selected clusters. Notwithstanding the substantial differences compared to our analysis, e.g., different definitions of $(\text{S/N})_\text{WL}$, different treatments of masks, photometric redshift uncertainty, member and foreground contamination, and LSS noise, we find that the total expected number of clusters agrees with their estimate.

Noise from LSS will be one of the main sources of uncertainty for WL mass calibration in Stage-IV surveys \citep{koh+al15}. If we consider both shape and LSS noise, the number of clusters with a large $(\text{S/N})_\text{WL}$ is significantly reduced. We find $\sim \num{3000}$, $\sim \num{260}$, $\sim \num{2}$ clusters with $(\text{S/N})_\text{WL} \geq 3$, 5, 10, respectively, in the redshift range $0.1 < z < 2.0$.

The previous analysis applies to massive, confirmed clusters, e.g., the redMaPPer or the PSZ2 sample. At a given mass and redshift, the measured $(\text{S/N})_\text{WL}$ can strongly differ from the mean value. Due to shape and LSS noise fluctuations, the $(\text{S/N})_\text{WL}$ measured along random line-of-sights that are empty of massive haloes can exceed 3 with a probability $\gs 10^{-1}$ percent. The cluster distribution is very steep and we expect much more low mass clusters to be upscattered above threshold due to noise than high mass clusters to be downscattered. Considering the population of haloes with $M_\text{200c} > 10^{12.5} M_\odot$ at $0.01 \le z \le 2$ in our reference cosmology, only $\sim 4\times 10^{-3}$ percent of the haloes have mass in excess of the mass threshold we considered, but $\sim 2\times 10^{-1}$ percent of them should have their signal boosted to $(\text{S/N})_\text{WL} \geq 3$ due to noise fluctuations.

A more precise forecast should account for additional impacting factors. For example, triaxiality and projection effects can either boost or reduce the signal. In addition, clusters are not uniformly distributed in space but more likely to be found at dense nodes and filaments of the cosmic web.


\subsection{Forecast from Stage-III surveys or precursors}
\label{sec_fore_stag}

Results from Stage-III surveys or precursors can be extrapolated to forecast the expected number of clusters with high $(\text{S/N})_\text{WL}$ in the Euclid Survey. Since LSS noise is negligible with respect to shape noise in Stage-III surveys, these extrapolations cannot account for LSS  and we can only compare to semi-analytical predictions with shape noise only.

The WL signal will be larger for Stage-IV surveys, and the noise smaller. A cluster detected with $(\text{S/N})_\text{WL} = 3$ in \Euclid would be detected with $(\text{S/N})_\text{Euclid, 3} < 3$ in a shallower survey. The shear signal is proportional to the inverse of the critical surface density $\Sigma_\text{cr}^{-1}$, see Eq.~(\ref{eq_Delta_Sigma_2}), and it is larger for higher redshift sources. Considering a lens at $z \sim 0.37$, see Sect.~\ref{sec_fore_semi}, and the redshift distribution of the surveys under consideration, see Table~\ref{tab_surv_lens}, the shear signal for \Euclid is expected to be approximately $\sim 24\,\%$, $32\,\%$, $37\,\%$, $57\,\%$, or $61\,\%$ larger than for HSC-SSP S16a, CFHTLenS, KiDS DR4, DES SV, or RCSLenS, respectively. 

Accounting for the larger number of sources, the shape noise for \Euclid measurements is expected to be approximately $15\,\%$ smaller than HSC-SSP S16a and $40\,\%$ smaller than CFHTLenS. It will be a factor of $\sim \num{2}$ smaller compared to KiDS DR4, DES SV, or RCSLenS.

As a result of increased signal and reduced noise, a halo detected with, e.g., $(\text{S/N})_\text{WL} \sim 1$ in KiDS will be detected with $(\text{S/N})_\text{WL} \sim 3$ in \Euclid. After rescaling both the signal and noise per survey to the \Euclid expectation, we estimate that $\sim\num{900}$ redMaPPer clusters from the subsample covered by the surveys considered here are expected to pass the $(\text{S/N})_\text{WL} > 3$ cut in \Euclid. 

The total expected number of clusters with $(\text{S/N})_\text{WL} > 3$ in \Euclid can be obtained by rescaling to both the larger survey area and the larger redshift baseline of the cluster sample. The area of each survey overlapping with that parsed in SDSS-DR8 by redMaPPer is proportional to the number of redMaPPer clusters that lie in the survey area. For forecasting, we re-scale to the area of the Euclid Survey.

The redMaPPer catalogue is nearly complete within the redshift range $0.1 \ls z \ls 0.6$, whereas \Euclid will successfully detect clusters up to higher redshifts \citep{sar+al16,euclid_ada+19}. Based on the results of Sect.~\ref{sec_fore_semi}, we expect that $\sim10\,\%$ of the massive clusters with $(\text{S/N})_\text{WL} > 3$ will lie at $z>0.6$. 

Taking the above into consideration, extrapolations from HSC-SSP S16a, CFHTLenS, KiDS DR4, DES SV, or RCSLenS provide a forecast of $\sim\num{4000}$, $\num{12000}$, $\num{13000}$, $\num{11000}$, or $\num{14000}$ \Euclid detected clusters with $(\text{S/N})_\text{WL} \ge 3$,  respectively. These individual forecasts can be finally combined taking the biweight estimators to give an expected number of $\num{13000}\pm \num{3000}$ \Euclid clusters with $(\text{S/N})_\text{WL} \ge 3$, in agreement with the semi-analytical result in Sect.~\ref{sec_fore_semi}.





\subsection{Forecast from AMICO XXL-HSC}

A third forecast can be based on the AMICO detected clusters in the XXL-North field covered by HSC SSP S16a. The equivalent $(\text{S/N})_\text{WL}$ threshold for \Euclid detection can be found as in Sect.~\ref{sec_fore_stag}. 

The catalogue is complete up to $z \sim 0.75$, and we expect $\sim 5\,\%$ more massive \Euclid clusters with $(\text{S/N})_\text{WL} \ge 3$ at larger redshift.

We find 14 clusters with $(\text{S/N})_\text{WL} > (\text{S/N})_\text{Euclid, 3}$ in around $30 \deg^2$ (unmasked) area. Rescaling to the Euclid Survey area and to the extended redshift baseline, we expect $\num{7000}\pm\num{2000}$ \Euclid clusters with  $(\text{S/N})_\text{WL} > 3$. The uncertainty accounts for the Poisson fluctuations on the actual number of detected clusters.

This figure is in agreement with, but lower than, the forecast based on the semi-analytical method, see Sect.~\ref{sec_fore_semi}, or the extrapolation based on redMaPPer clusters in Stage-III surveys or precursors, see Sect.~\ref{sec_fore_stag}. The forecast based on redMaPPer clusters in HSC-SSP S16a was also lower. For the forecast, we used the effective source density $n_\text{eff} \sim 23.4\,\text{arcmin}^{-2}$ estimated in Sect.~\ref{sec_surveys}, which is higher than the nominal survey value, $n_\text{eff} \sim 21.8\, \text{arcmin}^{-2}$, see Sect.~\ref{sec_HSC-SSP}. Using the nominal value, we would obtain $\sim \num{8000}$ expected clusters.

\subsection{Discussion}

Clusters with significant $(\text{S/N})_\text{WL}$ are still rare. Upon completion, the KiDS survey will cover $\sim 40$ clusters with $(\text{S/N})_\text{WL} >3$, as can be seen by rescaling results from Sect.~\ref{sec_mass_prec_large} to the final survey area. We can collect from literature a heterogeneous sample of $\num{375}$ clusters with $(\text{S/N})_\text{WL} \ge 3$ \citep{ser15_comalit_III}.

\Euclid will detect $\sim \num{13000}$ ($\num{3000}$) massive clusters with $(\text{S/N})_\text{WL} >3$ if we consider shape noise only (shape noise and LSS noise). In terms of quantity, this makes for a growth of nearly two orders of magnitude with respect to present samples. In terms of quality, the sample will be homogeneous and statistically complete and it will significantly extend the mass range (towards smaller haloes) and redshift range (to greater distances) of known clusters with direct WL mass measurement.

Some thousands may be a large number, but, out of the total number of groups and clusters expected to be detected from \Euclid data, this high $(\text{S/N})_\text{WL}$ sample only constitutes a small fraction. For the HSC-SSP survey, only 0.1--0.2$\,\%$ of all detected groups and clusters have $(\text{S/N})_\text{WL} \ge 3$, see Sect.~\ref{sec_mass_prec_small}. Considering a detection rate of $\sim \num{150}$ groups or clusters per $\deg^2$ (Euclid Collaboration: Cabanac et al., in prep.), the fraction of massive haloes  with $(\text{S/N})_\text{WL} \ge 3$ (shape noise only) will be 0.1--0.2\,\% for \Euclid too. \Euclid should uncover $\sim \num{200000}$ massive haloes with $M_\text{200c} > 10^{14}M_\odot$ \citep{sar+al16,euclid_ada+19}, and 1--2$\,\%$ of these massive haloes are expected to have $(\text{S/N})_\text{WL} \ge 3$ (shape noise only). Most of the clusters with measured $(\text{S/N})_\text{WL} \ge 3$ will be low mass haloes whose signal is upscattered due to noise. To assess the mass of most systems, WL stacking and optical proxies will be required \citep{bel+al19,mcc+al19,les+al22}.


\section{Systematics}

\begin{table*}
\caption{
Systematic error budget on the mass calibration for WL cluster analyses in Stage-III  surveys or precursors. Sources of systematics are listed in column 1. Uncertainties are given in percents.
}
\label{tab_sys}
\centering
\resizebox{\hsize}{!} {
\begin{tabular}[c]{l  r r r r r r}
\hline
\noalign{\smallskip}  
								&DES SV			&	DES Y1		&	KIDS DR3		&CFHTLenS/RSCLenS		& 	HSC 			\\
							&\citet{mel+al17}	&\citet{mcc+al19}    &\citet{bel+al19}		&\citet{ser+al17_psz2lens}	&	\citet{ume+al20}	\\
\hline
\noalign{\smallskip} 
Shear measurement				&	4			& 	1.7			&	1.3			&	5					&  	1.3				\\
Photometric redshifts			&	3			&	2.6			&	5.6			&	5					&	0.9				\\
\hline
Modelling systematics			& 	2			&	0.73			& 	3			&	1					&	3.3				\\
Cluster triaxiality				&	2			&	2.0			&	--			&	--					&	--				\\
Line-of-sight projections			&	2			&	2.0			&	--			& 	1					&	--				\\
Orientation and projections		&	--			&	--			&	3			&	--					&	--				\\
\hline
Background selection			&				&	--			&	2.7			&	3					&	3.1				\\	
selection / miscentring			&	$\ls 1$		&	0.78			&	--			&	--					&	--				\\
miscentring					&				&	--			&	--			& 	0.5					&	--				\\
\hline
Total							&6.1		&	4.3				&7.6				& 8						&	5				\\
\hline
	\end{tabular}
}
\end{table*}

Analyses of Stage-III surveys or precursors agree that the total level of systematic uncertainty affecting the mass calibration is of the order of 5--10\,\% \citep{sim+al17a,mel+al17,ser+al17_psz2lens,mcc+al19,bel+al19}. In Table~\ref{tab_sys} we summarise results from independent investigations. Main sources of systematic errors are residual biases in shear calibration or photo-$z$ estimations, contamination due to foreground or member galaxies, and modelling effects. 

Comparison between estimates from different surveys or by different analyses is not straightforward since the expected size of a systematic error may depend on the analysis or the data sample. For example, smoothing of the WL signal due to miscentring effects in \citet{ume+al20} is expected to be negligible thanks to X-ray centring information, whereas it was significant for optically selected groups in DES \citep{mel+al17,mcc+al19}. \citet{ume+al20} accounted for intrinsic variations of the lensing signal at fixed mass due to variations in halo concentrations, cluster asphericity, or correlated haloes by adding an additional term to the shear uncertainty covariance matrix, whereas, e.g., cluster triaxiality was treated as a systematic effect in other analyses \citep{sim+al17a,mel+al17,ser+al17_psz2lens,mcc+al19}. Sources of systematics can be taken as uncorrelated or correlated in different analyses \citep{mel+al17}. Notwithstanding the differences, different analyses agree on the main sources of systematics and on their size. Mass biases of the order of 5--10\,\% are consistent with results from our cross-comparison.

We perform an end-to-end analysis, and we compare the main results, which are affected by all potential systematics, which enter at different steps. However, cross-checks can be also used to assess specific systematic effects. In App.~\ref{sec_she_cal}, we show how residual effects in shear calibration for Stage-III surveys and precursors are negligible with respect to the statistical uncertainties by comparing the WL profiles of clusters extracted from matched source catalogues.


\section{Conclusions}
\label{sec_conc}

Both statistical uncertainty and unaccounted for systematics can limit the accuracy of the mass calibration. The main contributors to the systematic error budget are the calibration uncertainties of the shear measurements, the photo-$z$ performance, and the selection of the source galaxies. Furthermore, in the regime of cluster lensing, shape calibration has to be revised for the impact of stronger shears and increased blending \citep{her+al20}. \citet{koh+al15} explored statistical uncertainties and systematic errors in WL mass estimates of galaxy clusters in a \Euclid-like survey and showed that requirements for WL by clusters are similar but a bit weaker than what is needed for cosmic shear. In principle, proper calibration for cosmic shear should also meet requirements for WL clusters. 


To test the residual level of systematic effects in Stage-III surveys, we performed a cross-survey analysis. To assess the robustness of the cluster mass determinations, we performed a unified analysis exploiting the \texttt{COMB-CL} processing function to be used by the Euclid Collaboration for WL cluster mass measurements. On one hand, \texttt{COMB-CL} can deal with different data sets, or shear and photo-$z$ estimation algorithms, and, on the other hand, it can measure the WL signal and determine the cluster mass with uniform and consistent model assumptions, and the same inference pipeline. The analysis also served as a testing of the \texttt{COMB-CL} processing function.

Our choices and assumptions were meant to provide a consistent framework for comparison of very different data sets. 
The goal was to look for discrepancies, not to rank alternative methodologies for WL cluster analyses or surveys, or fully exploit their statistical power.
For example, we adopted the same secure colour-colour cut for all surveys. This made the background selection and the systematic effects related to member and foreground contamination quite consistent. Different cuts which were optimised per survey could improve the purity and completeness of background selection at a price of a less straightforward cross-survey comparison.

The comparison of the cluster lensing signal and of the inferred WL mass showed a remarkable agreement between independent imaging surveys. Differences in excess surface density profiles were consistent with the statistical uncertainties and well within survey defined systematic error budgets for Stage-III surveys and precursors. This is consistent with previous results on cosmic shear \citep{cha+al19,lon+al23}, or galaxy-galaxy lensing \citep{lea+al22}. Masses of individual clusters were found to agree with literature results. Our analysis suggests that unknown or residual systematic effects are subdominant with respect to statistical noise or known systematic errors for Stage-III surveys and precursors.

However, in order to improve WL mass calibration, difficult challenges remain. Shear calibration can be hampered by incomplete image simulations \citep{hsc_man+al18}. Sufficient archival Hubble Space Telescope data should be available to create realistic populations of mock galaxies for \Euclid-like simulations and achieve the required sub-percent level accuracy \citep{hoe+al17}. Performance could also improve with data-driven approaches such as metacalibration \citep{hu+ma17,sh+hu17,hoe+al21}.

Spectroscopic training samples used to calibrate photo-$z$s can be incomplete at high redshifts or for some galaxy types, resulting in systematic uncertainty in the photo-$z$ calibration. With these issues in mind, the Euclid Collaboration has been making a significant effort to enlarge the calibration sample \citep[see, e.g.,][]{euclid_pre_sag+al22} and properly train the photo-$z$ algorithms \citep[see, e.g.,][]{euclid_pre_bis+al23}.

In any analysis, one seeks to obtain the right balance between high statistical precision and a low level of systematic errors. For example, we favoured very conservative background galaxy selections, where we retained only the sources with high fidelity redshift estimates. On one hand, this results in a larger shape noise than in, e.g., analyses using the full source sample. On the other hand, systematic uncertainties are expected to be smaller.

Precise mass calibration is required to ensure the success of Stage-IV cluster cosmology measurements \citep{sar+al16}. Mass calibration at the level of a few percent is already within reach of Stage-III analyses. In a \Euclid-like survey, shear is going to be calibrated within an accuracy of $|\Delta m| \sim 2\times 10^{-3}$ \citep{cro+al13}. Given the exquisite space-based infrared photometry complemented by ground-based optical coverage and spectroscopic calibration samples \citep{euclid_pre_sag+al22}, photo-$z$ estimates will be very accurate, with an expected bias of the order of $|\Delta z_\text{p}| \sim 0.002 (1 + z)$ \citep{euclid_pre_bis+al23}. Such accurate photo-$z$s along with extended optical/infrared photometry will also strongly aid background selection. We expect to select nearly complete (up to $\sim 90\,\%$) and pure samples of background sources, with a contamination at the sub-percent level \citep{euclid_pre_les+al23}. Systematics due to analysis assumptions can be significantly reduced with lens modelling accounting for, e.g., miscentring, concentration and truncation, and the effect of correlated or uncorrelated matter \citep{ser+al18_psz2lens,bel+al19,mcc+al19}. Mass calibration at the percent level or better could be within the reach of Stage-IV surveys.


\begin{acknowledgements}

This work was made possible by public sharing of tools and data sets by the astronomical community, which we thanks in the following with the suggested formulations. MS acknowledges financial contributions from contract ASI-INAF n.2017-14-H.0, contract INAF mainstream project 1.05.01.86.10, INAF Theory Grant 2023: Gravitational lensing detection of matter distribution at galaxy cluster boundaries and beyond (1.05.23.06.17), and contract
Prin-MUR 2022 supported by Next Generation EU (n.20227RNLY3, The concordance cosmological model: stress-tests with galaxy clusters). CG thanks the support from INAF theory Grant 2022: Illuminating Dark Matter using Weak Lensing by Cluster Satellites, PI: Carlo Giocoli.
LM acknowledges support from the grants PRIN-MIUR 2017 WSCC32 and ASI n.2018-23-HH.0.

\AckEC

This research has made use of NASA's Astrophysics Data System (ADS) and of the NASA/IPAC Extragalactic Database (NED), which is operated by the Jet Propulsion Laboratory, California Institute of Technology, under contract with the National Aeronautics and Space Administration. 

The Hyper Suprime-Cam (HSC) collaboration includes the astronomical communities of Japan and Taiwan, and Princeton University. The HSC instrumentation and software were developed by the National Astronomical Observatory of Japan (NAOJ), the Kavli Institute for the Physics and Mathematics of the Universe (Kavli IPMU), the University of Tokyo, the High Energy Accelerator Research Organization (KEK), the Academia Sinica Institute for Astronomy and Astrophysics in Taiwan (ASIAA), and Princeton University. Funding was contributed by the FIRST program from Japanese Cabinet Office, the Ministry of Education, Culture, Sports, Science and Technology (MEXT), the Japan Society for the Promotion of Science (JSPS), Japan Science and Technology Agency (JST), the Toray Science Foundation, NAOJ, Kavli IPMU, KEK, ASIAA, and Princeton University. 



Based in part on data collected at the Subaru Telescope and retrieved from the HSC data archive system, which is operated by Subaru Telescope and Astronomy Data Center at National Astronomical Observatory of Japan.

This work is based in part on observations obtained with MegaPrime/MegaCam, a joint project of CFHT and CEA/IRFU, at the Canada-France-Hawaii Telescope (CFHT) which is operated by the National Research Council (NRC) of Canada, the Institut National des Sciences de l'Univers of the Centre National de la Recherche Scientifique (CNRS) of France, and the University of Hawaii. This research used the facilities of the Canadian Astronomy Data Centre operated by the National Research Council of Canada with the support of the Canadian Space Agency. CFHTLenS and RCSLenS data processing was made possible thanks to significant computing support from the NSERC Research Tools and Instruments grant program.

This work is based in part on observations made with ESO Telescopes at the La Silla Paranal Observatory under programme IDs 177.A-3016, 177.A-3017, 177.A-3018 and 179.A-2004, and on data products produced by the KiDS consortium. The KiDS production team acknowledges support from: Deutsche Forschungsgemeinschaft, ERC, NOVA and NWO-M grants; Target; the University of Padova, and the University Federico II (Naples).

This project used public archival data from the Dark Energy Survey (DES). Funding for the DES Projects has been provided by the U.S. Department of Energy, the U.S. National Science Foundation, the Ministry of Science and Education of Spain, the Science and Technology FacilitiesCouncil of the United Kingdom, the Higher Education Funding Council for England, the National Center for Supercomputing Applications at the University of Illinois at Urbana-Champaign, the Kavli Institute of Cosmological Physics at the University of Chicago, the Center for Cosmology and Astro-Particle Physics at the Ohio State University, the Mitchell Institute for Fundamental Physics and Astronomy at Texas A\&M University, Financiadora de Estudos e Projetos, Funda{\c c}{\~a}o Carlos Chagas Filho de Amparo {\`a} Pesquisa do Estado do Rio de Janeiro, Conselho Nacional de Desenvolvimento Cient{\'i}fico e Tecnol{\'o}gico and the Minist{\'e}rio da Ci{\^e}ncia, Tecnologia e Inova{\c c}{\~a}o, the Deutsche Forschungsgemeinschaft, and the Collaborating Institutions in the Dark Energy Survey.

The Collaborating Institutions are Argonne National Laboratory, the University of California at Santa Cruz, the University of Cambridge, Centro de Investigaciones Energ{\'e}ticas, Medioambientales y Tecnol{\'o}gicas-Madrid, the University of Chicago, University College London, the DES-Brazil Consortium, the University of Edinburgh, the Eidgen{\"o}ssische Technische Hochschule (ETH) Z{\"u}rich,  Fermi National Accelerator Laboratory, the University of Illinois at Urbana-Champaign, the Institut de Ci{\`e}ncies de l'Espai (IEEC/CSIC), the Institut de F{\'i}sica d'Altes Energies, Lawrence Berkeley National Laboratory, the Ludwig-Maximilians Universit{\"a}t M{\"u}nchen and the associated Excellence Cluster Universe, the University of Michigan, the National Optical Astronomy Observatory, the University of Nottingham, The Ohio State University, the OzDES Membership Consortium, the University of Pennsylvania, the University of Portsmouth, SLAC National Accelerator Laboratory, Stanford University, the University of Sussex, and Texas A\&M University.

Based in part on observations at Cerro Tololo Inter-American Observatory, National Optical Astronomy Observatory, which is operated by the Association of Universities for Research in Astronomy (AURA) under a cooperative agreement with the National Science Foundation.

\end{acknowledgements}


\bibliographystyle{aa}


\appendix


\section{COMB-CL Pipeline}
\label{sec_combcl}

\begin{figure*}
    \centering
    \includegraphics[width=1.0\linewidth]{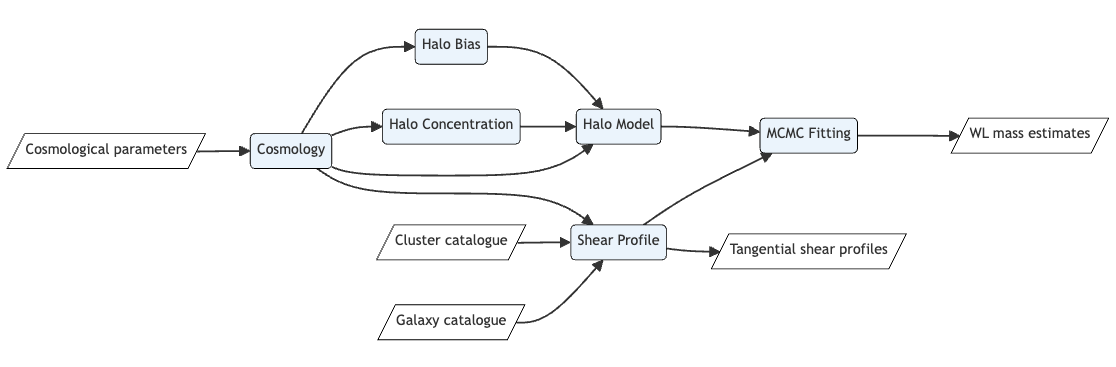}
    \caption{Structure of the \Euclid combined cluster and weak-lensing pipeline, \texttt{COMB-CL}. White parallelograms show input or output products and blue boxes show \texttt{COMB-CL} modules that perform different processing steps.}
    \label{fig:comb-cl_flow}
\end{figure*}

\begin{table*}
\caption{Survey specifics to instantiate the \texttt{COMB-CL} \texttt{Shear} class
}
\label{tab_comb_cl_shear_par}
\centering
\resizebox{\hsize}{!} {
\begin{tabular}[c]{l  r r r r r}
	\hline
	\noalign{\smallskip}  
	option & HSC-SSP S16a & CFHTLenS	& KiDS DR4 & DES SV1  	& RCSLenS		\\    	\noalign{\smallskip}  
	\hline
	\noalign{\smallskip}  
	\texttt{shape\_estimator}  & \texttt{distortion} & \texttt{ellipticity} & \texttt{ellipticity} & \texttt{ellipticity} & \texttt{ellipticity} \\
    \texttt{sky\_coordinates} & \texttt{eastward} & \texttt{westward} & \texttt{westward} & \texttt{westward} & \texttt{westward} \\    
    \texttt{selection\_photo\_z}   & \texttt{True} & \texttt{True} & \texttt{True} & \texttt{True} & \texttt{True} \\
     \texttt{z\_p\_fails} & \texttt{[-99,99]} & \texttt{[-99,99]} & \texttt{[-99,99]} & \texttt{[-99,99]} & \texttt{[-99,99]} \\
    \texttt{Delta\_z\_lens\_0} & \texttt{0.1} & \texttt{0.1} & \texttt{0.1} & \texttt{0.1} & \texttt{0.1} \\
    \texttt{Delta\_z\_lens\_z} & \texttt{0.0}  & \texttt{0.0}  & \texttt{0.0}  & \texttt{0.0}  & \texttt{0.0} \\
    \texttt{z\_p\_range\_min} & \texttt{0.2} & \texttt{0.2} & \texttt{0.2} & \texttt{0.2} & \texttt{0.4} \\
    \texttt{z\_p\_range\_max} & \texttt{1.5} & \texttt{1.2} & \texttt{1.2} & \texttt{1.2} & \texttt{1.2} \\
    \texttt{z\_p\_conf\_min} & \texttt{0.8} &\texttt{0.8} & \texttt{0.8} & \texttt{0.8} & \texttt{0.8} \\
    \texttt{colour\_cut\_method} & \texttt{'Medezinski18\_griz'} & \texttt{'Medezinski18\_griz'} & \texttt{'Medezinski18\_griz'} & \texttt{'Medezinski18\_griz'} & \texttt{'Medezinski18\_griz'} \\
    \texttt{mag\_fails} & \texttt{[-99,99]} & \texttt{[-99,99]} & \texttt{[-99,99]} & \texttt{[-99,99]} & \texttt{[-99,99]} \\
    \texttt{alpha} & \texttt{1.0} & \texttt{1.0} & \texttt{1.0} & \texttt{1.0} & \texttt{1.0} \\
    \texttt{delta\_alpha} & \texttt{1.0} & \texttt{1.0} & \texttt{1.0} & \texttt{1.0} & \texttt{1.0}\\
	\hline
	\end{tabular}
	}
\end{table*}

The analysis presented in this paper is performed with the combined cluster and weak-lensing pipeline, \texttt{COMB-CL}, that forms part of the global \Euclid data processing pipeline. In the following text, we provide a brief overview of the structure of \texttt{COMB-CL} as well as the specific configuration choices made for this work. A more comprehensive description of the code and methodology implemented in \texttt{COMB-CL} will be provided with the code public release.

The structure of \texttt{COMB-CL} is presented in Fig.~\ref{fig:comb-cl_flow}, where input and output products are shown in white parallelograms, while the various modules that perform different processing steps are shown in blue boxes. The input cluster  catalogue is expected to contain the positions, redshifts and, if available, richness estimates for the detected objects. The input galaxy catalogue is expected to contain the positions of source galaxies with measured shear values, photometric redshift point estimates and, if available, magnitudes or fluxes. 

The \texttt{Cosmology} module simply defines a shared class instance for the cosmological parameters provided ($H_0$, $\Omega_\text{b}$, $\Omega_\text{m}$, $n_\text{s}$ and $\sigma_8$). This instance provides consistent cosmological parameters and functions, such as the Hubble parameter, $H(z)$, and the comoving distance, $d_\text{C}(z)$, to the other \texttt{COMB-CL} modules.

The \texttt{Halo Bias}, and \texttt{Halo Concentration} modules implement various models for the halo bias and concentration as a function of mass, respectively. The models used for the work presented in this paper are the halo bias model of \citet{tin+al10} and the halo concentration model of \citet{di+jo19}.

The \texttt{Halo Module} module takes a given combination of halo and concentration models and defines various models of the 1-halo and 2-halo components of the surface mass density as a function of radius. These components can then be combined to provide a final model for the reduced differential density profile. The specific model used for the work presented in this paper is described in Sect.~\ref{sec:halo_model}.

The \texttt{Shear Profile} module reads in the galaxy and cluster catalogues provided. Then, for each cluster, background source galaxies are selected using a series of cuts in photometric redshift and/or colour as described in Sect.~\ref{sec_back_sel}. The tangential shear profile for each cluster is then derived as described in Sect.~\ref{sec_wl_sign} and provided as an output product of \texttt{COMB-CL}. In Table~\ref{tab_comb_cl_shear_par}, we list the specific argument values passed to the \texttt{Shear Profile} module for this work.

Finally, the \texttt{MCMC Fitting} module performs a Markov chain Monte Carlo (MCMC) analysis to fit the defined halo model to the measured tangential shear profiles to derive weak-lensing mass estimates, as well as other parameters describing the 1-halo term (e.g. concentration) or the 2-halo term (e.g., halo bias), for each of the clusters. The fitting procedure is described in Sect.~\ref{sec:mass_infer}. The specific package employed by \texttt{COMB-CL} to perform the MCMC analysis is \texttt{emcee} \citep{emcee:13}.


\section{\Planck clusters in multiple surveys}
\label{sec_planck_multi}

\begin{figure}
\begin{tabular}{c}
\resizebox{0.9\hsize}{!}{\includegraphics{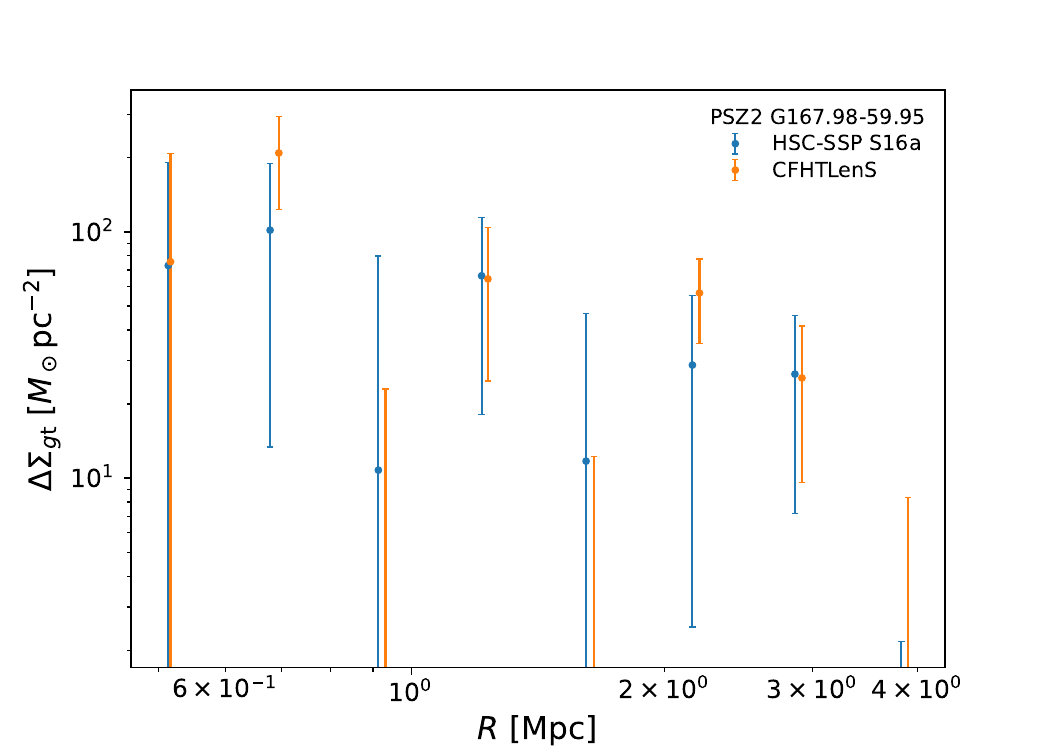}} \\
\noalign{\smallskip}  
\resizebox{0.9\hsize}{!}{\includegraphics{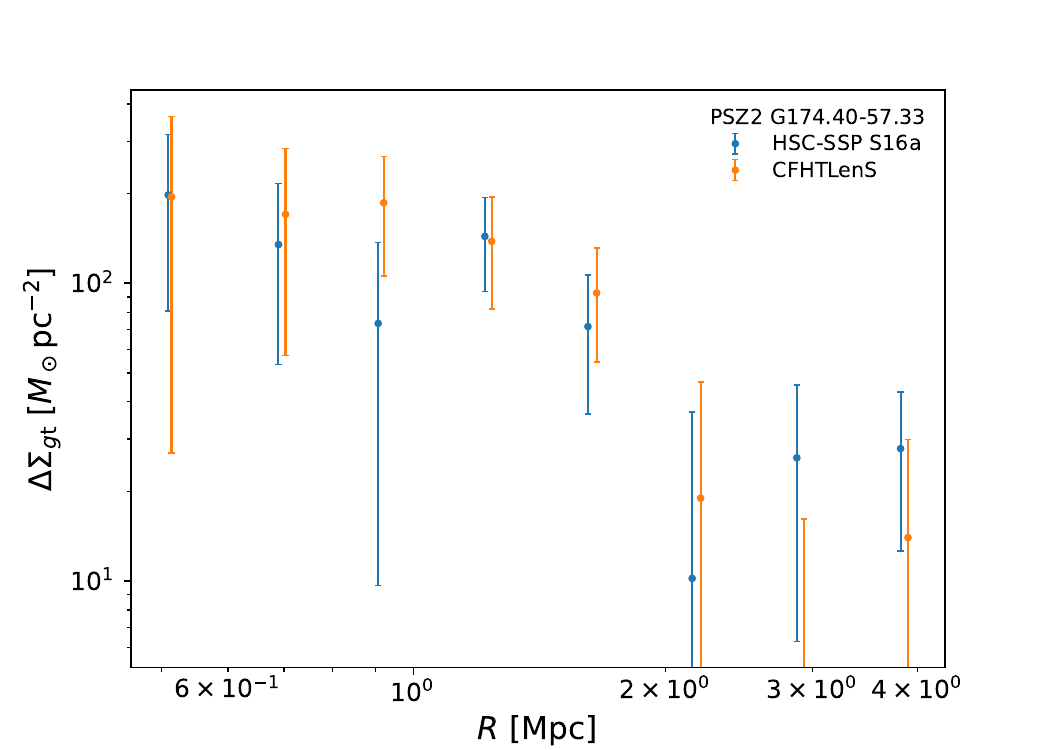}} \\
\noalign{\smallskip}  
\resizebox{0.9\hsize}{!}{\includegraphics{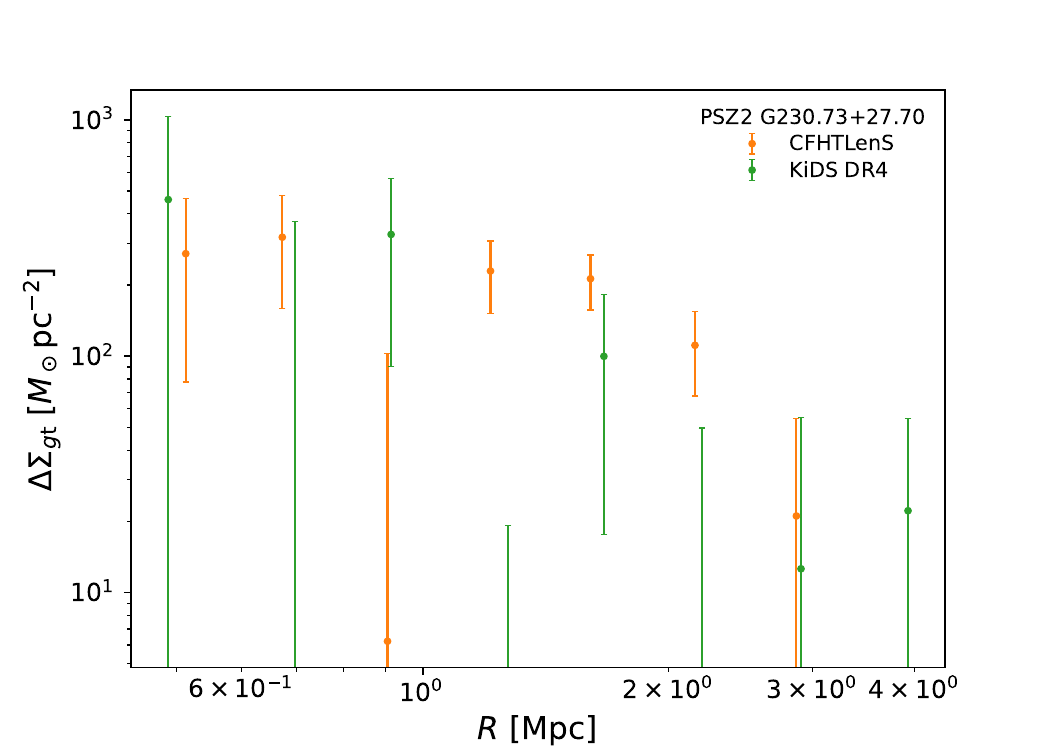}} \\
\noalign{\smallskip}  
\resizebox{0.9\hsize}{!}{\includegraphics{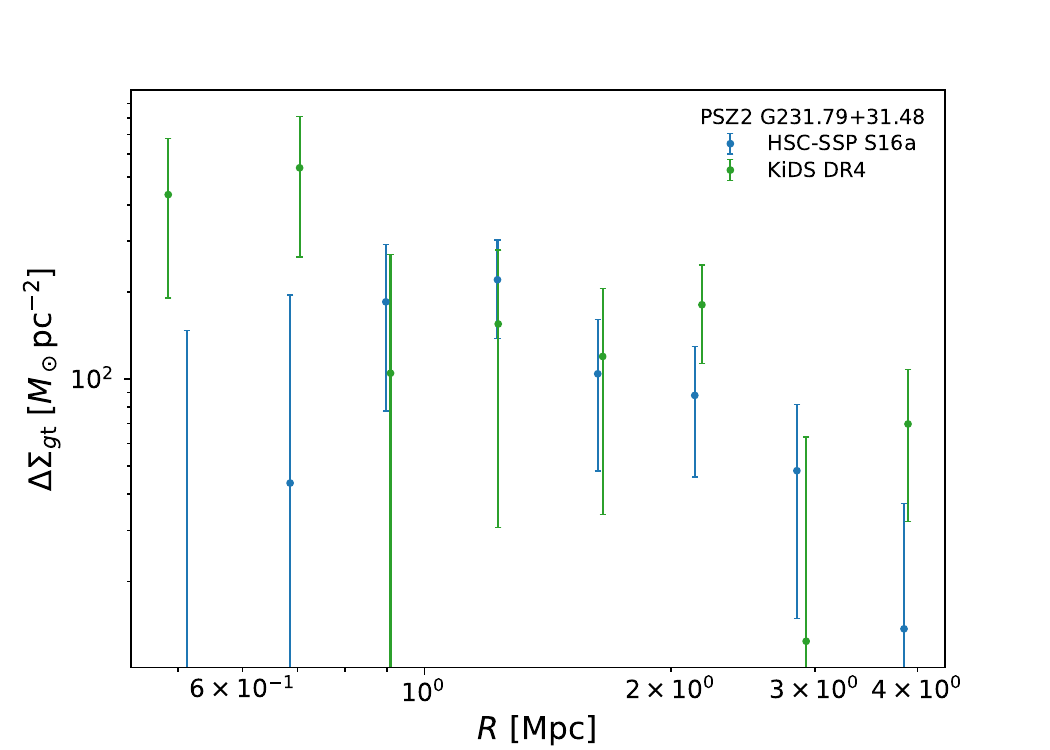}} \\
\end{tabular}
\caption{Reduced excess surface surface density profiles of \Planck clusters detected in multiple public surveys.}
\label{fig_planck_shear_profiles_singles}
\end{figure}

\begin{table}
\caption{\Planck clusters in multiple surveys. Comparison of the shear profiles as derived from survey $a$ (col. 2) or $b$ (col. 3); $\chi^2_{a,b}$ is the $\chi^2$ between the two profiles; $p_{a,b} = p\left(\chi^2 > \chi_{a,b}^2\right)$ is the probability to exceed this value.
}
\label{tab_she_pro_planck_single}
\centering
\resizebox{\hsize}{!} {
\begin{tabular}[c]{l  l l r r}
	\hline
	\noalign{\smallskip}  
	Cluster				&	survey a &	survey b 	&	 $\chi^2_{a,b}$		& $p_{a,b}$  \\
	\noalign{\smallskip}  
	\hline
	\noalign{\smallskip}  
	PSZ2~G167.98$-$59.95 & HSC-SSP S16a & CFHTLenS & 2.3 & 0.97 \\
	PSZ2~G174.40$-$57.33 & HSC-SSP S16a & CFHTLenS & 3.0 & 0.93 \\
	PSZ2~G230.73$+$27.70 & CFHTLenS & KiDS DR4 & 14.7 & 0.07 \\
	PSZ2~G231.79$+$31.48 & HSC-SSP S16a & KiDS DR4 & 8.6 &0.38 \\

	\hline
	\end{tabular}
	}
\end{table}

For four PSZ2 clusters, we retrieve data from two surveys. The measured reduced excess surface density profiles are shown in Fig.~\ref{fig_planck_shear_profiles_singles}, whereas the statistical comparison is summarised in Table~\ref{tab_she_pro_planck_single}. We find no evidence for disagreement.


\section{Shear calibration}
\label{sec_she_cal}

\begin{table}
\caption{
Systematic error budget on the shear calibration for WL cluster analyses in Stage-III surveys or precursors. Cluster samples are listed in column 1. The references and the comparison surveys are listed in cols. 2, and 3, respectively. The number of clusters in the matched catalogue is reported in col.~4. Differences between the reduced excess surface density, in percents, are reported either considering the same weights for both the reference and the comparison sample (col.~5), or using the respective survey weights (col.~6).
}
\label{tab_sys_shear}
\centering
\resizebox{\hsize}{!} {
\begin{tabular}[c]{l l l r r@{$\,\pm\,$}l r@{$\,\pm\,$}l}
\hline
\noalign{\smallskip}  
Cluster sample	&Survey	 ref.	&	Survey comp. &	$N_{\rm cl}$	&\multicolumn{2}{c}{weights ref}		& 	\multicolumn{2}{c}{weights comp}		\\
\hline
\noalign{\smallskip} 
CAMIRA-HSC-16A  &	HSC-SSP S16a	& 	KiDS DR4	&	$918$		&	$-9$	&  	$10$ & $-11$ & $10$	    \\
AMICO-KiDS-DR3	&	HSC-SSP S16a	& 	KiDS DR4	&	$1677$	 &	$-6$		&  	$10$ & $-16$ & $10$   \\
AMICO-HSC-16A-XMM	&	HSC-SSP S16a	& 	CFHTLenS	&	$2210$ &	$-3$	&  	$17$  &   $-9$  & $17$		\\
\hline
	\end{tabular}
}
\end{table}

\begin{figure}
\resizebox{\hsize}{!}{\includegraphics{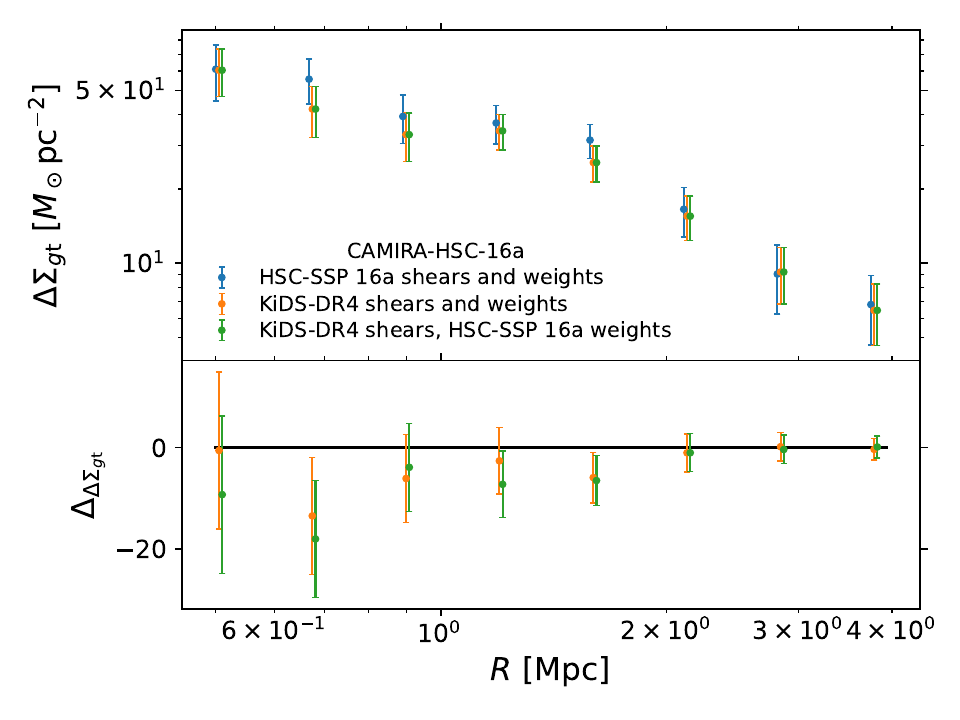}}
\caption{
Average reduced excess surface density profiles of clusters detected by CAMIRA in the fields covered by HSC-SSP S16a and KiDS DR4. Top panel: average profiles. Bottom: differences.}
\label{fig_camira_average_profiles}
\end{figure}

\begin{figure}
\resizebox{\hsize}{!}{\includegraphics{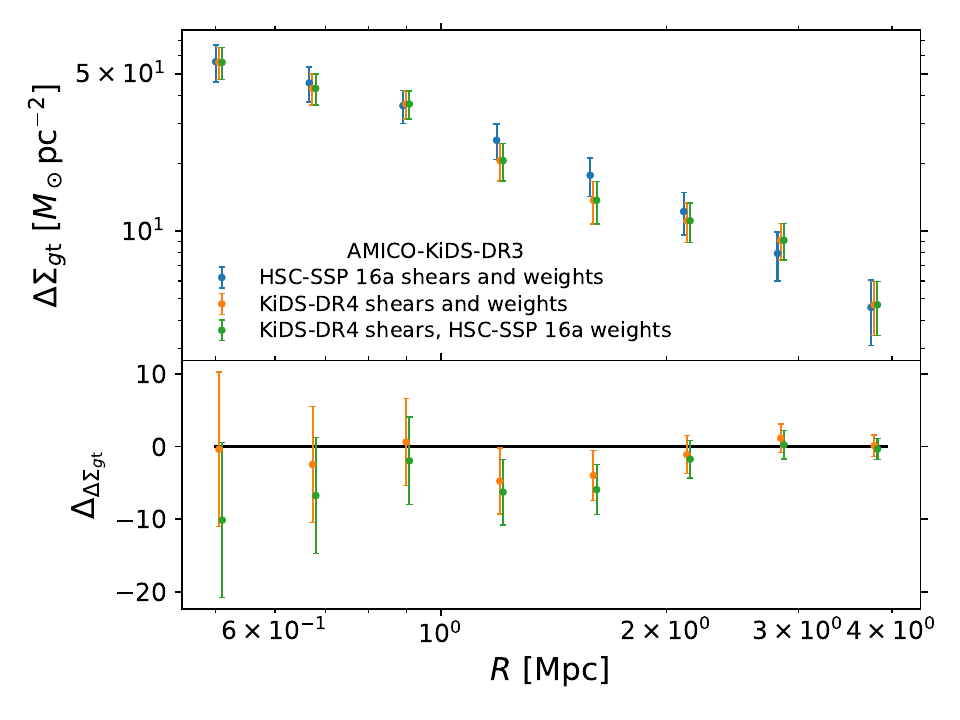}}
\caption{
Same as Fig.~\ref{fig_camira_average_profiles} but for AMICO-KiDS-DR3 clusters.}
\label{fig_amico_kids_dr3_average_profiles}
\end{figure}

\begin{figure}
\resizebox{\hsize}{!}{\includegraphics{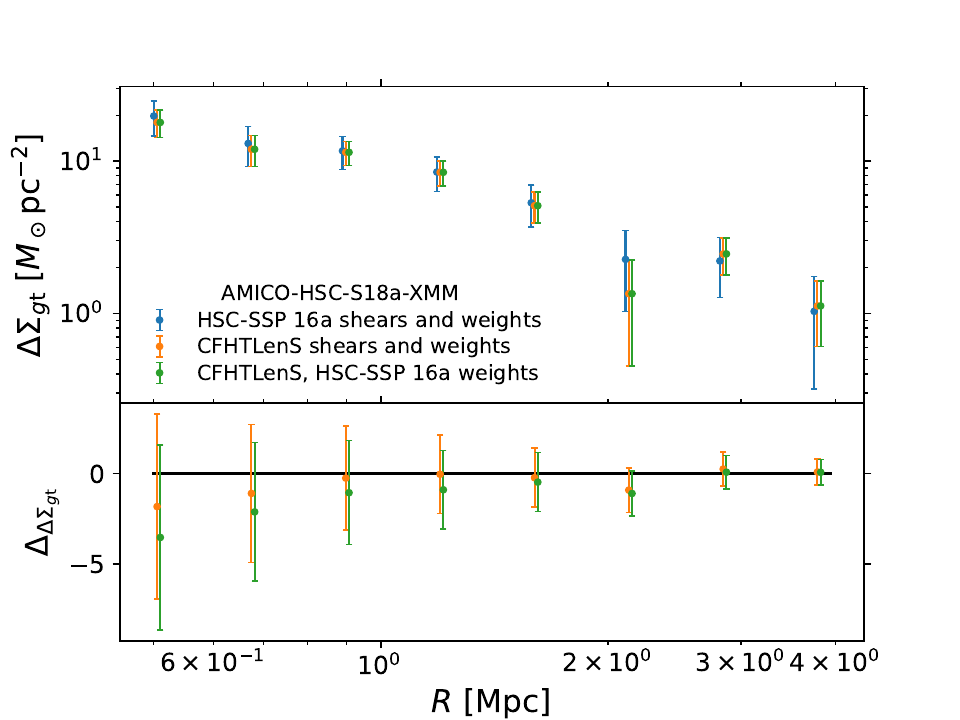}}
\caption{
Average reduced excess surface density profiles of clusters detected by AMICO in the XXL North field covered by HSC-SSP S16a and CFHTLenS. Top panel: average profiles. Bottom: differences.}
\label{fig_amico_hsc_s18a_xmm_average_profiles}
\end{figure}

Shear calibration is regarded as one of the main systematic effects for lensing and we can examine its effect by cross-checking shear estimates. We compare measurements from overlapping survey areas. Direct comparison of ellipticity estimates obtained with different resolution and/or in different band passes might be misleading, as the spatial distribution of the light emission may not be identical and different effective radial weight functions may be used \citep{sch+al18b}. However, the estimated reduced tangential cluster shear profile should be consistent when a matched shear catalogue is used \citep{sch+al18b}. 

In contrast with the main analysis in the paper, for this test we consider cluster samples detected from the same data-sets used for shear measurements. This maximises the relevant difference in signal. Source selection is performed based on the photo-$z$ or colours of the reference survey. Any difference in signal between the samples can be only due to differently measured and calibrated galaxy shear.

We compare ellipticity measurements from a reference survey with a comparison survey under two different schemes. We use either identical weights from the reference survey, or we apply each respective survey weights consistently for that survey. Results are summarised in Table~\ref{tab_sys_shear}. Agreement between HSC-SSP S16a and KiDS DR4, see Figs.~\ref{fig_camira_average_profiles},~\ref{fig_amico_kids_dr3_average_profiles}, or CHFTLenS, see Fig.~\ref{fig_amico_hsc_s18a_xmm_average_profiles}, is good, with differences consistent with zero within the statistical uncertainty.



\end{document}